\documentclass[ALICE,manyauthors]{cernphprep}
\usepackage[comma,square,numbers,sort&compress]{natbib}
\usepackage{hyperref}
\usepackage{cleveref} 
\usepackage{lineno}
\usepackage{xspace}
\usepackage[nohyperlinks, nolist]{acronym}
\usepackage{rotating, graphicx}
\usepackage{placeins}
\usepackage{color}
\usepackage{booktabs} 
\usepackage{isomath}
\usepackage{mathtools}
\usepackage{xargs}
\usepackage{afterpage}
\numberwithin{equation}{section}
\usepackage[T1]{fontenc}
\usepackage{orcidlink}
\usepackage{relsize}

\begin{document}
%

\newcommand{\pp}           {pp\xspace}
\newcommand{\ppbar}        {\mbox{$\mathrm {p\overline{p}}$}\xspace}
\newcommand{\XeXe}         {\mbox{Xe--Xe}\xspace}
\newcommand{\PbPb}         {\mbox{Pb--Pb}\xspace}
\newcommand{\pA}           {\mbox{pA}\xspace}
\newcommand{\pPb}          {\mbox{p--Pb}\xspace}
\newcommand{\AuAu}         {\mbox{Au--Au}\xspace}
\newcommand{\dAu}          {\mbox{d--Au}\xspace}

\newcommand{\s}            {\ensuremath{\sqrt{s}}\xspace}
\newcommand{\snn}          {\ensuremath{\sqrt{s_{\mathrm{NN}}}}\xspace}
\newcommand{\pt}           {\ensuremath{p_{\rm T}}\xspace}
\newcommand{\meanpt}       {$\langle p_{\mathrm{T}}\rangle$\xspace}
\newcommand{\ycms}         {\ensuremath{y_{\rm CMS}}\xspace}
\newcommand{\ylab}         {\ensuremath{y_{\rm lab}}\xspace}
\newcommand{\etarange}[1]  {\mbox{$\left | \eta \right |~<~#1$}}
\newcommand{\yrange}[1]    {\mbox{$\left | y \right |~<~#1$}}
\newcommand{\dndy}         {\ensuremath{\mathrm{d}N_\mathrm{ch}/\mathrm{d}y}\xspace}
\newcommand{\dndeta}       {\ensuremath{\mathrm{d}N_\mathrm{ch}/\mathrm{d}\eta}\xspace}
\newcommand{\avdndeta}     {\ensuremath{\langle\dndeta\rangle}\xspace}
\newcommand{\dNdy}         {\ensuremath{\mathrm{d}N_\mathrm{ch}/\mathrm{d}y}\xspace}
\newcommand{\Npart}        {\ensuremath{N_\mathrm{part}}\xspace}
\newcommand{\Ncoll}        {\ensuremath{N_\mathrm{coll}}\xspace}
\newcommand{\dEdx}         {\ensuremath{\textrm{d}E/\textrm{d}x}\xspace}
\newcommand{\RpPb}         {\ensuremath{R_{\rm pPb}}\xspace}

\newcommand{\nineH}        {$\sqrt{s}$~=~0.9~Te\kern-.1emV\xspace}
\newcommand{\seven}        {$\sqrt{s}$~=~7~Te\kern-.1emV\xspace}
\newcommand{\eight}        {$\sqrt{s}$~=~8~Te\kern-.1emV\xspace}

\newcommand{\twoH}         {$\sqrt{s}$~=~0.2~Te\kern-.1emV\xspace}
\newcommand{\twosevensix}  {$\sqrt{s}$~=~2.76~Te\kern-.1emV\xspace}
\newcommand{\five}         {$\sqrt{s}$~=~5.02~Te\kern-.1emV\xspace}
\newcommand{\sT}           {$\sqrt{s}$~=~13~Te\kern-.1emV\xspace}
\newcommand{\twosevensixnn}{$\sqrt{s_{\mathrm{NN}}}~=~2.76$~Te\kern-.1emV\xspace}
\newcommand{\fivenn}       {$\sqrt{s_{\mathrm{NN}}}~=~5.02$~Te\kern-.1emV\xspace}
\newcommand{\LT}           {L{\'e}vy-Tsallis\xspace}
\newcommand{\GeVc}         {\ensuremath{\text{Ge\kern-.1emV}/c}\xspace}
\newcommand{\MeVc}         {Me\kern-.1emV/$c$\xspace}
\newcommand{\TeV}          {Te\kern-.1emV\xspace}
\newcommand{\GeV}          {Ge\kern-.1emV\xspace}
\newcommand{\MeV}          {Me\kern-.1emV\xspace}
\newcommand{\GeVmass}      {Ge\kern-.1emV/$c^2$\xspace}
\newcommand{\MeVmass}      {Me\kern-.1emV/$c^2$\xspace}
\newcommand{\lumi}         {\ensuremath{\mathcal{L}}\xspace}

\newcommand{\ITS}          {\rm{ITS}\xspace}
\newcommand{\TOF}          {\rm{TOF}\xspace}
\newcommand{\ZDC}          {\rm{ZDC}\xspace}
\newcommand{\ZDCs}         {\rm{ZDCs}\xspace}
\newcommand{\ZNA}          {\rm{ZNA}\xspace}
\newcommand{\ZNC}          {\rm{ZNC}\xspace}
\newcommand{\SPD}          {\rm{SPD}\xspace}
\newcommand{\SDD}          {\rm{SDD}\xspace}
\newcommand{\SSD}          {\rm{SSD}\xspace}
\newcommand{\TPC}          {\rm{TPC}\xspace}
\newcommand{\TRD}          {\rm{TRD}\xspace}
\newcommand{\VZERO}        {\rm{V0}\xspace}
\newcommand{\VZEROA}       {\rm{V0A}\xspace}
\newcommand{\VZEROC}       {\rm{V0C}\xspace}
\newcommand{\TZERO}        {\rm{T0}\xspace}
\newcommand{\Vdecay} 	   {\ensuremath{\text{V}^{\text{0}}}\xspace}

\newcommand{\ee}           {\ensuremath{\text{e}^{+}\text{e}^{-}}\xspace} 
\newcommand{\pip}          {\ensuremath{\rm\pi^{+}}\xspace}
\newcommand{\pim}          {\ensuremath{\rm\pi^{-}}\xspace}
\newcommand{\kap}          {\ensuremath{\rm{K}^{+}}\xspace}
\newcommand{\kam}          {\ensuremath{\rm{K}^{-}}\xspace}
\newcommand{\pbar}         {\ensuremath{\rm\overline{p}}\xspace}
\newcommand{\kzero}        {\ensuremath{{\rm K}^{0}_{\rm{S}}}\xspace}
\newcommand{\kzerol}        {\ensuremath{{\rm K}^{0}_{\rm{L}}}\xspace}
\newcommand{\lmb}          {\ensuremath{\rm\Lambda}\xspace}
\newcommand{\almb}         {\ensuremath{\rm\overline{\Lambda}}\xspace}
\newcommand{\Om}           {\ensuremath{\rm\Omega^-}\xspace}
\newcommand{\Mo}           {\ensuremath{\rm\overline{\Omega}^+}\xspace}
\newcommand{\X}            {\ensuremath{\rm\Xi^-}\xspace}
\newcommand{\Ix}           {\ensuremath{\rm\overline{\Xi}^+}\xspace}
\newcommand{\Xis}          {\ensuremath{\rm\Xi^{\pm}}\xspace}
\newcommand{\Oms}          {\ensuremath{\rm\Omega^{\pm}}\xspace}
\newcommand{\degree}       {\ensuremath{^{\rm o}}\xspace}


\newcommand{\pT}{\ensuremath{p_{\mbox{\tiny T}}}\xspace}
\newcommand{\qT}{\ensuremath{q_{\mbox{\tiny T}}}\xspace}
\newcommand{\xT}{\ensuremath{x_{\mbox{\tiny T}}}\xspace}
\newcommand{\mT}{\ensuremath{m_{\mbox{\tiny T}}}\xspace}
\newcommand{\minv}{\ensuremath{m_{\mbox{\tiny inv}}}\xspace}

\newcommand{\piz}{${\pi^{0}}$\xspace}
\newcommand{\et}{$\eta$\xspace}
\newcommand{\om}{$\omega$\xspace}
\newcommand{\g}{$\gamma$\xspace}

\newcommand{\pTPiz}{$p_{\mbox{\tiny T, $\pi^{0}$}}$\xspace}
\newcommand{\pTEta}{$p_{\mbox{\tiny T, $\eta$}}$\xspace}

\newcommand{\etopi}{$\eta/\pi^{0}$\xspace}

\newcommand{\pioneq}{\pi^{\text{\tiny 0}}}
\newcommand{\pTeq}{p_{\mbox{\tiny T}}}
\newcommand{\ShowerShape}  {\ensuremath{\sigma^{2}_{\rm long}}\xspace}

\newcommand{\AACol}{{\mbox{AA}}\xspace}
\newcommand{\Vo}{\ensuremath{\text{V}^{\text{0}}}\xspace}

\newcommand{\RConv}{R_{\mbox{\tiny conv}}}

\newcommand{\pTg}{$p_{\mbox{\tiny T}}^{\gamma}$\xspace}

\newcommand{\ZConv}{Z_{\mbox{\tiny conv}}}

\newcommand{ \la }{\langle}
\newcommand{ \ra }{\rangle}

\newcommand{\Figure}[1]{\hyperref[#1]{Figure~\ref*{#1}}}
\newcommand{\Fig}[1]{\hyperref[#1]{Fig.~\ref*{#1}}}
\newcommand{\Figures}[2]{\hyperref[#1]{Figure~\ref*{#1}} and \hyperref[#2]{\ref*{#2}}}
\newcommand{\Figs}[2]{\hyperref[#1]{Figs.~\ref*{#1}} and \hyperref[#2]{\ref*{#2}}}
\newcommand{\Section}[1]{\hyperref[#1]{Section~\ref*{#1}}}
\newcommand{\Sec}[1]{\hyperref[#1]{Sec.~\ref*{#1}}}
\newcommand{\Sections}[2]{\hyperref[#1]{Sections~\ref*{#1}} and \hyperref[#2]{\ref*{#2}}}
\newcommand{\Chapter}[1]{\hyperref[#1]{Chapter~\ref*{#1}}}
\newcommand{\App}[1]{\hyperref[#1]{App.~\ref*{#1}}}
\newcommand{\Appendix}[1]{\hyperref[#1]{Appendix~\ref*{#1}}}
\newcommand{\Table}[1]{\hyperref[#1]{Table~\ref*{#1}}}
\newcommand{\Tab}[1]{\hyperref[#1]{Tab.~\ref*{#1}}}
\newcommand{\TabsThree}[3]{\hyperref[#1]{Tab.~\ref*{#1}}, \hyperref[#2]{\ref*{#2}}, and \hyperref[#3]{\ref*{#3}}}
\newcommand{\Tables}[2]{\hyperref[#1]{Table~\ref*{#1}} and \hyperref[#2]{\ref*{#2}}}
\newcommand{\TablesT}[2]{\hyperref[#1]{Tables~\ref*{#1}}-{\ref*{#2}}}
\newcommand{\TablesThree}[3]{\hyperref[#1]{Tables~\ref*{#1}}, \hyperref[#2]{\ref*{#2}}, and \hyperref[#3]{\ref*{#3}}}
\newcommand{\Equation}[1]{\hyperref[#1]{Equation~\ref*{#1}}}
\newcommand{\Eq}[1]{\hyperref[#1]{Eq.~\ref*{#1}}}
\newcommand{\Eqs}[2]{\hyperref[#1]{Eq.~\ref*{#1}} and \hyperref[#2]{\ref*{#2}}}

	\counterwithout{figure}{subsubsection}
	
	\begin{titlepage}
		\PHyear{2024}       
		\PHnumber{304}      
		\PHdate{13 November}  
		
		
		\title{Light neutral-meson production in pp collisions at $\mathbf{\sqrt{s}}$ = 13 \TeV}
		
		\ShortTitle{Neutral-meson production in pp at $\sqrt{s}$ = 13 \TeV}   
		
		\Collaboration{ALICE Collaboration\thanks{See Appendix~\ref{app:collab} for the list of collaboration members}}
		\ShortAuthor{ALICE Collaboration} 
		
		\begin{abstract}
			The momentum-differential invariant cross~sections of \piz and \et mesons are reported for \pp collisions at \sT at midrapidity ($|y|$~$<$~0.8). The measurement is performed in a broad transverse-momentum range of 0.2~$<$~\pT~$<$~200~\GeVc and 0.4~$<$~\pT~$<$~60~\GeVc for the \piz and \et, respectively, extending the \pT coverage of previous measurements. Transverse-mass-scaling violation of up to 60\% at low transverse momentum has been observed, agreeing with measurements at lower collision energies. Transverse Bjorken $x$ (\xT) scaling of the \piz cross sections at LHC energies is fulfilled with a power-law exponent of $n$ = 5.01$\pm$0.05, consistent with values obtained for charged pions at similar collision energies.
			The data are compared to predictions from next-to-leading order perturbative QCD calculations, where the \piz spectrum is best described using the CT18 parton distribution function and the NNFF1.0 or BDSS fragmentation function.
			Expectations from PYTHIA8 and EPOS LHC overestimate the spectrum for the \piz and are not able to describe the shape and magnitude of the \et spectrum.
			The charged-particle multiplicity dependent \piz and \et \pT~spectra show the expected change of the spectral shape, characterized by a flatter slope with increasing multiplicity. This is demonstrated across a broad transverse-momentum range and up to events with a charged-particle multiplicity exceeding five times the mean value in minimum bias collisions.
			The \etopi~ratio depends on the charged-particle multiplicity for \pT~$<$~4~\GeVc. PYTHIA8 and EPOS LHC qualitatively explain this behavior with an increasing contribution from the feed-down of heavier particles to the \piz spectrum.
		\end{abstract}
	\end{titlepage}
	
	\setcounter{page}{2} 

	\begin{acronym}
		\acro{LHC}[LHC]{Large Hadron Collider}
		\acro{V0}[V0]{Neutral particle decay into two charged particles}
		\acro{ALICE}[ALICE]{A Large Ion Collider Experiment}
		\acro{ATLAS}[ATLAS]{A Toroidal LHC ApparatuS}
		\acro{CMS}[CMS]{Compact-Muon-Solenoid}
		\acro{LHCb}[LHCb]{Large Hadron Collider beauty}
		\acro{PHOS}[PHOS]{Photon Spectrometer}
		\acro{QA}[QA]{Quality Assurance}
		\acro{PID}[PID]{Particle Identification}
		\acro{QCD}[QCD]{Quantum Chromo Dynamics}
		\acro{QGP}[QGP]{quark--gluon plasma}
		\acro{CERN}[CERN]{European Council for Nuclear Research}
		\acro{MB}[MB]{minimum bias}
		\acro{MC}[MC]{Monte Carlo}
		\acrodef{MCSim}[MC simulation]{Monte Carlo simulation}
		\acro{pT}[\pT]{transverse momentum}
		\acro{IR}[IR]{Interaction Rate}
		\acro{TOF}[TOF]{Time of Flight}
		\acro{SPD}[SPD]{Silicon Pixel Detector}
		\acro{SDD}[SDD]{Silicon Drift Detector}
		\acro{SSD}[SSD]{Silicon Micro-Strip Detector}
		\acro{ITS}[ITS]{Inner Tracking System}
		\acro{TPC}[TPC]{Time Projection Chamber}
		\acro{TRD}[TRD]{Transition Radiation Detector}
		\acro{MWPC}[MWPC]{Multi Wire Proportional Chamber}
		\acro{DCA}[DCA]{Distance of Closest Approach}
		\acro{CPV}[CPV]{Charged Particle Veto}
		\acro{TRD}[TRD]{Transition Radiation Detector}
		\acro{EMC}[EMCal]{Electromagnetic Ca\-lo\-ri\-me\-ter}
		\acro{mEMC}[mEMC]{merged EMCal technique}
		\acro{mPHOS}[mPHOS]{merged PHOS technique}
		\acro{DC}[DCal]{Di-Jet Calorimeter}
		\acro{PHOS}[PHOS]{Photon Spectrometer}
		\acro{SM}[SM]{Super module}
		\acro{PCM}[PCM]{Photon Conversion Method}
		\acro{KF}[KF]{Kalman Filter}
		\acro{pyt}[PYTHIA8]{ PYTHIA8 Monash 2013 Tune}
		\acro{TCM}[TCM]{two-component model}
	\end{acronym}

	\section{Introduction}
	Over the last few decades, our understanding of particle production in high-energy hadronic collisions has increased significantly due to the experimental results obtained from the CERN Intersection Storage Rings (ISR), the CERN Super Proton-Antiproton Synchrotron (Sp$\bar{\mathrm{p}}$S), Tevatron at Fermilab, RHIC and the LHC~\cite{Jacob:156665,UA5:1987rzq,UA1:1989bou,CDF:2009cxa,CDF:1988evs,STAR:2006xud,PHENIX:2007kqm,PHENIX:2008sgl,PHENIX:2011rvu,Levin:2010dw,Moriggi:2020zbv,dEnterria:2011twh,dEnterria:2013sgr,ALICE:2022wpn,LHCb:2021vww,LHCb:2022tjh},
	as well as the ongoing development of theoretical and phenomenological models~\cite{Bierlich:2022pfr,Brambilla:2014jmp,Pierog:2013ria,Fedynitch:2115393,nCTEQ:2021,Hou:2019jgw,deFlorian:2014xna,Bertone:2017tyb}.
	Theoretical models typically separate the particle production into the soft and hard regime, which describe processes with a small and large momentum transfer, respectively.
	Hard-scattering processes can be calculated using perturbative quantum chromodynamics (pQCD). These calculations rely on input from parton distribution functions (PDF) (f($x$)) and fragmentation functions (FF) (D($z$)), where Bjorken $x$ represents the fraction of the proton's longitudinal momentum carried by a parton and $z$ is the fraction of the final-state hadron momentum to the parton momentum. As \piz and \et mesons are among the most abundant particles produced at LHC energies, with average production rates of $\text{d}N_{\pi^{0}}/\text{d}y\rvert_{y \approx 0} \approx 2.5$ and $\text{d}N_{\eta}/\text{d}y\rvert_{y \approx 0} \approx 0.2$~\cite{ALICE:2017ryd,ALICE:2020jsh}, measurements of these mesons provide valuable constraints on these models~\cite{FF_BDSS,deFlorian:2014xna}. With increasing collision energy, the measurement of a final-state hadron at fixed transverse momentum probes smaller and smaller values of $x$, where the gluon contribution becomes dominant~\cite{nCTEQ:2021}. Hence, the measurement of the neutral meson production cross section at \sT gives further insights into the gluon to meson fragmentation. In addition, measurements of the \et meson allow the investigation of a possible dependence of FFs on hidden strangeness~\cite{Aidala:2010bn}. A precise comparison of the differences in the hadronization process between two particles can be achieved by investigating the ratios of the production cross section of identified particles, such as the \etopi~ratio. As the initial state of the collision is identical for both particles, the ratio is primarily sensitive to the effects arising from the differences in the parton-to-hadron fragmentation. 
	
	Global analyses based on collections of experimental data are used to determine, and regularly update, PDFs~\cite{Hou:2019jgw} and FFs~\cite{deFlorian:2014xna,Bertone:2017tyb}.
	For example, first neutral pion measurements at the LHC~\cite{ALICE:2012wos} are included in the global analysis of parton-to-pion fragmentation functions~\cite{deFlorian:2014xna}, or 
	similarly, the parton-to-pion FF reported in~\cite{Bertone:2017tyb} include the \acs{ALICE} charged pion measurements. 
	On the other hand, global analyses of the \et FF do not yet include data from the LHC~\cite{Aidala:2010bn},
	although \et meson measurements are already available at various LHC collision energies in wide transverse momentum (\pT) ranges~\cite{ALICE:2012wos,Abelev:2014ypa,Acharya:2017hyu,ALICE:2017ryd,ALICE:2018vhm,ALICE:2018mdl,ALICE:2021est,LHCb_Eta}.
	At low transverse momenta, where the production cross section of \piz and \et mesons has its maximum, particle production is driven by soft processes. Details of the description of particle production at low \pT, which is not calculable perturbatively and relies on phenomenological models, can be further improved by comparing experimental data with theoretical models and event generators such as PYTHIA8~\cite{Pythia:2015} or EPOS~\cite{EPOS:2005}, which are tuned to data from \ee collisions as well as to early LHC data, depending on the generator. For PYTHIA8, the Monash 2013 tune~\cite{PythiaMonash:2014} is commonly used at LHC energies. The PYTHIA8 Ropes variant~\cite{PythiaRopes:2015,PythiaRopes2:2015} was recently used to describe effects arising in high-multiplicity pp collisions. While PYTHIA8 relies solely on string fragmentation, EPOS is based on a model exploiting multiple scattering with pomerons, effectively assuming the formation of a \ac{QGP}. The EPOS LHC tune~\cite{Pierog:2013ria} is based on early LHC data.
	In addition, improved knowledge of the different processes involved in particle production and more constrained parameters in hadronic models are of great importance in astrophysics to achieve a deeper understanding of ultra-high energy cosmic ray physics~\cite{dEnterria:2011twh}.
	
	Furthermore, precise knowledge of the neutral meson production cross section is important for analyses of rare probes, including direct photons, dielectrons or electrons from heavy flavour decays. 
	Due to the large abundance of the \piz and \et mesons in hadronic collisions, their decay photons account for more than 97\% of all decay photons.
	Therefore, a high-precision measurement of the \piz and \et cross section is mandatory to attempt a direct photon measurement in pp collisions, especially at low transverse momenta where a thermal photon signal, possibly produced by a \ac{QGP} droplet, is expected to be of the order of less than 2\% compared to the decay photon background
	\cite{ALICE:2018mjj,Shen:2016zpp,ALICE:dirG13}.
	
	Phenomenological scaling models are often used to predict \pT~spectra for particles for which there is no exact measurement at a given center-of-mass energy in the desired \pT or rapidity interval. Transverse mass (\mT = $\sqrt{m^{2}_{0} + p_{\text{\tiny T}}^{2}}$, with $m_{0}$ being the rest mass of a given particle) scaling is typically used to obtain the \pT spectrum of a heavy particle, taking the \pT spectrum of a lighter particle, measured in the same collision system, as an input~\cite{Ren:2021pzi,Altenkamper:2017qot}. However, a violation of this scaling, especially at low transverse momenta, is reported in~\cite{Agakishiev:1998mw} and this violation was confirmed in~\cite{ALICE:2012wos,Abelev:2014ypa,Acharya:2017hyu,ALICE:2017ryd,ALICE:2018vhm,ALICE:2018mdl,Altenkamper:2017qot,Ren:2021pzi}. Furthermore, measurements of identified charged particles~\cite{ALICE:2020jsh} also show \mT scaling violation. In this work, we extend the studies of the validity of \mT scaling to neutral mesons in pp collisions at \sT. 
	Transverse Bjorken $x$ scaling (\xT scaling, with \xT~$= 2p_{\mbox{\tiny T}}/\sqrt{s}$ at midrapidity where $y~\approx~0$), can be used to predict hadron spectra for \pT $\gtrapprox$~3~\GeVc at collision energies where no measurement is available yet. This scaling relies on the power-law behavior of particle spectra in ultra-relativistic collisions at high transverse momenta~\cite{Arleo:2009ch,Arleo:2010kw,Sassot:2010bh}. This scaling was verified by experimental data at the Tevatron~\cite{CDF:2009cxa,CDF:1988evs}, RHIC~\cite{STAR:2006xud,PHENIX:2007kqm,PHENIX:2008sgl,PHENIX:2011rvu}, CERN Sp$\bar{\mathrm{p}}$S~\cite{UA1:1989bou}, and CERN LHC energies
	\cite{CMS:2011mry,ALICE:2020jsh}. 
	However, scaling violations are expected due to the running of $\alpha_{\mathrm S}$ and the scale evolution of PDFs and FFs. 
	The broad \pT coverage of neutral meson measurements at the LHC allows a test of the \xT scaling over a very large \xT range.
	
	Further understanding of the underlying particle-production mechanisms can be obtained by analyzing the \pT~spectra of identified particles as a function of the event charged--particle multiplicity per pseudorapidity interval ($\text{d}N_{\text{ch}}/\text{d}\eta$). Recent studies at LHC energies in pp, p--Pb, and Pb--Pb collisions have revealed a smooth transition from small to large collision systems as a function of $\text{d}N_{\text{ch}}/\text{d}\eta$ for observables such as strangeness enhancement, elliptic flow, and modifications in the meson to baryon ratio~\cite{Dusling:2015gta,Nagle:2018nvi,ALICE:2019Mul7,ALICE:2014Mul5}. These results suggest a common underlying mechanism, defining the chemical composition of the produced particles in all collision systems. The dependence of \pT~spectra and particle ratios on $\text{d}N_{\text{ch}}/\text{d}\eta$ at LHC energies was studied for most of the light flavor hadrons~\cite{Dusling:2015gta,Nagle:2018nvi,ALICE:2019Mul7,ALICE:2014Mul5} and has not yet been published for neutral mesons. Hence, this article provides the first constraints for the dependence of neutral meson production on the charged--particle multiplicity. These results offer valuable input for the tuning of phenomenological models and \ac{MC} generators, as such constraints are not accessible through inclusive spectra alone.

	In this article, the measurements of the \piz and \et meson production cross section for inelastic collisions at \sT with ALICE are reported. These cross sections include all primary produced \piz and \et, including those from feed down from strong and electromagnetic decays, but excluding those from weak decays~\cite{alicePrimPart}. This is especially important for the \piz, where the fraction of feed-down from strong decays ranges between about 80\% for $\pt~<~1~\GeVc$ to  40\% for $\pt~>~10~\GeVc$~\cite{Altenkamper:2017qot}. Tests of \mT and \xT scaling are performed including measurements at lower collision energies. Moreover, the dependence of light neutral-meson production on the event charged-particle multiplicity is reported.
	Results are compared to pQCD calculations and to PYTHIA8 and EPOS LHC predictions. 
	This article is organized as follows: the detectors relevant to the measurement are described in \Sec{sec:exp}; details of the event selection and data samples are given in  \Sec{sec:EventSelection}; the analysis methods are explained in \Sec{sec:analysis}, followed by a summary of the systematic uncertainties evaluation in \Sec{sec:sys};
	the results, as well as comparison to theoretical models are presented and discussed in \Sec{sec:results}; finally, the conclusions of the paper are given in \Sec{sec:conclusions}.

	\section{Experimental setup}
	\label{sec:exp}
	
	The \piz and \et mesons are reconstructed via their decays into two photons, \piz~(\et)~$\rightarrow$~\g\g, with a branching ratio of $\text{BR}$~=~98.823~$\pm$~0.034\% and $\text{BR}$~=~39.41~$\pm$~0.20\%, respectively, and via their Dalitz decay, \mbox{\piz~(\et)~$\rightarrow$~\g\g$^*$~$\rightarrow$\g~e$^{+}$e$^{-}$}, with a branching ratio of $\text{BR}$~=~1.174~$\pm$~0.035\% and $\text{BR}$~=~0.69~$\pm$~0.04\%, respectively~\cite{ParticleDataGroup:2024cfk}. 
	The reconstruction of the real photons is done using three fully independent reconstruction techniques: via energy deposits in the \ac{EMC}~\cite{ALICE:2022wpn}, or in the \ac{PHOS}~\cite{ALICE:2022qhn}, and by using the \ac{PCM} utilizing \ee pairs from converted photons reconstructed with the tracking detectors.
	The detectors relevant to this measurement, including tracking detectors and calorimeters, are briefly described in this section.
	These detectors are situated inside the L3 magnet which provides a homogeneous magnetic field of B~=~0.5~T or B~=~0.2~T.
	A detailed description of the ALICE experiment during Run 1 and Run 2 and its performance can be found in~\cite{Aamodt:2008zz,Abelev:2014ffa,ALICE:2022wpn}.
	
	The \ac{ITS}~\cite{Aamodt:2010aa} consists of six tracking layers, covering the full azimuthal angle ($\varphi$) and a pseudorapidity range of at least $|\eta| < 0.9$. Its main purpose is the precise estimation of the collision point, referred to as the primary vertex in the following.
	Additionally, the \ac{ITS} information is used for pileup rejection and \ac{PID} utilizing the specific energy loss (\dEdx).
	The two innermost layers consist of the \ac{SPD}, followed by two layers of the \ac{SDD}, and two layers of the \ac{SSD}.
	The \ac{SDD} was absent in parts of the B = 0.2 T field data taking to limit the total dead time of \acs{ALICE} and, therefore, maximize the number of collected events.
	
	The \ac{TPC}~\cite{Alme:2010ke} consists of a large cylindrical drift volume that covers a pseudorapidity range of $|\eta| < 0.9$ and $2\pi$ azimuthal angle.
	The TPC allows for the reconstruction of the momentum for charged particles as well as providing \ac{PID} based on \dEdx.
	In the analysis reported in this article, the \ac{ITS} and \ac{TPC} are used to reconstruct photons converting in the detector material. The total material budget in the pseudorapidity range $|\eta| < 0.9$ up to $R~=$~180~cm, including the material of the beam pipe, the \ac{ITS} and the \ac{TPC} is (11.9 $\pm$ 0.3)\% in units of radiation lengths ($X/X_{\mathrm{0}}$)~\cite{ALICE:MBW}; $R$ is calculated in the transverse plane to the beam axis. The data-driven calibration of the ALICE material budget~\cite{ALICE:MBW} is applied in the presented analysis, reducing the systematic uncertainty on the material budget from 4.5\% to 2.5\% per photon.

	The \ac{TRD}~\cite{ALICE:2017ymw} has a modular structure and its basic component is a multiwire proportional chamber (MWPC). 
	It consists of 522 chambers arranged in 6 layers surrounding the TPC in full azimuth at a radial distance of 2.90 m to 3.68 m from the interaction point, and along the longitudinal direction in 5 stacks covering the pseudorapidity interval $|\eta| <$ 0.84.  A drift region of 3 cm precedes each chamber to allow the reconstruction of a local track segment, which can be used for matching TRD information with tracks reconstructed with ITS and TPC, or TPC only.
	
	The \ac{EMC} detector is a sampling calorimeter used for photon and electron detection, as well as for event triggering. It consists of 17664 individual cells arranged in ten full-sized, six 2/3-sized, and four 1/3-sized supermodules, covering $|\eta| <$ 0.7 for 80$^{\circ} < \varphi <$ 187$^{\circ}$ and 260$^{\circ} < \varphi < 327^{\circ}$ with the exception of the PHOS hole ( $|\eta| <$ 0.22 for $260^{\circ} < \varphi < 320^{\circ}$). The cells have a size of about 6 $\times$ 6 cm$^{2}$ corresponding to a coverage of $\Delta \eta \times \Delta \varphi$ $\approx$ 0.0143 $\times$ 0.0143. Each cell consists of 77 alternating layers of lead absorber and plastic scintillator with a total depth of about 20 radiation lenghts ($X_{0}$). The energy resolution of the \ac{EMC} is characterized by $\sigma_E /E = 2.9\%/E \oplus 9.5\%/ \sqrt{E} \oplus 1.4\%$ with the energy $E$ in units of \GeV. 
	A detailed description of the \ac{EMC} and its performance can be found in 
	\cite{Cortese:1121574,Allen:1272952,ALICE:2022qhn}.

	The \ac{PHOS} is a homogeneous electromagnetic calorimeter made of lead tungstate crystals (PbWO$_4$)~\cite{PHOS_TDR}.
	It covers a pseudorapidity range of $|\eta| < 0.12$ over $\Delta \varphi~=~70^{\circ}$ in azimuth.
	The detector is segmented into cells with $2.2 \times 2.2$ cm$^2$  corresponding to a coverage of $\Delta \eta \times \Delta \varphi$ $\approx$ 0.0048 $\times$ 0.0048. The high granularity and homogeneous design results in a better energy resolution compared to the \ac{EMC}. The energy resolution of the \ac{PHOS} can be parameterized with $\sigma_E~/E~=~1.3\%/E~\oplus~3.6\%/~\sqrt{E}~\oplus~1.1\%$, where $E$ is in units of \GeV. The total depth of the \ac{PHOS} is about 20 $X_{0}$.
	Details of the PHOS and its performance are described in~\cite{ALEKSANDROV2005169,PHOS_TDR,ALICE:2019cox}.

	The V0 detectors consist of two plastic scintillator arrays (V0A and V0C, referred to as V0M in combination) covering $2.8~<~\eta~<~5.1$ and $-3.7~<~\eta~<~-1.7$, respectively. The detectors provide a fast charged-particle multiplicity measurement in the forward region and as such are used to provide the \ac{MB} trigger, requiring a hit in both the V0A and V0C, in coincidence with a bunch crossing,  as well as a multiplicity trigger. Additionally, the V0 detector is used as a multiplicity estimator in the forward and backward regions.

	\section{Event selection and data sample}
	\label{sec:EventSelection}
	The data used in the analysis were collected in pp collisions at \sT from 2016 to 2018.
	While the majority of the data was recorded with a magnetic field of B\,=\,0.5\,T, a minimum-bias dataset with B\,=\,0.2\,T was additionally taken in three dedicated data-taking periods, one per year.
	Only events that fulfill the \ac{MB} trigger condition are used in the analysis.
	Pileup events, where more than one collision occurs in the same bunch crossing, are rejected using the SPD layers to identify multiple vertices~\cite{Abelev:2014ffa} as well as by rejecting events based on the correlation between the number of clusters and tracklets reconstructed with the SPD layers. To reduce the fraction of out-of-bunch pileup, only collisions where the neighboring 4 bunch crossings, occurring every 25 ns for most of the data taking, do not contain a triggered collision are considered in the analysis.
	Additionally, only events with a reconstructed $z$-vertex position of $|z| <$~10~cm with respect to the intended collision point are accepted in the analysis.
	
	A dedicated high-multiplicity trigger was used during most of the B\,=\,0.5\,T data taking to select pp collisions at the highest multiplicities. It triggers on a fixed amplitude in the V0 detector system corresponding to events with the 0.17\% highest charged-particle multiplicities within the V0 detector acceptance. Only events within the saturation region of the trigger are selected for the analysis.
	
	\begin{figure}[b]
		\centering
		\includegraphics[width=0.5\textwidth]{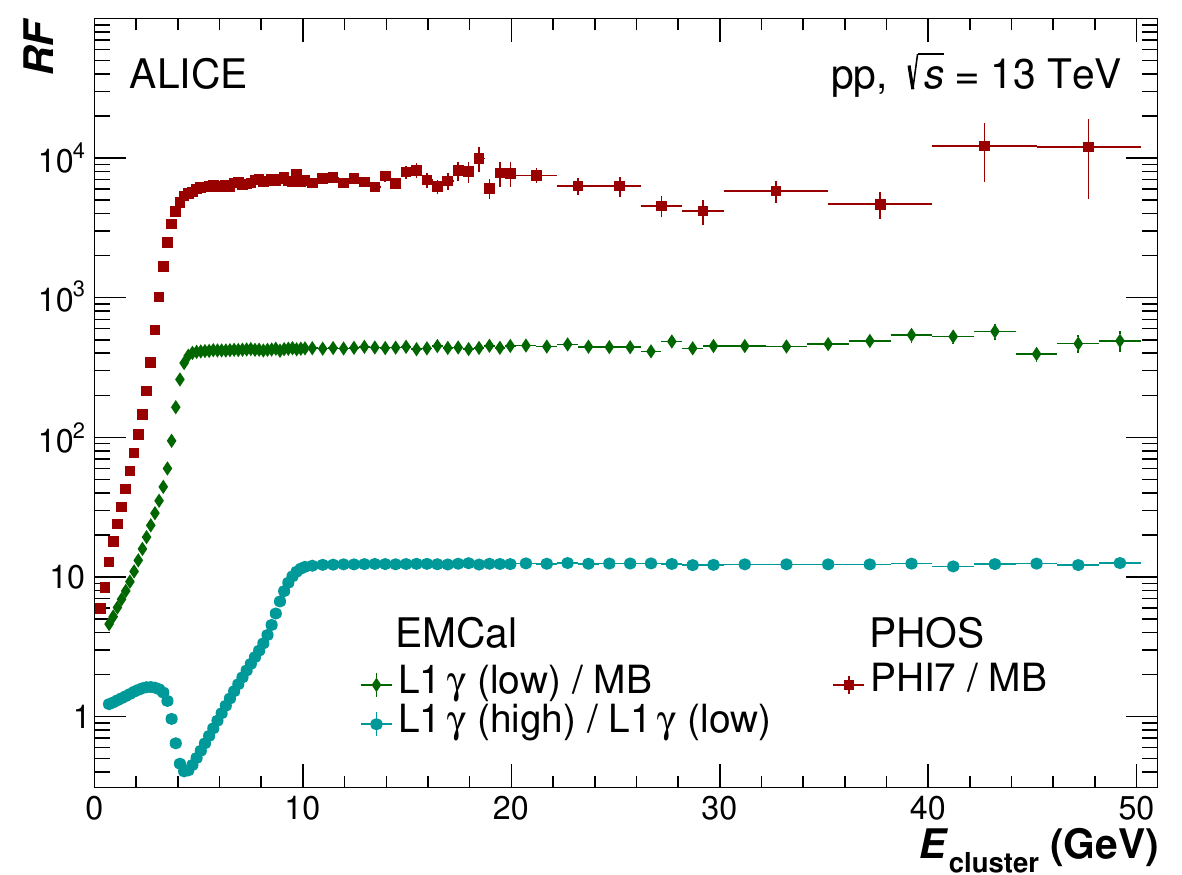}
		\caption{Trigger rejection factors for the \ac{EMC}-L1 $\gamma$ (low) (green), \ac{EMC}-L1 $\gamma$ (high) (cyan) and the PHOS-PHI7 (red) trigger as a function of the cluster energy. The dip in the \ac{EMC}-L1 $\gamma$ (high)/\ac{EMC}-L1 $\gamma$ (low) ratio arises due to the trigger turn-on of the \ac{EMC}-L1 $\gamma$ (low) trigger of about $4$~\GeV as explained in \Sec{sec:EventSelection}.}
		\label{fig:RF}
	\end{figure}

	To enhance the spectrum at high transverse momenta, the \ac{EMC} and \ac{PHOS} detectors can provide their own level 0 (L0) trigger while the EMCal additionally provides a set of level 1 (L1) triggers. The L0 trigger is set to an approximate threshold of $E^{\text{thr}}_{\text{L0, EMCal}}~\approx~$2.5~\GeV for the \ac{EMC} and $E^{\text{thr}}_{\text{L0, PHOS}}~\approx~$4~\GeV for the \ac{PHOS}, referred to as PHI7 trigger in the following. The \ac{EMC} provides L1 triggers for single particles (L1-$\gamma$) for which two different thresholds were configured during the data taking: $E^{\text{thr}}_{\text{L1} \gamma\text{-low}} \approx $~4~\GeV for the lower threshold trigger and $E^{\text{thr}}_{\text{L1} \gamma\text{-high}} \approx $~9~\GeV for the higher threshold trigger. A detailed description of the trigger system of the \ac{EMC}, as well as its performance, is given in~\cite{ALICE:2022qhn}.
	The enhancement achieved with these triggers compared to minimum bias collisions can be extracted from the ratio of the cluster energy ($E_{\text{cluster}}$) (compare \Sec{sec:PhotonSelection}) spectra normalized per event recorded with the \ac{EMC} and \ac{PHOS} calorimeter triggers to the normalized cluster spectra obtained from minimum bias collisions with the respective calorimeter, as shown in \Fig{fig:RF}. For the \ac{EMC}-L1 $\gamma$ (high) trigger, the ratio to the spectra of the \ac{EMC}-L1 $\gamma$ (low) triggered clusters was chosen in order to minimize statistical fluctuations. 
	The trigger turn-on is not sharp but smeared out as the cluster energy can differ from the input to the trigger logic~\cite{ALICE:2022qhn}.
	Above the trigger threshold, the ratio is not constant but exhibits a slight slope, which has been found to be an effect of acceptance holes from dead or masked trigger regions. To extract the trigger enhancement factors, the ratios shown in \Fig{fig:RF} are corrected for acceptance effects using a trigger emulation in the \ac{MC} simulation, after which a constant fit to the plateau region gives the trigger efficiency-corrected enhancement factor.

	\begin{table}[t!]
		\begin{minipage}[t]{.62\textwidth}
			\vspace{-\topskip}
			\centering
			\caption{Number of events recorded for each event class together with the corresponding integrated luminosity. Except for parts of the minimum bias trigger class, the data was recorded at B~=~0.5~T.}
			\begin{tabular}{l | l  l}
				\toprule
				Trigger class & $N_{\text{evt}}$ ($10^{6}$) & $\mathcal{L}_{\text{int}}$ \\ \hline
				Minimum bias (B~=~0.5~T) & 1516 & 26.23 $\pm$ 0.42 nb$^{-1}$ \\
				Minimum bias (B~=~0.2~T) & 581 & 10.05 $\pm$ 0.16 nb$^{-1}$ \\  
				\ac{EMC}-L1 $\gamma$ (low) & 116 & 0.84 $\pm$ 0.03 pb$^{-1}$ \\
				\ac{EMC}-L1 $\gamma$ (high) & 90.5 & 8.24 $\pm$ 0.26 pb$^{-1}$ \\
				\ac{PHOS} PHI7 & 48.6 & 9.93 $\pm$ 0.60 pb$^{-1}$ \\
				V0M high multiplicity & 261 & 7.09 $\pm$ 0.39 pb$^{-1}$
				\\ \bottomrule
			\end{tabular}
			\label{tab:events}
		\end{minipage}%
		\hspace{0.3cm}
		\begin{minipage}[t]{.36\textwidth}
			\vspace{-\topskip}
			\centering
			\caption{Definition of multiplicity classes used in the presented analysis together with the corresponding mean charged-particle multiplicity measured in $|\eta| < 0.5$~\cite{PseudoRapMult}.}
			\begin{tabular}{c|c}
				\toprule
				$\Delta\sigma/\sigma_{\text{MB}_{\text{AND}>\text{0}}}$ &  $\langle\text{d}N_{\text{ch}}/\text{d}\eta\rangle_{|\eta| < 0.5}$\\ \hline
				0--0.01\%         &  $35.82 \pm 0.47$\\
				0.01--0.05\%         &  $32.21 \pm 0.41$\\
				0.05--0.1\%         &  $ 30.13\pm 0.38$\\
				0--1\%         &  $26.01 \pm 0.34$\\
				1--5\%         &  $19.99 \pm 0.24$\\
				5--10\%         & $16.18 \pm 0.20$ \\
				10--20\%         &  $12.90 \pm 0.17$\\
				20--30\%         &  $10.03 \pm 0.13$\\
				30--50\%         &  $7.14 \pm 0.10$\\
				50--70\%         &  $4.49 \pm 0.06$\\
				70--100\%         &  $2.54 \pm 0.04$\\ \midrule
				0--100\%         &  $6.93 \pm 0.09$\\ \bottomrule
			\end{tabular}
			\label{tab:multClass}
		\end{minipage}%
	\end{table}

	The number of analyzed MB events as well as the corresponding integrated luminosity ($\mathcal{L}_{\text{int}}$) for both magnetic field configurations, together with the high-multiplicity triggered data sample in the 0--0.1\% multiplicity class, and the calorimeter-triggered samples are listed in \Tab{tab:events}. Events that are used for the multiplicity-dependent analyses require one charged particle in the ALICE acceptance of $|\eta| < 1$ (INEL$>$0). At least one SPD tracklet is required in these events in order to approximate this condition in the data. The multiplicity is determined using the integrated amplitude of the V0A and V0C detectors~\cite{ALICE:2013axi}. 
	\Table{tab:multClass} summarizes the multiplicity classes, given as the fraction of the visible cross~section of the \ac{MB} trigger ($\Delta\sigma/\sigma_{\text{MB}_{\text{AND}>\text{0}}}$), used in the analysis together with the corresponding mean charged-particle density $\langle\text{d}N_{\text{ch}}/\text{d}\eta\rangle_{|\eta| < 0.5}$ taken from~\cite{PseudoRapMult}. 
	
	\section{Analysis method}
	\label{sec:analysis}
	
	\subsection{Photon and virtual photon reconstruction}
	\label{sec:PhotonSelection}

	Photons and electrons hitting an electromagnetic calorimeter produce an electromagnetic shower that typically spreads over multiple cells. To retrieve the full energy of the original particle, the energy of adjacent cells is combined using a clusterization algorithm as described in~\cite{ALICE:2022qhn} for the \ac{EMC} and in~\cite{PHOS_TDR} for the \ac{PHOS}. The clusterization thresholds for the seed $E_{\rm{seed}}$ and aggregation $E_{\rm{agg}}$ cell energy are based on previous studies~\cite{ALICE:2022qhn,ALICE:2019cox} and are given in  \Tab{tab:CutCalo}.
	The calibration procedures of the cells and the resulting clusters of the \ac{EMC} and \ac{PHOS} are given in ~\cite{ALICE:2022qhn,ALICE:2019cox}. 
	To select clusters originating from photons, several selection criteria, listed in \Tab{tab:CutCalo}, are applied. 
	The selection criteria for \ac{EMC} and \ac{PHOS} clusters have the same motivation; however, the values differ because of differences in electronic noise, cell sizes, and the materials used in the two calorimeters.
	A minimum cluster energy $E_{\text{cluster}}$ is required to minimize contributions from hadronic clusters and electronic noise. The number of cells $N_{\text{cell}}$ is only used to select clusters for \ac{PHOS} above $E_{\text{cluster}} >$~1~\GeV. Below that energy threshold, a significant fraction of the photon clusters are expected to consist of only one cell.
	Out-of-bunch pileup is rejected by selecting clusters within a strict time window around the selected collision. Losses due to the width of the cluster-time ($t_{\text{cluster}}$) distribution are negligible for \ac{EMC}; however, a sizable loss for \ac{PHOS} at low and high cluster energies was found. This effect was emulated in the \ac{MC} simulation in order to be corrected. Furthermore, the purity of \g clusters is enhanced by rejecting clusters likely to be produced from charged particles. This is accomplished through the use of a geometrical track-cluster matching approach, comparing the cluster and estimated track position in the $\eta$ and $\varphi$ direction, while additionally considering the disparity between the track momentum $p_{\text{track}}$ and $E_{\text{cluster}}$. Only matched clusters that satisfy the condition $E_{\text{cluster}}/p_{\text{track}}~<$~1.75\,$c$ are rejected. The cluster shape $\sigma^{2}_{\mathrm{long}}$ is used to further reduce the hadronic background as well as to differentiate between clusters consisting of one \g and clusters consisting of multiple \g originating mainly from the two-photon \piz decays at high transverse momenta.
	The parameter \ShowerShape is defined as the larger eigenvalue of the covariance matrix of a cluster's energy distribution. For the \ac{EMC}, the covariance matrix is calculated in the $\eta$ and $\varphi$ directions~\cite{ALICE:2022qhn}, while for the \ac{PHOS}, it is evaluated in the $x$ and $z$ directions on the front plane~\cite{PHOS_TDR}. Consequently, \ShowerShape is dimensionless for the \ac{EMC} but measured in cm$^2$ for the \ac{PHOS}.

	\begin{table}[t!]
		\centering
		\caption{Collection of selection criteria applied to \ac{EMC} and \ac{PHOS} clusters to select photons and merged \piz candidates~\cite{ALICE:2022qhn,ALICE:2019cox}. For the charged particle veto $p_{\mbox{\tiny T, track}}$  is given in units of \GeVc. }
		\begin{tabular}{l | l l}
			\toprule
			&  EMCal & PHOS \\  \hline
			Seed threshold  &  $E_{\rm{seed}}$ $>$ 300 \MeV   & $E_{\rm{seed}}$ $>$ 50 \MeV \\
			Aggregation threshold  &  $E_{\rm{agg}}$ $>$ 100 \MeV   &  $E_{\rm{agg}}$ $>$ 15 \MeV \\
			Cluster energy  & $E_{\text{cluster}}$ $>$ 700 \MeV   &    $E_{\text{cluster}}$ $>$ 300 \MeV \\
			
			Number of cells  & $N_{\text{cell}}$ $\geq$ 1 &  $N_{\text{cell}}$ $\geq$ 1 \\
			&  & $N_{\text{cell}}$ $\geq$ 2 for $E_{\text{cluster}} >$ 1 \GeV  \\
			Cluster time  & -20 $< t_{\text{cluster}} < $ 25 ns  &   -30  $< t_{\text{cluster}} < $ 30 ns\\
			Cluster shape $\gamma$  &   0.1 $<$ $\sigma^{2}_{\rm{long}} < $ 0.7 &  $ \sigma^{2}_{\rm{long}}>$ 0.1 \\
			Cluster shape merged \piz  &   $\sigma^{2}_{\rm{long}} > $ 0.27 &   $\sigma^{2}_{\rm{long}} > $ $1.2+35.7/(\frac{E_{\text{cluster}}}{\text{GeV}}-12.9)$ \\
			Charged particle veto   &  $|\Delta\eta| < 0.01 + (p_{\mbox{\tiny T, track}} +4.07)^{-2.5}$ &  $|\Delta\eta| < 0.01 + (p_{\mbox{\tiny T, track}} +4.37)^{-2.5}$  \\
			& $|\Delta\varphi| < 0.015 + (p_{\mbox{\tiny T, track}} +3.65)^{-2}$ & $|\Delta\varphi| < 0.015 + (p_{\mbox{\tiny T, track}} +3.78)^{-2}$\\
			& $E_{\text{cluster}}/p_{\text{track}} < 1.75\,c$   &   \\
			\bottomrule
		\end{tabular}
		\label{tab:CutCalo}
	\end{table}

	\begin{table}[t!]
		\centering 
		\caption{Selection criteria of the converted photon reconstruction with \ac{PCM}. }
		\begin{tabular}{p{3.8cm}|c|c}
			\toprule
			\multicolumn{1}{c|}{} & B = 0.5 T & B = 0.2 T \\ 
			\hline
			
			\multicolumn{1}{l}{\textbf{Track reconstruction }} & \multicolumn{2}{c}{} \\ \hline
			$|\eta|$    & $< 0.8$ & same\\
			\pT       &  $> 0.05$~\GeVc  &  \pT $> 0.02$~\GeVc \\ 
			$N_{\mbox{\tiny clusters}}^{\mbox{\tiny TPC}}/N_{\mbox{\tiny findable clusters}}^{\mbox{\tiny TPC}}$ &  $> 60\%$ & same \\ 
			Conversion radius      & $5 <\RConv<$ 55 cm $||$ 72 $< \RConv<180$~cm & same\\
			
			Line cut                & $\RConv > |\ZConv| \times ZR_{\mbox{\tiny Slope}} - Z_{0}$  & same  \\
			&  $ZR_{\mbox{\tiny Slope}} = \tan{(2 \times \arctan(\exp(-\eta_{\mbox{\tiny cut}})))}$ & same  \\
			&  $Z_{0}$~=~7~cm, $\eta_{\mbox{\tiny cut}}$~=~0.8  & same \\ \hline

			\multicolumn{1}{l}{\textbf{Track identification }} & \multicolumn{2}{c}{} \\ \hline
			e$^{\pm}$ selection ($n \sigma_{\text{e}}$ TPC)  & $-3  < n \sigma_{\text{e}} <4$ & same 	 	                \\ 
			$\pi^{\pm}$ rejection ({$n$$\sigma_{\pi}$ TPC})      & $n \sigma_{\pi} <1$ at $0.4<p<3.5$ \GeVc & same \\
			& $n \sigma_{\pi} <0.5$ at $p>3.5$~\GeVc		& same\\
			TRD tracklet or ITS hit  & PCM-$\gamma\gamma$: $\geq$ 1, else: $\geq$ 0 & same  \\ \hline
			\multicolumn{1}{l}{\textbf{Conversion $\gamma$ topology }} & \multicolumn{2}{c}{} \\ \hline
			$q_{\mbox{\tiny T}}$        & $ < q_{\mathrm{T}}^{\tiny \mathrm{MAX}} \cdot \sqrt{1-(\alpha/0.95)^2}$~\GeVc &  same\\ 
			& $q_{\mathrm{T}}^{\tiny \mathrm{MAX}}=$  Min(0.125 \pTg, 0.05) &   $q_{\mathrm{T}}^{\tiny \mathrm{MAX}}=$ Min(0.2 \pTg, 0.035) \\
			$\psi_{\rm pair}$, $\chi^{2}_{\gamma}$  &  $|\psi_{\text{pair}}| < 0.18\cdot \exp(-0.055\cdot\chi^2_{\gamma})$ 
			& $|\psi_{\text{pair}}| < 0.35\cdot \exp(-0.075\cdot\chi^2_{\gamma})$ \\ 
			&$\chi^2_\gamma$/ndf~$<$~50  & same \\
			cos($\theta_{\rm PA}$)  & $>$ 0.85 & same\\ 
			Reject too close \Vo's  & $\Delta R <$\,6 cm \&\& 
			($\ensuremath{\text{V}^{\text{0}}}\sphericalangle) <$\,0.02 rad  &  same \\ \bottomrule
		\end{tabular}
		\label{tab:CutValuesPCM}
	\end{table}

	Photons that convert into \ee\,pairs in the detector material are reconstructed using a \Vo finder method~\cite{Abelev:2014ffa} that pairs secondary oppositely-charged tracks from a common neutral vertex. Charged tracks are reconstructed primarily with the ITS and the TPC. 
	The \Vo candidates comprise \kzero, \lmb, \almb decays and $\gamma$ conversions, as well as random combinations not originating from the same parent particle.
	Different selection criteria were applied for the photon reconstruction: quality of the charged tracks, particle identification, and photon conversion topology. Details on the selection of converted photons and virtual photons can be found in~\cite{Abelev:2014ypa,Acharya:2017hyu,ALICE:2018vhm,ALICE:2012wos,ALICE:2018mdl,ALICE:2017ryd,ALICE:2021est}.
	The complete list of applied selection criteria on the converted photons is summarized in \Tab{tab:CutValuesPCM}.
	Charged tracks are required to be within the kinematic limits of $|\eta| < 0.8$ and $\pT~>~0.05~\GeVc$. Furthermore, they are required to have at least 60\% of the expected track points in the TPC ($N_{\mbox{\tiny clusters}}^{\mbox{\tiny TPC}}/N_{\mbox{\tiny findable clusters}}^{\mbox{\tiny TPC}}$). The position of the neutral vertex in the transverse plane, also referred to as conversion radius ($R_{\mbox{\tiny conv}}$), is restricted to be larger than 5 cm to reject tracks from the primary vertex, and restricted to be smaller than 180 cm in order to ensure the track being reconstructable in the TPC. Tracks assigned to a secondary neutral vertex between 55 cm and 72 cm are rejected to minimize systematic uncertainties as found in the analysis presented in~\cite{ALICE:MBW}. In addition, the longitudinal position of the neutral decay vertex $Z_{\mbox{\tiny conv}}$ is used to restrict $R_{\mbox{\tiny conv}}$ to ensure that the secondary charged tracks are within the geometrical detector limits.
	For tracks that pass the quality criteria, electron selection and pion rejection are performed utilizing the specific energy loss \dEdx in the TPC. Selection and rejection criteria use the number of standard deviations around the expected electron and pion hypothesis ($n\sigma_{\text{e}}$, $n\sigma_\pi$), where $\sigma$ is the standard deviation of the \dEdx measurement. Electron tracks are accepted only if they satisfy the requirement for $n\sigma_{\text{e}}$ while falling outside the expected $\pi^{\pm}$ region $n\sigma_\pi$. Photon selection and further rejection of weak decays are based on
	the $\alpha$--\qT plane known as Armenteros Podolanski plot~\cite{podolanski1954iii}; $\alpha$ is the longitudinal momentum asymmetry between the positive and negative tracks, $\alpha = (p_{\text{L}}^+ - p_{\text{L}}^-)/(p_{\text{L}}^+ + p_{\text{L}}^-)$, and \qT is the transverse momentum of the decay particle with respect to the \Vo momentum. 
	The $\psi_{\rm{pair}}$ angle between the plane that is perpendicular to the magnetic field ($x-y$ plane) and the plane defined by the opening angle of the pair is additionally used to select photon conversions, as they have a preferred emission orientation, in contrast to the distribution for virtual photons of Dalitz decays or random combinations. 
	Furthermore, information on the photon $\chi^{2}_{\gamma}$ of the Kalman filter fit~\cite{KalmanFilter} in combination with the $\psi_{\rm{pair}}$ is used to enhance the photon purity.
	A selection based on the cosine of the pointing angle of cos($\theta_{\rm PA}$) $>$ 0.85 is performed, with $\theta_{\rm PA}$ being the angle between the reconstructed photon momentum vector and the vector connecting the conversion point and the collision vertex. 
	Moreover, it was found that \Vo's with similar conversion points and a small angle $\sphericalangle$ between their momentum axes likely originate from the same photon. To avoid double counting, only the \Vo with the best $\chi^{2}_{\gamma}$ value is considered for the analysis.
	To reduce the out-of-bunch pileup contribution, a TRD tracklet or at least one hit in the ITS ($N_{\text{hits}}^{\mbox{\tiny ITS}}$)~\cite{ALICE:2017ymw} was required for at least one track in the \ac{PCM}-$\gamma\gamma$ analysis. Thus, \Vo's with TPC-only tracks are not used in this analysis. This is especially important for the low multiplicity event class in the multiplicity-dependent analysis, where otherwise a large correction would need to be applied.
	
	\begin{table}[htb!]
		\centering
		\caption{Selection criteria for primary e$^{+}$ (e$^{-}$) tracks and e$^{+}$e$^{-}$ pairs from virtual photons ($\gamma^*$).}
		\begin{tabular}{p{4.3cm}|c|c}
			\toprule 
			Primary $\text{e}^+$ ($\text{e}^-$) & B = 0.5 T & B = 0.2 T \\ \hline 
			\multicolumn{1}{l}{\textbf{Track reconstruction }} & \multicolumn{2}{c}{} \\ \hline
			$|\eta|$   &   0.9  &  same \\
			\pT & $> 0.125$ \GeVc  &   $> 0.05$ \GeVc \\
			DCA$_{\text{xy}}$  & $< 0.105$~mm  $+$ 0.35~mm /$(\frac{p_{\mbox{\tiny T}} \cdot c}{\text{GeV}})^{1.1}$    & same  \\
			DCA$_{\text{z}}$  & $< 2$~cm   & same  \\
			$N_{\text{hits}}^{\mbox{\tiny ITS}}$, $N_{\text{hits}}^{\mbox{\tiny SPD}}$   &   $\geq 2$, $= 2$  &   same \\
			
			$N^{\mbox{\tiny TPC}}_{\mbox{\tiny clusters}}$,   $N^{\mbox{\tiny TPC}}_{\mbox{\tiny clusters}}/N^{\mbox{\tiny TPC}}_{\mbox{\tiny findable clusters}}$ & $\geq 70$, $> 60\%$    &  same \\ 
			$\chi^2_{\mbox{\tiny ITS}}/N_{\rm hits}$  &   $< 36$  &  same \\ 
			$\chi^2_{\mbox{\tiny TPC}}/N_{\rm clusters}$  &   $< 4$  &  same \\ \hline
			
			\multicolumn{1}{l}{\textbf{Track identification }} & \multicolumn{2}{c}{} \\ \hline
			
			e$^{\pm}$ selection  ($n \sigma_{\text{e}}$ TPC)      & $-4  < n \sigma_{\text{e}} <5$    &  same  \\
			$\pi^{\pm}$ rejection ($n \sigma_{\pi} $ TPC)   & $n \sigma_{\pi} <2$ if $0.4<p<3.5$ \GeVc  &  same \\
			& $n \sigma_{\pi} <0.5$ if $p>3.5$ \GeVc  &  same \\ \hline
			\multicolumn{1}{l}{\textbf{\boldmath $\gamma^*$ identification}} (e$^{+}$e$^{-}$ pairs) & \multicolumn{2}{c}{} \\ \hline
			$M_{{\gamma^*}}$    & $<$ 0.015 \GeVmass if \pT $<$1 \GeVc &  same \\
			&  $<$ 0.035 \GeVmass if \pT $>$1 \GeVc &  same \\
			&    &    \\
			$|\psi_{\rm{pair}}|$  &  $<$  0.5~ if $0.00<\Delta \varphi<0.02~\text{rad}$  &   $ <$  0.98 if $0.00<\Delta \varphi<0.02~\text{rad}$     \\
			rejection        &  $<$  0.44 if $0.02<\Delta \varphi<0.04~\text{rad}$  &   $ <$  0.11 if $0.02<\Delta \varphi<0.04~\text{rad}$     \\
			&  $<$  0.07 if $0.04<\Delta \varphi<0.06~\text{rad}$  &      \\ \bottomrule
		\end{tabular}
		\label{tab:CutGstar}
	\end{table}
	
	The virtual photons of the Dalitz decay ($\gamma^*$) are reconstructed from pairs of primary electron and positron tracks. 
	To be considered primary, tracks are required to originate from the primary vertex, with a maximum \ac{DCA} to the primary vertex in the longitudinal direction (DCA$_{\text{z}}< 2$~cm) and in the transverse plane \ac{DCA}$_{\text{xy}} < 0.105$~mm~$+$~0.35 mm /($p_{\mbox{\tiny T}}\,c/\text{GeV}))^{1.1}$. In order to reduce the contamination from photon conversions further, tracks are required to have a hit in each SPD layer ($N_{\text{hits}}^{\mbox{\tiny SPD}}$). 
	The primary tracks are also required to have both ITS and TPC segments with at least 70 \ac{TPC} clusters (with a maximum of 158) and the fraction of TPC clusters to the number of findable clusters has to be larger than 60\%. The fit quality for the ITS and TPC track points should satisfy
	the conditions  $\chi^2 _{\mbox{\tiny ITS}}/N_{\rm hits}^{\mbox{\tiny ITS}} < 36$ and  $\chi^2 _{\mbox{\tiny TPC}}/N_{\rm clusters}^{\mbox{\tiny TPC}}< 4$. Electron candidates are selected using tracks with a $n\sigma_e$ value in the range $-4 < n\sigma_e < 5$ around the electron hypothesis, but are rejected if they are consistent with the $\pi^{\pm}$ hypothesis.
	Pion contamination in the electron sample is further reduced by a selection criterion on the $\gamma^*$ invariant mass $M_{\gamma^*}< 0.015$ \GeVmass and $M_{\gamma^*}< 0.035$ \GeVmass for \pT $<$ 1 \GeVc and  \pT $>$ 1 \GeVc, 
	respectively. Contamination from $\gamma$ conversions in the $\gamma^*$ sample is suppressed by a $\Delta \varphi$ dependent selection on $\psi_{pair}$, and a $\Delta \varphi$ dependent selection on $0<\Delta \varphi<0.12$~rad where $\Delta \varphi = \varphi(\text{e}^+)-\varphi(\text{e}^-) $\cite{ALICE:2018vhm}. The complete list of selection criteria is given in \Tab{tab:CutGstar}.

	\subsection{Neutral meson reconstruction}
	\label{sec:NeutralMesons}
	
\begin{figure}[t!]
		\centering
		\includegraphics[width=0.45\textwidth]{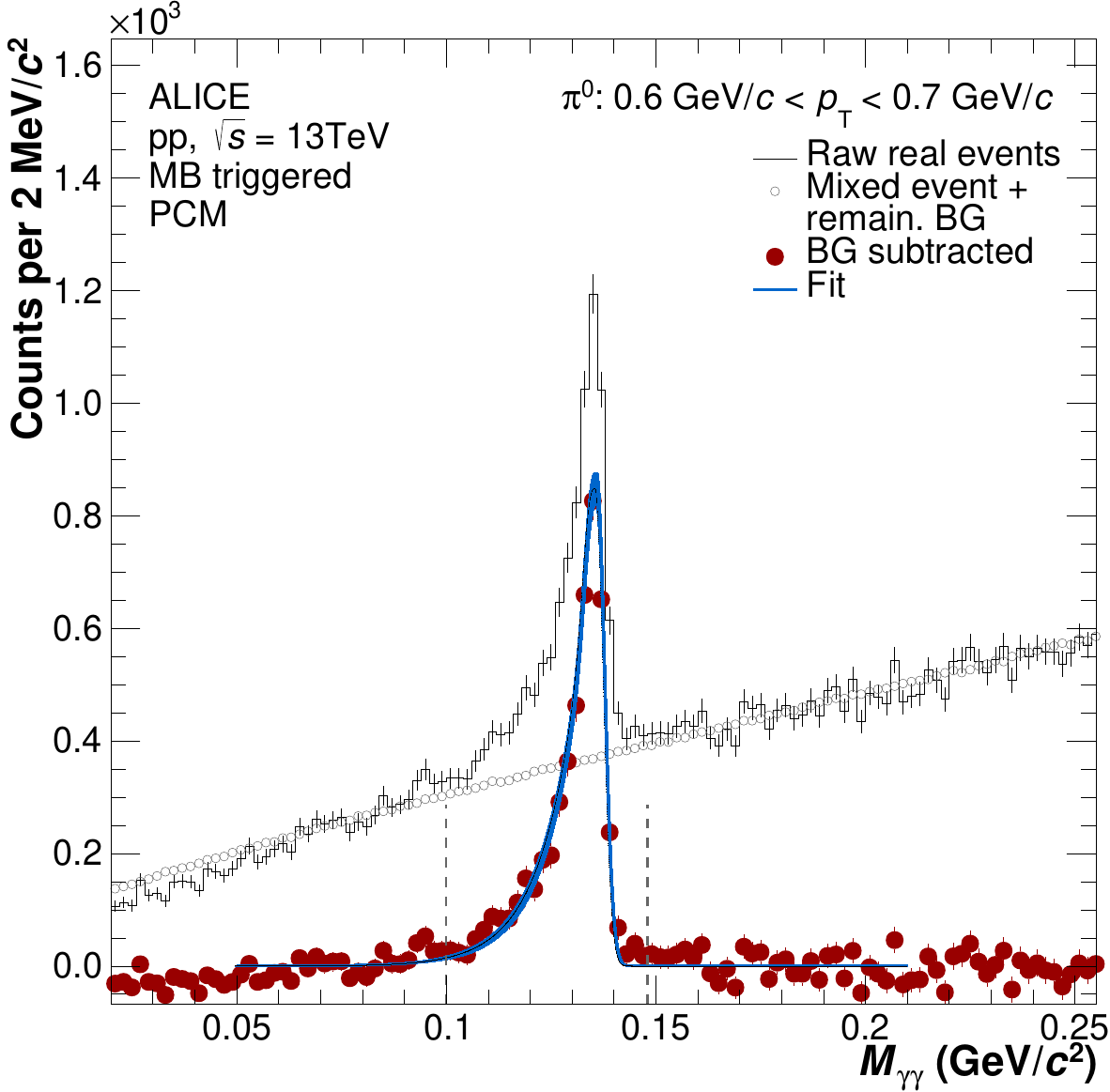}
		\includegraphics[width=0.45\textwidth]{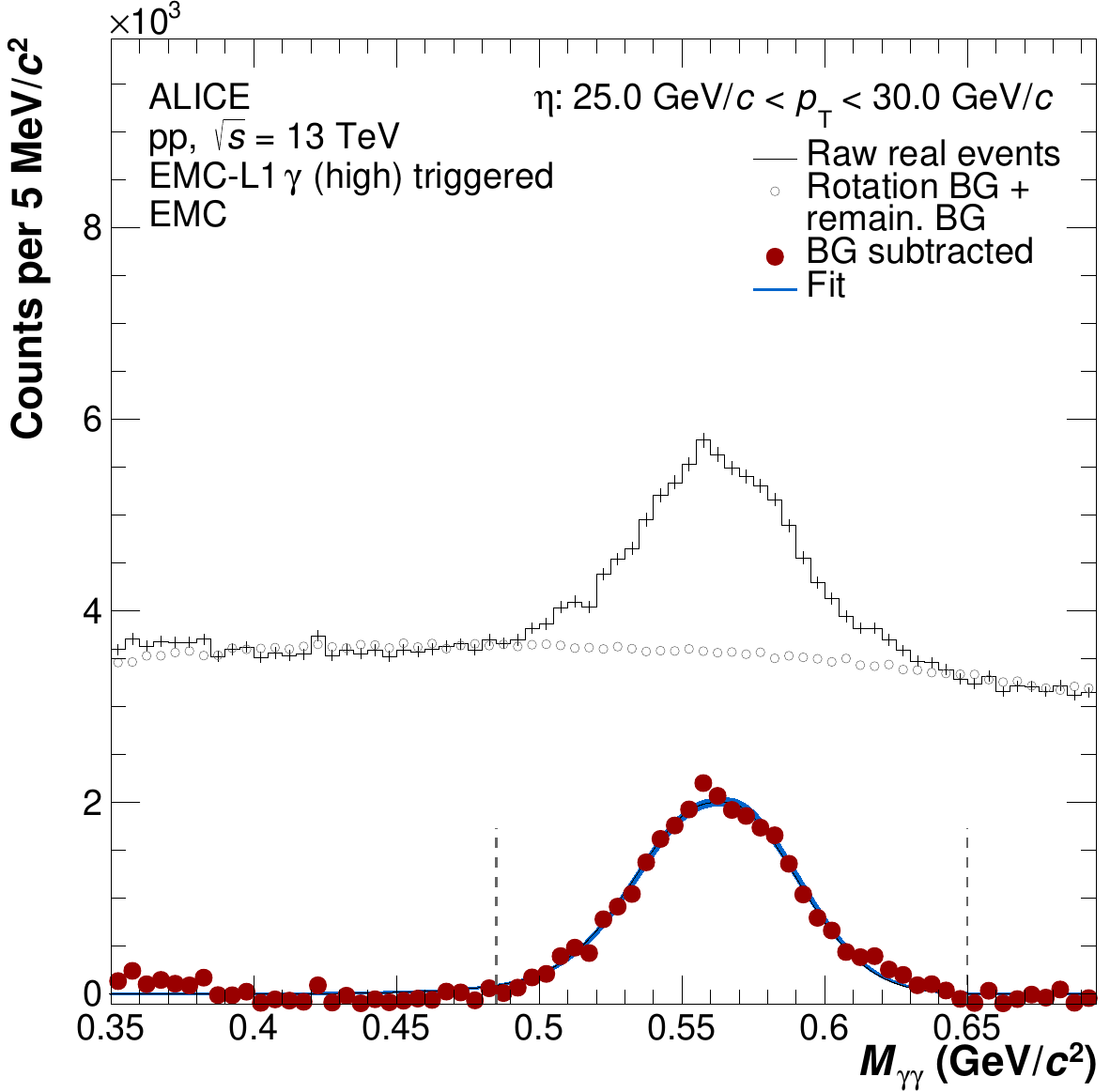}
		\caption[width=0.5\textwidth]{Invariant mass distribution of $\gamma\gamma$ pairs around the \piz (left) and \et (right) rest mass for the \ac{PCM} and \ac{EMC} reconstruction methods, respectively. The extracted meson peak is shown in red, with the parametrization in blue and its uncertainty represented by a blue band. The vertical dashed lines correspond to the limits of the signal integration window. }
		\label{fig:Pi0InvMass}
	\end{figure}

	In the presented measurement, the \piz and \et are reconstructed making use of an invariant mass analysis accompanied by two purity-based analyses for the \piz at high \pT. ALICE has reported both reconstruction methods at lower collision energies in~\cite{ALICE:2012wos,ALICE:2017ryd,Acharya:2017hyu}, where additional details on the reconstruction can be found. In the invariant mass analysis, photons from the same or different (hybrid-method) reconstruction techniques are paired within the same event. The following photon combinations are considered, with the corresponding reconstruction method abbreviations used in this paper given in brackets: EMCal-EMCal (EMC), PHOS-PHOS (PHOS), PCM-$\gamma\gamma$ (PCM) as well as the hybrid methods PCM-EMCal (PCM-EMC) and PCM-PHOS (PCM-PHOS). The PCM-$\gamma\gamma^*$ decay-based analysis is performed using photons from the PCM method, while the $\gamma^*$ is treated like a real $\gamma$ except with a non-zero mass. A restriction on the opening angle between two-photon candidates is applied in the calorimeter-based and hybrid analyses. For the calorimeter-based methods, the opening angle is limited by the granularity of the detectors, as the distance between two clusters cannot fall below one cell. This is especially relevant for the \ac{EMC} due to the large cell size. To ensure the same behavior in the reconstructed signal and the estimated background, which will be described in the course of this section, the opening angle has to be greater than 17~mrad for \ac{EMC} while for \ac{PHOS} the value is set to 5 mrad, which approximately corresponds to the length of a cell diagonal for both calorimeters.
	The resulting number of meson candidates as a function of the invariant mass ($M_{\text{inv}}$) and \pT contains both the \piz and \et signal at their respective rest mass as well as background coming from \g combinations that do not originate from the same \piz or \et decay. Examples of invariant mass distributions are shown in \Fig{fig:Pi0InvMass} for selected \pT ranges for the \piz and \et, respectively. The invariant mass distributions for all reconstruction techniques can be found in~\cite{ALICE:LNM}.

	Two different approaches, event mixing and photon rotation, are used to describe the background.
	The mixed event technique is a well-established method to describe the background in invariant mass analyses~\cite{Kopylov:1974th,Jancso:1977dz}. In this method, photon candidates from different events are paired, leading to totally uncorrelated candidates. Only events with similar $z$-vertex position and similar photon multiplicity were considered for the mixing. Eight $z$-vertex event classes and 4 multiplicity classes, each with a mixing depth of 80 (i.e., the number of events combined in the mixing pool), were used in the analysis.
	For the EMC and PCM-EMC reconstruction methods, the mixed event distribution does not describe the background below the \piz and \et meson peaks. Correlations between the reconstructed photon candidates cannot be described by the event mixing method as this method by definition breaks all correlations. 
	In order to only take out first-order correlations, meaning the correlation coming from the same parent particle (e.g. the photons from the neutral pions and \et mesons), an in-event particle rotation approach was developed for the presented analysis. In this approach, it is assumed that a pair of photon candidates originates from the same parent particle. These photon candidates are rotated around the momentum vector of their reconstructed parent particle by 90$^{\circ}$, keeping the combined momentum intact. The rotated candidates are paired with all other photon candidates in the event like in the same-event method, but excluding the combination of the two rotated candidates. Assuming the candidates come from the same parent particle decaying in two photons, this process simulates a possible decay of the particle and, therefore, results in an accurate description of the background. Collisions where two or fewer photon candidates were found cannot be used to estimate the rotation background as at least three candidates are required. The shape of the background as a function of the number of photon candidates was found to vary only slightly, and hence, the rotation technique is suited to describe the background in the same-event distribution.
	
	To extract the number of measured \piz and \et mesons, the estimated background has to be subtracted from the same-event distribution in each \pT interval.
	The background is scaled to the same-event distribution, either in a signal-free region to the left or right side of the \piz (\et) meson peak or by including a signal shape template from \ac{MC} simulation in the fit, thus including the signal region in the fit. The latter is used for reconstruction methods involving \ac{PHOS} or \ac{EMC} clusters as the signal can have significant tails due to cluster overlaps and contributions from photon conversions in the material between the \ac{TPC} and the calorimeters~\cite{ALICE:2022qhn}. The scaling function is a constant for most reconstruction methods, while for the PHOS and PCM-PHOS methods, a second-order polynomial is used to account for a slight mismatch in the shape of the estimated background.
	The scaled background is then subtracted from the same-event distribution. The remaining signal is parameterized with a three-component function consisting of a Gaussian component to describe the distribution of \g\g pairs, an exponential component at low invariant masses to describe bremsstrahlung of the \ac{PCM} photons and energy loss of late-conversions in front of the calorimeter in case of the \ac{EMC} and \ac{PHOS} photons. Furthermore, for the \ac{EMC} and \ac{PHOS} triggered data, an exponential tail for $M_{\text{inv}} > M_{\pi^{0} (\eta)}$ is used to account for overlapping clusters in these triggered events with a high cluster occupancy. For the \ac{PCM} and \ac{PCM}-$\gamma\gamma^*$ methods, an additional first-order polynomial function is included in the fit to correctly estimate any remaining background that was not described by the scaled mixed-event method.
	\begin{figure}[t]
		\centering
		\includegraphics[width=0.49\textwidth]{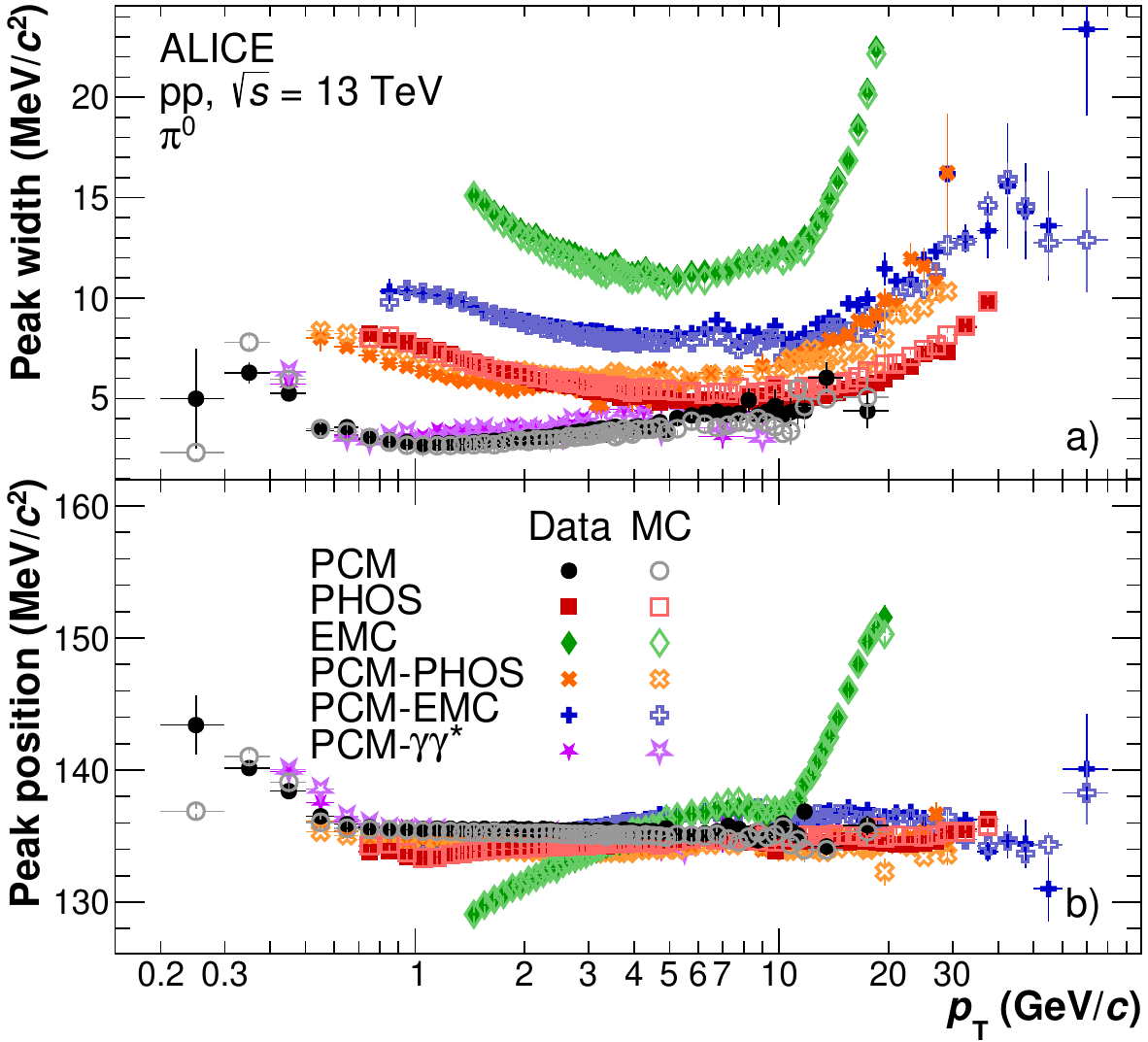}
		\includegraphics[width=0.49\textwidth]{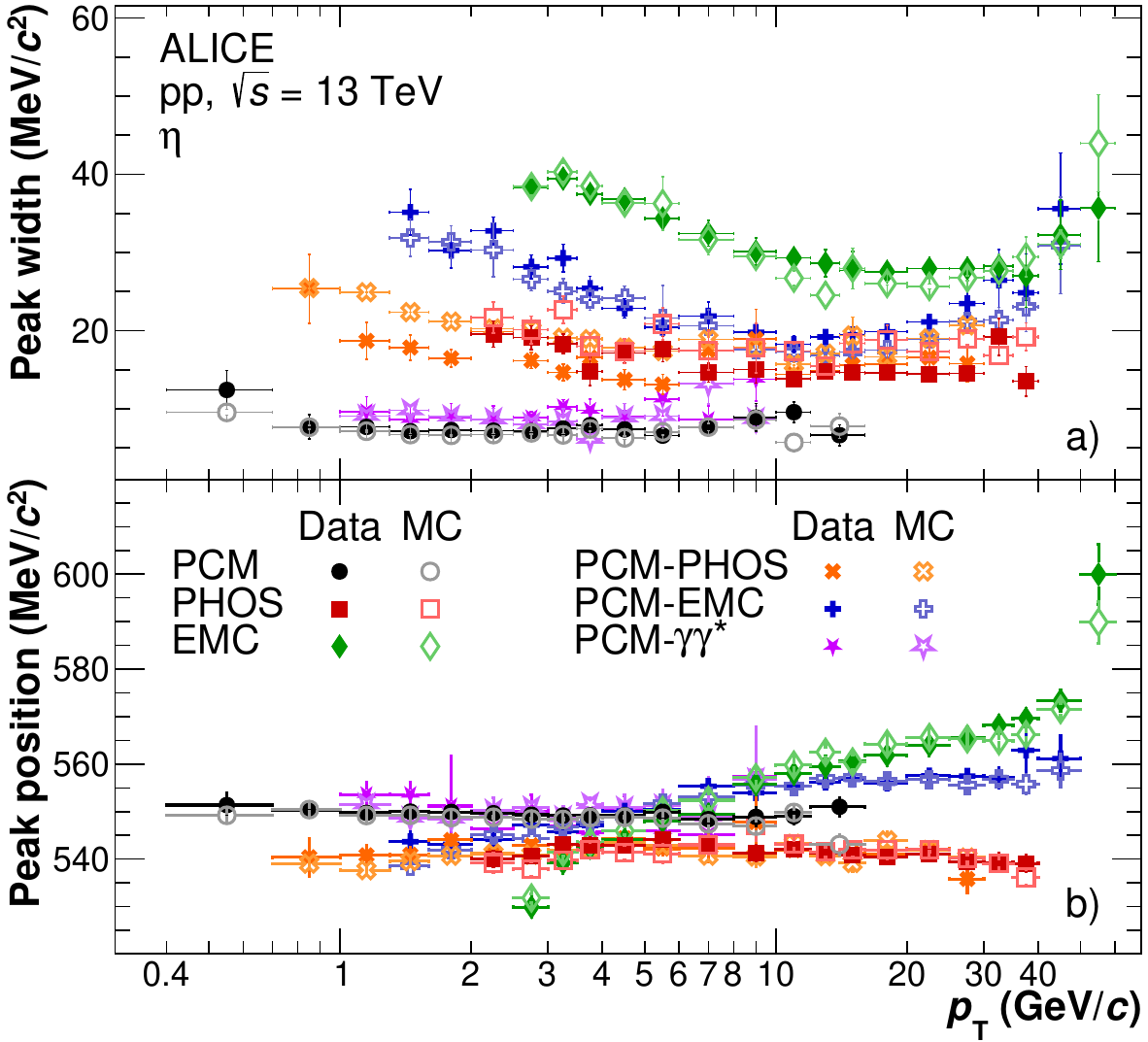}
		\caption{Peak width (a) and peak position (b) as a function of \pT for the \piz (left) and \et (right) for all reconstruction techniques and for data in closed markers and \ac{MC} simulation in open markers. The values are extracted from a fit of the meson peak.}
		\label{fig:MassAndWidth}
	\end{figure}
	The mean of the Gaussian component of the parametrization gives an estimate for the peak position and is presented for the \piz and \et mesons in \Fig{fig:MassAndWidth}{ (lower panels)} as a function of \pT for the different reconstruction techniques.
	A good agreement between data and \ac{MC} simulation is observed, indicating a good calibration of all included detectors. The peak width shown in \Fig{fig:MassAndWidth}{ (top panels)} is estimated by the full-width half maximum (FWHM) of the parametrization divided by 2.35, giving an approximation of the standard deviation of the combined parametrization.
	To obtain the raw \piz and \et-meson counts, the signal is integrated within a fixed window, whose size depends on the peak width of the specific reconstruction method, around the estimated peak position 
	shown in \Fig{fig:MassAndWidth}.
	
\begin{figure}[t]
		\centering
		\includegraphics[width=0.49\textwidth]{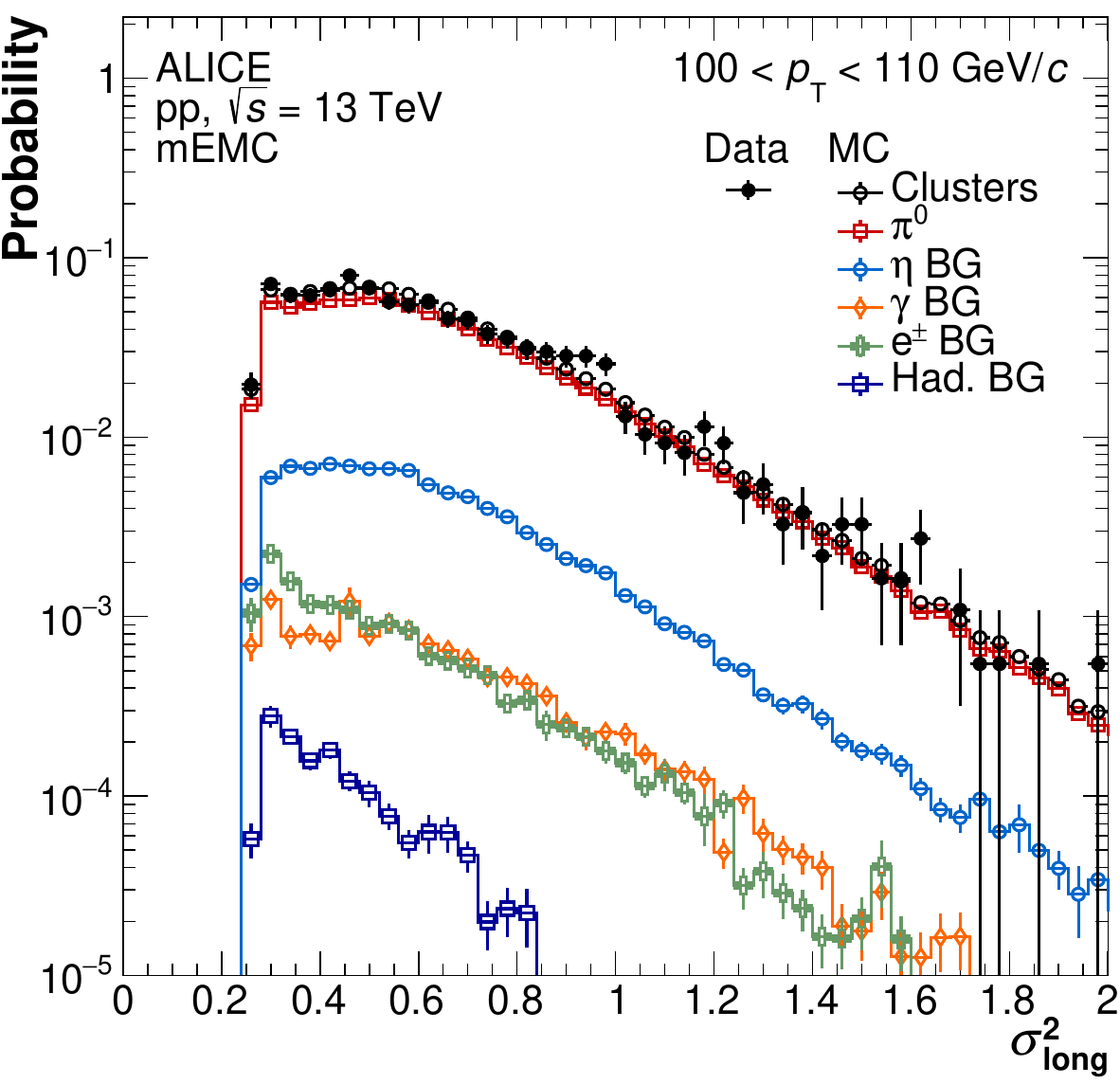}
		\includegraphics[width=0.49\textwidth]{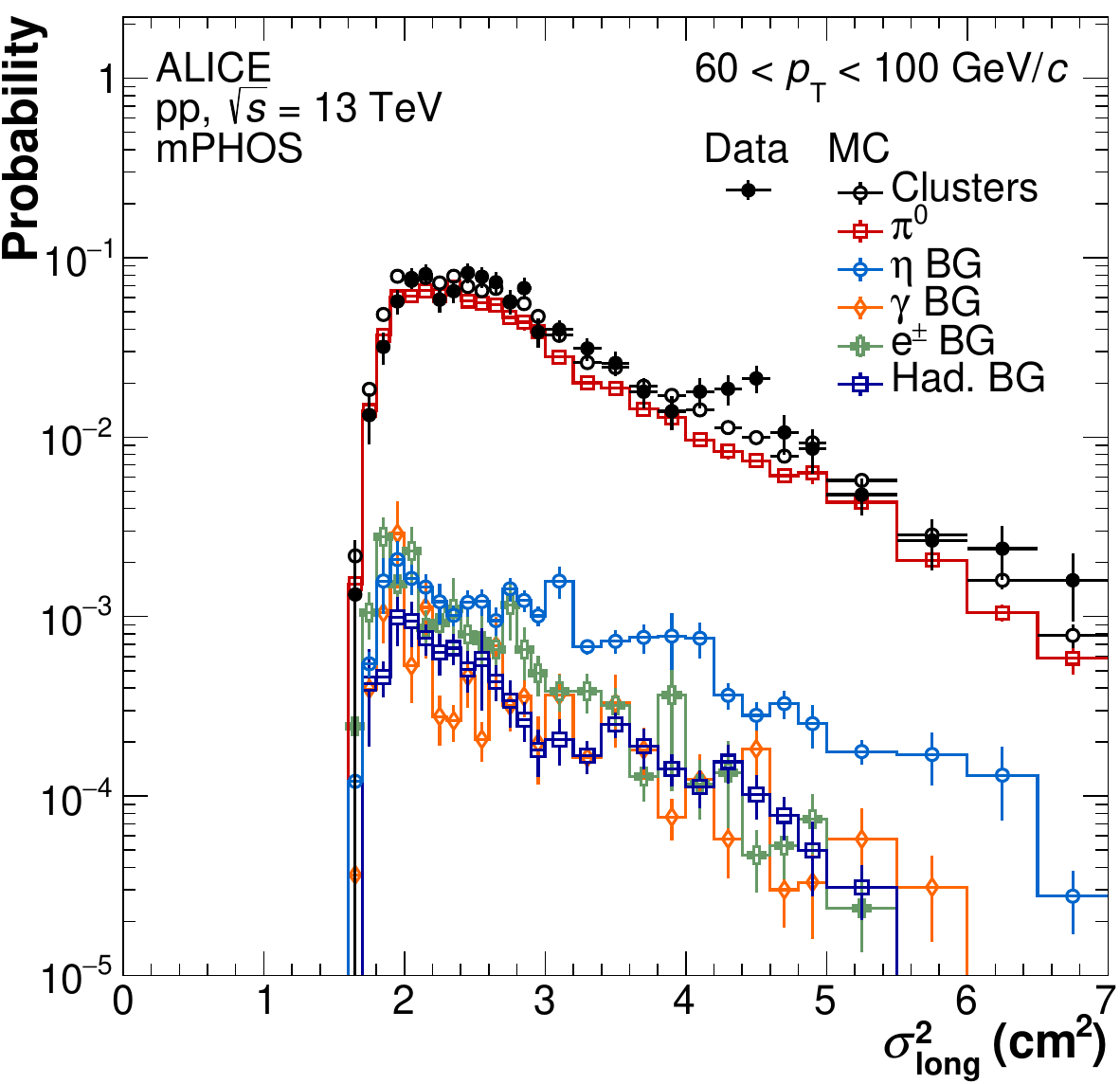}
		\caption{Distribution of \ShowerShape for \ac{EMC} (left) and for \ac{PHOS} (right) clusters for data (full markers) and for \ac{MC} simulation (open markers). Clusters with a leading contribution from photons from \piz decays are shown in black together with different background contributions in colored markers.}
		\label{fig:mergedPi0M02}
	\end{figure}

	At sufficiently high transverse momenta, the opening angle between the two photons from meson decays becomes smaller than the cluster size. 
	In this regime, the classical invariant-mass technique is no longer applicable for the EMC and PHOS reconstruction method. 
	The resulting cluster contains both decay photons and is as such elongated, which can be characterized by the cluster shape elongation parameter \ShowerShape. 
	For the \ac{EMC}, cluster merging of photons originating from \piz decays becomes dominant at \pTPiz~$\approx$~16~\GeVc~\cite{ALICE:2022qhn}, while for \ac{PHOS}, due to higher granularity and a different cluster splitting algorithm, cluster merging starts to become relevant above \pTPiz~$\approx$~30~\GeVc.
	Clusters above \pT = 16 \GeVc in \ac{EMC} and \pT = 30 \GeVc in \ac{PHOS} were used to reconstruct the neutral pions using a purity-driven analysis technique called \ac{mEMC} and \ac{mPHOS}~\cite{Acharya:2017hyu,ALICE:2021est}. \Figure{fig:mergedPi0M02} shows the probability distribution of clusters from data and \ac{MC} simulation as a function of \ShowerShape in the transverse momentum ranges 100~$<$~\pT~$<$~110~\GeVc and 60~$<$~\pT~$<$~100~\GeVc for \ac{EMC} and \ac{PHOS}, respectively. Additionally, the contributions of different particle species to the cluster spectrum, as obtained from the \ac{MC} simulation, are shown. All selected clusters are shown as black markers, while clusters with a leading contribution from a photon originating from a \piz decay are shown as red markers. Clusters originating from \et decays, the largest background component in this analysis, are shown as blue markers. Additional background components are clusters from photons that do not originate from \piz or \et decays, shown in orange, clusters from electrons shown in green, and hadronic clusters that are represented by the open blue markers. To precisely estimate the number of \piz mesons, the different background components have to be precisely understood and their relative abundance in the \ac{MC} simulation has to be tuned to match their abundance in data. This will be discussed in \Sec{sec:Corrections}. 
	For the \ac{mEMC} method, the \piz raw yield is obtained by the integral of all clusters with a value of \ShowerShape~$>$~0.27, while for \ac{PHOS} this value is energy dependent (see \Tab{tab:CutCalo}) to allow for a better separation between single photon clusters and merged \piz clusters.
	
	The raw meson spectra are obtained for each dataset listed in \Tab{tab:events} that is available for the respective reconstruction method. For the EMC, PCM-EMC and \ac{mEMC} methods, this includes both \ac{EMC}-L1 \g triggers, where \piz and \et mesons are used from \pT~=~8~(16)~\GeVc onwards for the low (high) threshold. For the PHOS and PCM-PHOS methods, the \ac{PHOS} PHI7 triggered data is used from \pT~=~10~\GeVc onwards, while the \ac{mPHOS} method starts from \pT~=~30~\GeVc. Furthermore, for the PCM, PCM-EMC, and PCM-\g\g$^{*}$ method, data from the B~=~0.2~T data-taking period is used, allowing the \ac{PCM} method to reach down to \pT~=~0.2~\GeVc in case of the \piz. The raw spectra are individually corrected for detector effects, as discussed in the next section, and combined using the same combination method as described in \Sec{sec:results}.
	
	\subsection{Corrections}
	\label{sec:Corrections}
	
	As outlined in the previous section, the raw meson yields obtained with the different reconstruction methods must be corrected for detector effects and contamination from secondary particles. These corrections are presented in this section, starting with the correction for \piz from weak decays, followed by the out-of-bunch pileup correction for the \ac{PCM} method and the correction for contamination in the \ac{PCM}-$\gamma\gamma^*$ measurement. Furthermore, the acceptance correction and reconstruction efficiency for all methods are presented, followed by the purity correction for the merged-cluster-based analyses. The section concludes with a discussion on the correction of inefficiencies of the triggers used for the analysis.
	
	In the presented measurement, the production of primary neutral mesons is reported, and hence secondary neutral pions from weak decays of \kzero, \kzerol, and \lmb as well as from hadronic interactions with the detector material are subtracted~\cite{alicePrimPart}. A data-driven approach~\cite{ALICE:2018mjj} is used to estimate the contributions from weak decays. In this approach, the measured spectra of these particles in MB events~\cite{ALICE:2020jsh} and as a function of multiplicity~\cite{ALICE:2019avo} are used as input.
	As the measurement of \kzero, \kzerol, and \lmb does not cover the highest multiplicity intervals of the presented measurement, an extrapolation of these spectra from the 0--1\% multiplicity interval to the multiplicity intervals above 0.1\% was performed. To obtain the raw yield of \piz from these decays, the detection efficiency and acceptance of secondary neutral pions are taken from \ac{MC} simulation. 
	The fraction of mesons from hadronic interactions is estimated using \ac{MC} simulation including GEANT3~\cite{Brun:1987ma,Brun:1994aa} for the description of the interaction between the traversing particles from the collision and the detector material.
	The correction is of the order of up to about 8\% at low \pT, depending on the reconstruction method, while at high \pT it is about 3\% at maximum.
	
	The \ac{PCM} analysis needs a correction to account for \piz and \et mesons produced in bunch crossings other than the triggered one, referred to as out-of-bunch pileup. The fraction of out-of-bunch pileup is obtained by using the \ac{DCA} distribution of reconstructed photons. Photons originating from neighboring collisions but assigned to the current one, have a wider \ac{DCA} distribution compared to photons from the triggered collision. 
	In contrast to previous analyses, no TPC-only tracks are considered for the PCM analysis, and tracks always have a constraint by either the ITS or the TRD giving a much better timing resolution than for TPC-only tracks. TPC-only tracks typically contribute as the largest fraction to the out-of-bunch pileup. Therefore, the contribution in the presented analysis is much smaller than in previous publications. The fraction of out-of-bunch pileup is estimated by fitting the underlying distribution. For the MB events, the correction is largest at \pT~$\approx$~1.5~\GeVc and amounts to about 7\%, decreasing to 4\% at 20 \GeVc. For the multiplicity event class 70--100\% the correction is 35\% at \pT~$\approx$~2~\GeVc while for the 0.0--0.01\% the correction is about 3\%. The strong dependence of the fraction of out-of-bunch pileup on the multiplicity originates from the difference in the number of mesons produced in the triggered collision, while the number of mesons from out-of-bunch pileup is constant on average.
	
	Contamination in the virtual photon sample for the PCM-$\gamma\gamma^*$ measurement is kept to a minimum as laid out in \Sec{sec:PhotonSelection}. The remaining contamination of photon conversion electrons, misidentified as primary particles, is estimated using \ac{MC} simulation: for the B~=~0.5~T data, the contamination is between approximately 1\% at around \pT~=~2~\GeVc and 7\% for high and low \pT while for the B~=~0.2~T data, the contamination is slightly higher, from about 2.5\% to 15\%, with a similar \pT dependence as for the data collected with the nominal magnetic field.
	
	Corrections for the geometrical acceptance and the reconstruction efficiency as well as impurities in the extracted signal are done using the PYTHIA8 Monash event generator in combination with a full GEANT3 detector simulation. While for the correction of the \ac{MB} data, PYTHIA8 simulations with \ac{MB} processes are used, the calorimeter-triggered data are corrected using PYTHIA8 simulations, generated in intervals of the transverse momentum of the initial hard scattering, with two jets in the final state. These simulations allow for small statistical uncertainties of the correction factor up to high \pT.
	The acceptance correction is performed by calculating the fraction of \piz (\et) produced within $|y| < $ 0.8, where all decay products are within the geometrical limits of the detector used for the reconstruction.
	For the reconstruction efficiency, the same photon selection and signal extraction procedures as described in \Sec{sec:NeutralMesons} are performed using the respective output of the \ac{MC} simulation. The resulting peak properties of the extracted \piz (\et) are compared to data, as shown in \Fig{fig:MassAndWidth}, to verify the description of the data by the MC.
	The meson reconstruction efficiency $\varepsilon_{\text{rec}}$ is calculated using the \ac{MC} simulation by comparing the extracted raw \piz (\et) yield to all generated \piz (\et) within the geometrical acceptance. Hence, $\varepsilon_{\text{rec}}$ contains loss effects due to the photon selection criteria (\Sec{sec:PhotonSelection}), energy resolution effects, as well as impurities in the signal extraction. To reduce statistical fluctuations, the PCM-$\gamma\gamma$ and PCM-$\gamma\gamma^\star$ use verified reconstructed mesons for the $\varepsilon_{\text{rec}}$ calculation, and hence no signal extraction is involved. It was checked that, within statistical fluctuations, both approaches result in the same correction factor. 
	The reconstruction efficiency was studied as a function of multiplicity for each reconstruction method using the same mean charged-particle multiplicity in the simulation as in the data. A relative reduction of the efficiency of up to about 7\% was found for the calorimeter-based methods. These discrepancies can be attributed to the differences in the spectral shape, particularly at low \pT, and the increasing fraction of clusters rejected by the charged-particle veto with increasing multiplicity. However, to decrease the statistical uncertainty, the multiplicity-integrated efficiency correction is used and scaled with a parametrization to the ratio of the multiplicity-dependent efficiency to the integrated efficiency. As the difference in efficiency originates from differences in the photon reconstruction efficiency, the change in efficiency is assumed to be identical for both the \piz and the \et meson.
	
	A purity correction is employed in the merged cluster analysis to account for clusters from \g originating from \et decays as well as prompt photons and electrons. The relative abundance of these background components is evaluated using data and MC-driven approaches: the relative abundance of the \et meson compared to the \piz is estimated with the constant fit to the \etopi~ratio at high \pT shown in \Fig{fig:EtaToPi}. The prediction from the simulation is then subsequently scaled to match the data. The additional contribution from prompt photons is estimated using PYTHIA8 $\gamma$--jet processes simulations where Compton scattering $\text{q} \text{g} \rightarrow \text{q}\gamma$, quark--antiquark annihilation $\text{q} \overline{\text{q}} \rightarrow \text{g} \gamma$ and, with a much smaller contribution, $\text{q} \overline{\text{q}} \rightarrow \gamma \gamma$ events are generated. Furthermore, electrons from decays of $\text{W}^{\pm}$ and Z are not included in the simulation and their contribution is therefore estimated using \acs{MC} based on the POWHEG~\cite{Nason:2004rx,Frixione:2007vw} event generator. Again the same efficiency as for primary electrons is assumed to estimate the relative abundance in the raw merged cluster yield of these electrons. The \piz purity ranges from 88\% (90\%) at \pT $\approx$ 30 \GeVc to 78\% (82\%) at the highest \pT for \ac{mEMC} (\ac{mPHOS}).
	
	\begin{figure}[t]
		\centering
		\includegraphics[width=0.49\textwidth]{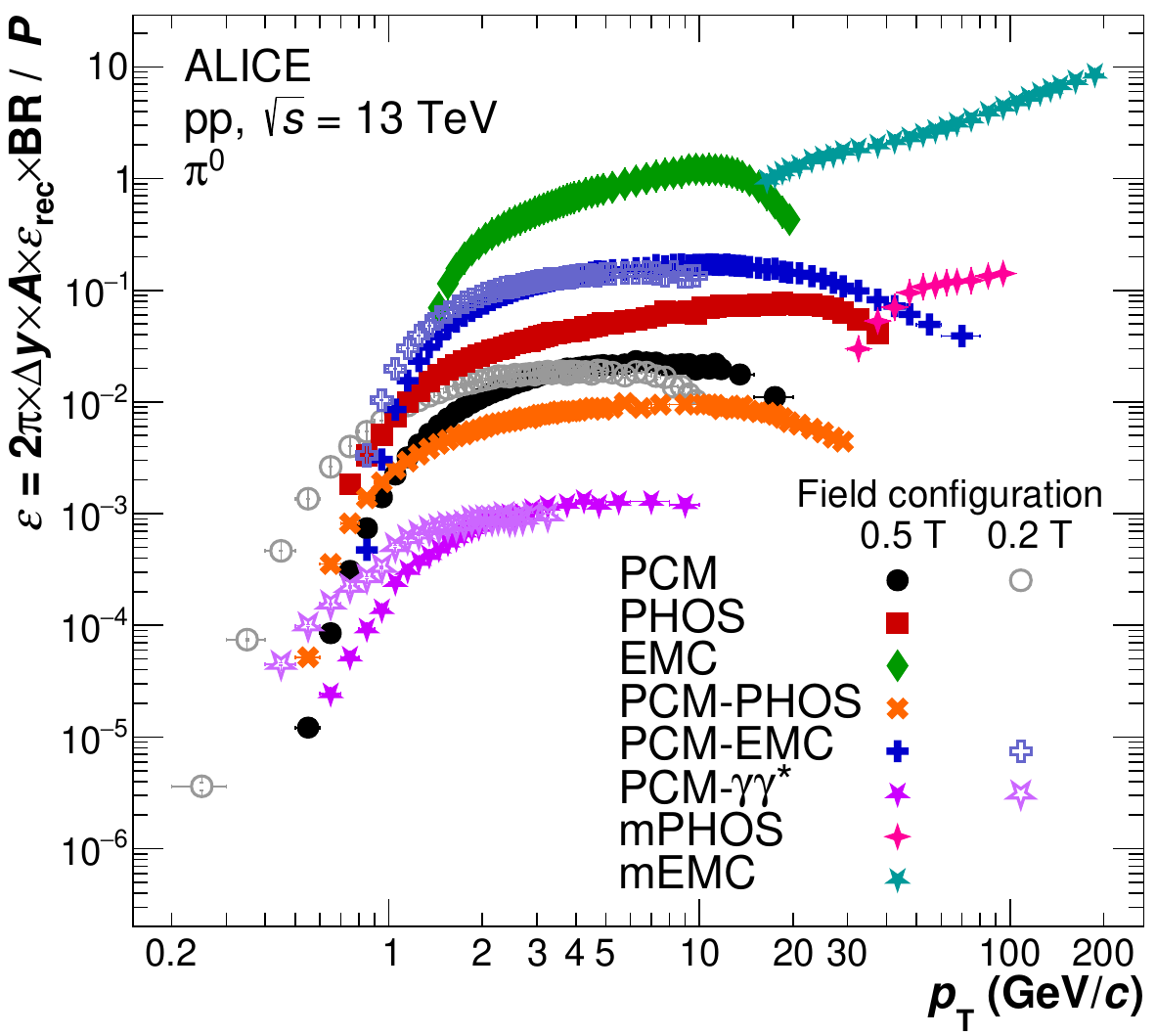}
		\includegraphics[width=0.49\textwidth]{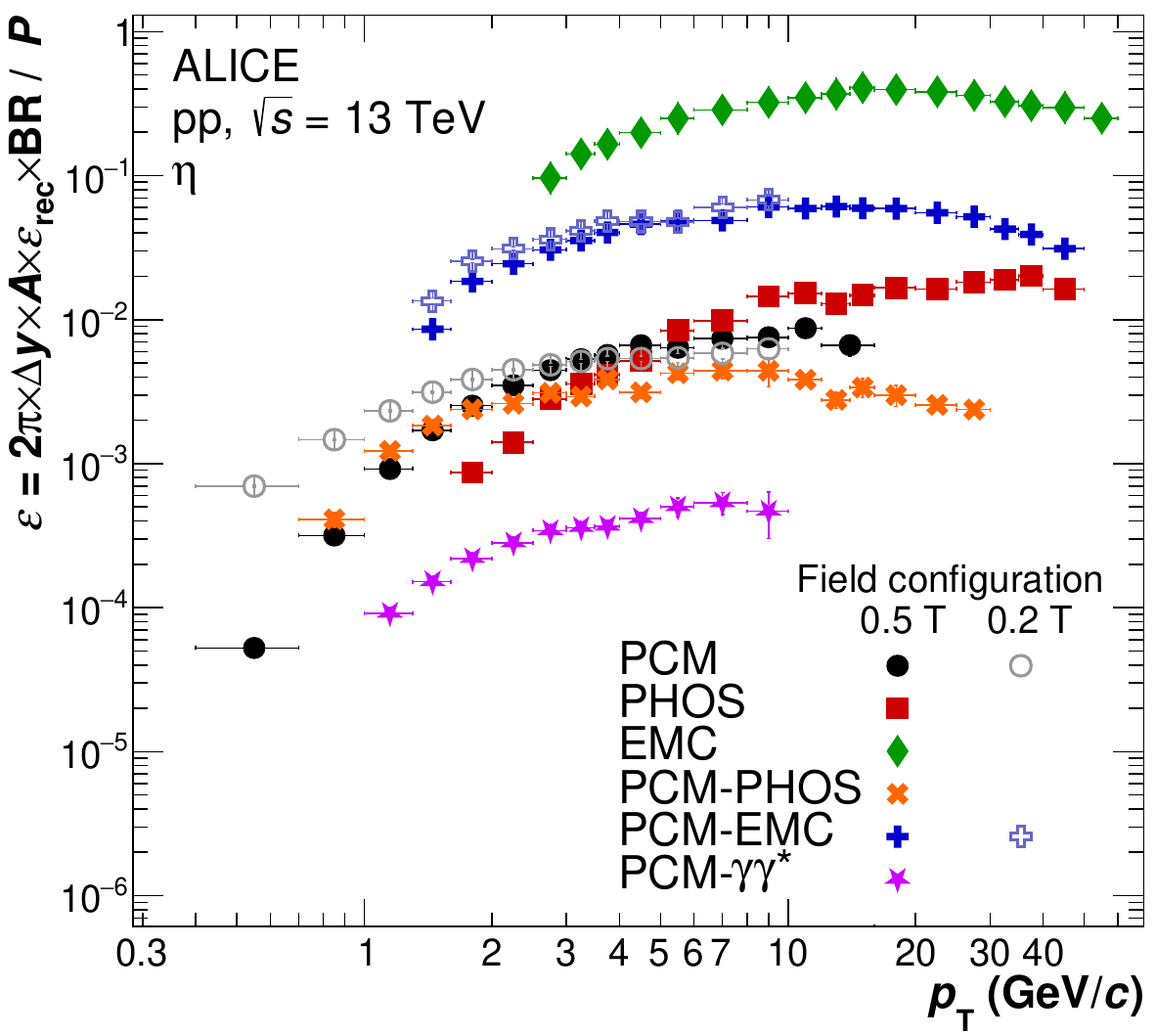}
		\caption{Correction factor for the \piz (left) and \et (right), including the reconstruction efficiency, acceptance, purity, branching-ratio and normalization for the azimuthal angle $\varphi$ and rapidity $y$ coverage, as a function of \pT for the different reconstruction methods.}
		\label{fig:AcceptanceTimesEff}
	\end{figure}
	
	The combined correction factors are presented in \Fig{fig:AcceptanceTimesEff} for \piz and \et mesons. The correction factors include the geometrical acceptance ($A$), reconstruction efficiency ($\varepsilon_{\text{rec}}$), and purity ($P$) for the merged cluster analyses, as well as normalizations for the rapidity ($\Delta y$) and $\varphi$ coverage ($2\pi$) and the branching ratio ($\text{BR}$), however excluding the \ac{EMC} and \ac{PHOS} trigger efficiency. An initial increase with \pT can be seen, as it becomes more likely that both photons are within the detector acceptance, and because the single-photon reconstruction efficiency increases with photon energy. At high transverse momenta, the \piz meson correction factor for the \ac{EMC} method drops due to the merging of both decay photons into the same cluster. For the \ac{PHOS} method, this decrease is much less pronounced and the onset occurs at higher transverse momentum as the granularity of the calorimeter is much finer, and the effect is only expected above \pT~$\approx$~30~\GeVc. 
	
	The minimum bias trigger used for the presented analysis is not able to trigger on all events where \piz or \et mesons were produced. This results in a loss of signal, which is estimated using PYTHIA8 Monash 2013 while additionally taking PHOJET 1.12~\cite{PHOJET} and EPOS LHC into account for the estimation of the systematic uncertainty. Below \pT~=~1~\GeVc the \piz and \et meson signal loss is up to 2\%, while it is zero at high transverse momenta. The systematic uncertainty of this correction is estimated using different event generators and is of the order of 0.4\% at maximum.
	The fraction of events lost due to the minimum bias trigger efficiency is compensated in the measurement of the visible cross section ($\sigma_{\text{MB}_{\text{AND}}}$) based on the V0M detector system~\cite{ALICE-PUBLIC-2021-005}. The integrated luminosity is then calculated as $\mathcal{L}_{\text{int}} = \frac{N_{\text{evt}}}{\sigma_{\text{MB}_{\text{AND}}}}RF$, with $N_{\text{evt}}$ being the number of inspected events and $RF$ being the trigger rejection factor.
	
	The \piz and \et meson \pT~spectra obtained using the calorimeter-triggered data have to be scaled down to correct for the higher integrated luminosity of the triggered data. The increase in luminosity can be derived by comparing the cluster spectra in triggered data to the cluster spectrum in minimum bias collisions as shown in \Fig{fig:RF}. Additionally, a correction based on a trigger emulation in \ac{MC} simulation has to be applied to the cluster spectra of the triggered data to correct for trigger inefficiencies due to masked trigger regions, resulting in an additional correction on the order of 5--7\% for \ac{EMC} and about 40\% for \ac{PHOS}. As the same trigger emulation is also used for the calculation of the reconstruction efficiency $\varepsilon_{\text{rec}}$, the absolute magnitude of the correction cancels for the calculation of the \piz (\et) cross section. However, a mild \pT dependence affects the shape of the triggered cluster spectra, thereby modifying the estimated trigger-rejection factor. Details on this correction for the \ac{EMC} can be found in~\cite{IsoPhotonDraft:2024}.

	\section{Systematic uncertainties}
	\label{sec:sys}
	
	\begin{table}[h!t]
		\fontsize{9}{10.5}\selectfont
		\vspace{-0.5cm}
		\centering
		\caption{Relative systematic uncertainties $(\%)$ of the \piz spectrum for the invariant-mass-based reconstruction methods for $1.4 < \pT < 1.5~\GeVc$ and $6.5 < \pT < 7.0~\GeVc$ and for the merged-cluster-based analyses for $50 < \pT < 55~\GeVc$ and $90 < \pT < 100~\GeVc$. If a source of uncertainty is negligible for a specific reconstruction method, it is labeled as 'negl'; if it is not considered for that method, it is indicated with a dash.}
		\begin{tabular}{p{2.4cm}|p{0.38cm}p{0.38cm}|p{0.38cm}p{0.38cm}|p{0.38cm}p{0.38cm}|p{0.38cm}p{0.38cm}|p{0.38cm}p{0.38cm}|p{0.38cm}p{0.38cm}|p{0.38cm}p{0.38cm}|p{0.38cm}p{0.38cm}}
			
			\toprule
			\textbf{Rec. method} 
			
			&\multicolumn{2}{c |}{\textbf{PCM}}
			&\multicolumn{2}{c |}{\textbf{EMC}}
			&\multicolumn{2}{c |}{\textbf{PHOS}}
			&\multicolumn{2}{c |}{\textbf{PCM-}}
			&\multicolumn{2}{c |}{\textbf{PCM-}}
			&\multicolumn{2}{c |}{\textbf{PCM-}}
			&\multicolumn{2}{c |}{\textbf{mEMC}}
			&\multicolumn{2}{c }{\textbf{mPHOS}}\\[-12pt]
			
			& \multicolumn{2}{c }{ }& \multicolumn{2}{c }{ } & \multicolumn{2}{c }{ } & \multicolumn{2}{c }{ \textbf{{$\gamma \gamma^{*}$}}} & \multicolumn{2}{c }{ \textbf{EMC}} & \multicolumn{2}{c }{ \textbf{PHOS}} & \multicolumn{2}{c }{ } & \multicolumn{2}{c }{ } \\
			
			\pT(\GeVc) & 1.45 & 6.75 & 1.45 & 6.75 & 1.45 & 6.75 & 1.45 & 6.75 & 1.45 & 6.75 & 1.45 & 6.75 & 52.5 & 95 & 52.5 & 95\\
			\hline
			Yield extraction & 1.7 & 2.5 & 2.9 & 2.8 & 2.2 & 2.7 & 2.7 & 4.6 & 2.1 & 2.0 & 3.7 & 3.7 &  - & - & - & -\\
			Cluster description &  - & - &2.2 & 2.1 & 3.6 & 2.8 &  - & - &2.1 & 2.2 & 3.3 & 2.0 & 6.5 & 7.5 & 2.0 & 4.2 \\ 
			Cluster $E$ calib. &  - & - &2.2 & 2.2 & 2.0 & 2.0 &  - & - &2.1 & 1.6 & 2.0 & 2.0 & 2.1 & 2.1 & 3.2 & 3.2 \\ 
			Ch. particle veto &  - & - &0.4 & 0.7 & 0.7 & 1.1 &  - & - &0.4 & 0.7 & 0.4 & 0.7 & 1.5 & 1.5 & 2.0 & 2.6 \\ 
			e$^{\pm}$ track rec. & 0.0 & 0.1 &  - & - & - & - &2.0 & 2.0 & 0.3 & 0.3 & 0.3 & 0.3 &  - & - & - & -\\
			e$^{\pm}$ PID & 0.5 & 1.7 &  - & - & - & - &1.2 & 1.2 & 0.8 & 1.0 & 0.8 & 1.0 &  - & - & - & - \\
			PCM photon PID & 0.3 & 1.0 &  - & - & - & - &2.0 & 2.0 & 0.4 & 0.8 & 1.0 & 1.3 &  - & - & - & - \\
			Efficiency &  negl. & negl. &2.0 & 2.0 & 1.5 & 1.5 & 0.5 & 0.5 &  2.0 & 2.0 &1.5 & 1.5 & 4.8 & 4.8 & 6.8 & 8.6 \\ 
			Outer material &  - & - &4.2 & 4.2 & 2.0 & 2.0 &  - & - &2.1 & 2.1 & 1.0 & 1.0 & 4.2 & 4.2 & 1.7 & 1.7 \\ 
			Inner material & 5.0 & 5.0 &  - & - & - & - &2.5 & 2.5 & 2.5 & 2.5 & 2.5 & 2.5 &  - & - & - & - \\
			Norm. \& pileup & 3.3 & 3.2 &  negl. & negl. & negl. & negl. & negl. & negl. & negl. & negl. & negl. & negl. &2.8 & 2.8 & negl. & negl. \\ 
			Branching ratio &  negl. & negl. & negl. & negl. & negl. & negl. &3.0 & 3.0 &  negl. & negl. & negl. & negl. & negl. & negl. & negl. & negl. \\ \hline
			Total systematic & 6.2 & 6.7 & 6.3 & 6.3 & 5.3 & 5.1 & 6.1 & 7.1 & 5.3 & 5.3 & 6.3 & 5.8 & 9.9 & 10.5 & 8.2 & 10.6 \\ 
			Total statistical & 0.6 & 3.1 & 0.5 & 0.6 & 0.6 & 2.4 & 3.2 & 11.6 & 0.8 & 1.4 & 1.2 & 4.9 & 0.9 & 1.8 & 6.4 & 12.2 \\  \bottomrule

		\end{tabular}
		\label{tab:SysErrCombPi0}

		\vspace{0.3cm}
		\caption{Relative systematic uncertainties $(\%)$ of the \et spectrum for the different reconstruction methods for $2.5 < \pT < 3.0~\GeVc$ and $6.0 < \pT < 8.0~\GeVc$. If a source of uncertainty is negligible for a specific reconstruction method, it is labeled as 'negl'; if it is not considered for that method, it is indicated with a dash.}
		\centering
		\begin{tabular}{p{2.4cm}|ll|ll|ll|ll|ll|ll}
			\toprule
			\textbf{Rec. method} &\multicolumn{2}{c|}{\textbf{PCM}}&\multicolumn{2}{c|}{\textbf{EMC}}&\multicolumn{2}{c|}{\textbf{PHOS}}&\multicolumn{2}{c|}{\textbf{PCM$-\gamma \gamma^{*}$}}&\multicolumn{2}{c|}{\textbf{PCM-EMC}}
			&\multicolumn{2}{c}{\textbf{PCM-PHOS}}\\
			\pT(\GeVc) & 2.75 & 7.0 & 2.75 & 7.0 & 2.75 & 7.0 & 2.75 & 7.0 & 2.75 & 7.0 & 2.75 & 7.0 \\ \hline
			Yield extraction & 1.5 & 6.7 & 5.7 & 3.1 & 6.5 & 6.9 & 7.5 & 17.3 & 4.1 & 7.8 & 6.4 & 10.0 \\ 
			Cluster description &  - & - &2.4 & 2.5 & 3.9 & 3.7 &  - & - &2.6 & 2.7 & 2.7 & 2.7 \\ 
			Cluster $E$ calib. &  - & - &2.2 & 2.2 & 1.5 & 1.5 &  - & - &2.1 & 1.6 & 2.0 & 2.0 \\ 
			Ch. particle veto &  - & - &0.5 & 0.7 & 0.8 & 1.2 &  - & - &0.5 & 0.7 & 0.5 & 0.7 \\ 
			e$^{\pm}$ track rec. & 0.3 & 0.3 &  - & - & - & - &2.0 & 2.0 & 0.3 & 0.3 & 0.3 & 0.3 \\ 
			e$^{\pm}$ PID & 1.0 & 2.3 &  - & - & - & - &1.2 & 1.2 & 0.8 & 1.0 & 0.8 & 1.0 \\ 
			PCM photon PID & 0.5 & 1.5 &  - & - & - & - &2.0 & 2.0 & 0.5 & 0.9 & 1.1 & 1.3 \\ 
			Efficiency &  negl. & negl. &2.0 & 2.0 &  1.0 & 1.0 &0.5 & 0.5 &  2.0 & 2.0 & 1.0 & 1.0 \\
			Outer material &  - & - &4.2 & 4.2 & 2.0 & 2.0 &  - & - &2.1 & 2.1 & 1.0 & 1.0 \\ 
			Inner material & 5.0 & 5.0 &  - & - & - & - &2.5 & 2.5 & 2.5 & 2.5 & 2.5 & 2.5 \\ 
			Norm. \& pileup & 3.0 & 2.9 &  negl. & negl. & negl. & negl. & negl. & negl. & negl. & negl. & negl. & negl. \\
			Branching ratio &  negl. & negl. & negl. & negl. & negl. & negl. &5.8 & 5.8 &  negl. & negl. & negl. & negl.  \\ \hline
			Total systematic & 6.1 & 9.3 & 8.0 & 6.5 & 8.0 & 8.3 & 10.3 & 18.7 & 6.6 & 9.3 & 7.9 & 11.0 \\ 
			Total statistical & 2.9 & 6.9 & 1.7 & 2.1 & 6.9 & 9.2 & 20.8 & 45.1 & 2.0 & 4.5 & 7.5 & 17.5 \\
			\bottomrule
		\end{tabular}
		\label{tab:SysErrCombEta}
		\vspace{0.3cm}
		\caption{Relative systematic uncertainties $(\%)$ of the \etopi~ratio for the different reconstruction methods for $2.5 < \pT < 3.0~\GeVc$ and $6.0 < \pT < 8.0~\GeVc$. If a source of uncertainty is negligible for a specific reconstruction method, it is labeled as 'negl'; if it is not considered for that method, it is indicated with a dash.}
		\begin{tabular}{p{2.4cm}|ll|ll|ll|ll|ll|ll}
			\toprule
			\textbf{Rec. method} & \multicolumn{2}{c|}{\textbf{PCM}}&\multicolumn{2}{c|}{\textbf{EMC}}&\multicolumn{2}{c|}{\textbf{PHOS}}&\multicolumn{2}{c|}{\textbf{PCM$-\gamma \gamma^{*}$}}&\multicolumn{2}{c|}{\textbf{PCM-EMC}}
			&\multicolumn{2}{c}{\textbf{PCM-PHOS}}\\
			\pT(\GeVc) & 2.75 & 7.0 & 2.75 & 7.0 & 2.75 & 7.0 & 2.75 & 7.0 & 2.75 & 7.0 & 2.75 & 7.0 \\ \hline
			Yield extraction & 2.1 & 7.2 & 6.5 & 4.1 & 6.5 & 6.9 & 8.1 & 17.5 & 4.6 & 8.1 & 6.1 & 12.0 \\ 
			Cluster description &  - & - &1.6 & 2.0 & 3.7 & 3.5 &  - & - &1.9 & 1.9 & 2.2 & 2.2 \\ 
			Cluster $E$ calib. &  - & - & negl. & negl. &  negl. & negl. & - & - & negl. & negl. & negl. & negl. \\ 
			Ch. particle veto &  - & - &0.5 & 0.5 & 0.5 & 0.5 &  - & - &0.5 & 0.5 & 0.5 & 0.5 \\ 
			e$^{\pm}$ track rec. & 0.2 & 0.3 &  - & - & - & - &2.0 & 2.0 & 0.3 & 0.3 & 0.3 & 0.3 \\ 
			e$^{\pm}$ PID & 1.0 & 2.2 &  - & - & - & - &1.2 & 1.2 & 0.8 & 1.0 & 0.8 & 1.0 \\ 
			PCM photon PID & 0.6 & 1.8 &  - & - & - & - &2.0 & 2.0 & 0.5 & 0.9 & 1.1 & 1.3 \\ 
			Efficiency &  negl. & negl. & 0.9 & 0.9 &  0.6 & 0.6 &0.2 & 0.2 &  0.4 & 0.4 & 0.4 & 0.4 \\
			Outer material &  - & - & negl. & negl. & negl. & negl. & - & - & negl. & negl. & negl. & negl. \\
			Inner material &  negl. & negl. & - & - & - & - & negl. & negl. & negl. & negl. & negl. & negl. \\
			Norm. \& pileup & 2.6 & 2.8 &  negl. & negl. & negl. & negl. & negl. & negl. & negl. & negl. & negl. & negl. \\
			Branching ratio &  negl. & negl. & negl. & negl. & negl. & negl. &6.5 & 6.5 &  negl. & negl. & negl. & negl. \\ \hline
			Total systematic & 3.5 & 8.2 & 6.8 & 4.7 & 7.5 & 7.8 & 10.8 & 18.9 & 5.1 & 8.4 & 6.7 & 12.3 \\ 
			Total statistical & 3.0 & 7.1 & 1.7 & 2.1 & 6.9 & 9.3 & 21.4 & 46.7 & 2.0 & 4.6 & 7.6 & 17.9 \\  \bottomrule
			
		\end{tabular}
		\label{tab:SysErrCombEtaPi0}

	\end{table}
	\afterpage{\clearpage}
	The basis for the evaluation of the systematic uncertainties of the meson spectra and the \etopi~ratio are variations of all selection criteria presented in \Sec{sec:analysis}: the selection of photon candidates, meson candidates, and the signal extraction. Subsequently, the systematic uncertainties are estimated by comparing the corrected meson spectra obtained with the default setting to those obtained by using variations of each selection criterion.
	Furthermore, as done in previous measurements of the neutral-meson production, additional sources of systematic uncertainties (cluster energy calibration, inner material, outer material, and efficiency)  were estimated and a more detailed breakdown of these sources is presented in~\cite{ALICE:2022qhn,ALICE:2019cox, ALICE:MBW,Acharya:2017hyu,ALICE:2017ryd}. Hence, in this paper, only the dominant sources of uncertainties will be discussed in detail.
	\TablesThree{tab:SysErrCombPi0}{tab:SysErrCombEta}{tab:SysErrCombEtaPi0} show all the sources of systematic uncertainties considered in the analysis and their magnitude 
	in two representative \pT intervals for \piz, \et mesons, and the \et/\piz ratio, respectively. The total systematic uncertainty, given at the bottom of the respective tables, is calculated by adding the single sources in quadrature, as no correlation between the different sources is expected. In the following, these components will be discussed briefly.
	
	One of the leading sources of uncertainty of the \ac{PCM} reconstruction method as well as the hybrid and \ac{PCM}-$\gamma\gamma^{\star}$ reconstruction methods is the uncertainty arising from the material budget. It represents the precision to which the material of the ALICE experiment between the collision vertex and the outer wall of the \ac{TPC} is implemented in the \ac{MC} simulation. This uncertainty was reduced compared to previous publications on neutral mesons~\cite{ALICE:MBW}. An uncertainty of 2.5\% per photon is assigned, leaving the \ac{PCM} method with a total of 5\% and the hybrid reconstruction methods with 2.5\%.
	
	As for the inner material, the uncertainty on the outer material describes the precision to which the detector material that lies between the TPC and the calorimeters is described in the \ac{MC}. This includes both the material of the TRD and TOF detectors and their support structures, accounting for up to about 54\% $X_{0}$. For the EMCal, this was studied in~\cite{ALICE:2017ryd} where \ac{EMC} modules with and without TRD modules in front were present. An uncertainty of 4.2\% for the \ac{EMC} was assigned, which is one of the leading uncertainties. For the PHOS detector, the uncertainty is estimated by comparing data with and without magnetic field, which affects the converted photons in the detector material differently. An uncertainty of 2\% was found, which is much lower compared to the \ac{EMC} due to less material in front of the \ac{PHOS}.

	Systematic uncertainties on the e$^{\pm}$ track reconstruction, selection, and the subsequent \ac{PCM}-photon PID are small compared to the material budget and the signal extraction, due to the excellent tracking performance of \acs{ALICE}. In the case of the \ac{PCM}-$\gamma\gamma^{\star}$ measurement, the e$^{\pm}$ track reconstruction and selection also includes primary e$^{\pm}$ tracks. Consequently, the systematic uncertainties are about 2\%, which is slightly larger compared to all other methods ($< 0.5\%$).
	
	For the calorimeter-based invariant-mass reconstruction methods, the uncertainty of the cluster description, which comprises variations of the selection criteria given in \Tab{tab:CutCalo}, is one of the leading uncertainties. For \ac{EMC}, the uncertainty originating from the description of the \ShowerShape is the largest source within this category, while for the \ac{PHOS}, the uncertainty related to the cluster timing is dominant at low and high \pT. Furthermore, the uncertainty on the cluster energy calibration, explained in detail in~\cite{ALICE:2022qhn} for the \ac{EMC}, is estimated to be about 2\% for both \ac{EMC} and \ac{PHOS}. For \ac{PHOS} the uncertainty is obtained by comparing the energy calibration estimated using \ee track to cluster matching and by using the \piz peak position. 
	
	For the purity-based calorimeter methods, the largest source of uncertainty is the description of the \ShowerShape in the \ac{MC}, which is listed in the cluster description for the \ac{mEMC} and the efficiency for the \ac{mPHOS}. Furthermore, overlapping showers from two or more \piz might not be correctly described by the \ac{MC}, as it depends on the jet fragmentation. By varying the fraction of clusters with overlapping particles, an uncertainty of 5--8\% was found.
	
	To estimate the uncertainty for the signal extraction, the integration window, the fit range for the background as well as the background description is varied. For the \piz, the uncertainty related to the signal extraction is rather small as the \piz significance is high. However, for the \et it becomes one of the dominating uncertainties at low and high \pT.
	
	The multiplicity dependence of the different sources of systematic uncertainties was studied and found to be negligible within the statistical uncertainties for most of the sources, including variations of the cluster, conversion photon, and meson selection as listed in \TabsThree{tab:CutCalo}{tab:CutValuesPCM}{tab:CutGstar}. Multiplicity-dependent systematic uncertainties are assigned for the contribution from out-of-bunch pileup, the signal extraction as well as for the efficiency correction as outlined in the following.
	The contribution from out-of-bunch pileup was found to be up to 7\% for the lowest multiplicity and negligible for the highest multiplicity intervals. 
	Uncertainties assigned for the signal extraction for both the \piz and \et meson are evaluated for each multiplicity interval separately.
	Furthermore, the signal and event losses, by default estimated using PYTHIA8, are highly generator dependent. Thus, results from EPOS LHC and PHOJET are compared to those obtained with PYTHIA8 to estimate the systematic uncertainty.
	Additionally, a systematic uncertainty that accounts for the efficiency correction, which slightly varies with multiplicity depending on the reconstruction method, is assigned by taking half the size of the estimated shift of the efficiency.
	
	Systematic uncertainties for the ratios of the \piz and \et meson \pT~spectra to the inclusive \pT~spectrum, as presented in \Sec{sec:MultDep} cancel for all sources of uncertainties that are not multiplicity dependent. All multiplicity-dependent uncertainties are propagated to the ratios.
	
	\section{Results}
	\label{sec:results}

	\subsection{${\pi^{0}}$~and ${\eta}$~meson inelastic differential cross section} 
	\label{sec:cross-section}
	
	The invariant differential \piz and \et cross sections $E \frac{\mbox{d}^3 \sigma}{\mbox{d}p^3}$ are obtained for each reconstruction method
	\begin{equation}
		E \frac{\mbox{d}^3 \sigma}{\mbox{d}p^3} = \frac{1}{2\pi} \frac{1}{\pTeq} \frac{1}{ \mathcal{L_{\text{int}}} } \frac{P}{ A^{\pi^0(\eta)}  \varepsilon_{\mbox{\tiny rec }}^{\pi^0(\eta)}  } \frac{1}{BR} \frac{N^{\pi^0(\eta)}-N^{\pi^{0}}_{\rm sec}}{\Delta y \Delta \pTeq}. 
		\label{eq:InvXsection}
	\end{equation}
	The corrections for the geometrical acceptance ($A$), reconstruction efficiency ($\varepsilon_{\mbox{\tiny rec }}$), branching ratio ($BR$) and purity ($P$) to the raw meson yield ($N^{\pi^{0} (\eta)}$) discussed in \Sec{sec:Corrections} are applied, including the correction of the raw \piz yield for secondary \piz from weak decays ($N_{\text{sec}}^{\pi^{0}}$). Furthermore, the normalization of the integrated luminosity ($\mathcal{L_{\text{int}}}$) is taken from~\cite{ALICE-PUBLIC-2021-005}, while for the calorimeter-triggered data the enhancement factor is used in addition (see \Tab{tab:events}). Finally, normalizations on the inspected rapidity interval ($\Delta y$) and the \pT interval width ($\Delta$\pT) are applied.
	
	The individual spectra obtained with the different reconstruction methods are combined via a weighted average~\cite{ParticleDataGroup:2024cfk}, taking into account the correlation of the systematic uncertainties between the measurements using the Best Linear Unbiased Estimate (BLUE) method~\cite{Lyons:1988rp,Valassi:2003mu}. 
	The different reconstruction methods are statistically uncorrelated, and it is assumed that there is no correlation of systematic uncertainties between the \ac{PCM}, the EMC, and the \ac{PHOS} measurements.
	However, the hybrid methods \ac{PCM}-EMC and \ac{PCM}-\ac{PHOS} are correlated to the \ac{PCM} method as well as the respective calorimeter method, while the \ac{PCM}-$\gamma\gamma$ and \ac{PCM}-$\gamma\gamma^*$ are only correlated via the PCM-$\gamma$. Uncertainties related to the reconstruction of the conversion photons are fully correlated between the \ac{PCM}-method with respect to the \ac{PCM}-Calo methods or \ac{PCM}-$\gamma\gamma^*$ method. In addition, the systematic uncertainty assigned for the cluster description and cluster energy calibration uncertainty is also fully correlated between the \ac{PCM}-Calo and the Calo methods as well as for the EMC (PHOS) and merged \ac{mEMC} (mPHOS) analyses. The uncertainty related to the signal extraction is assumed to be fully independent between the different methods.
	
	Due to the finite bin width and the steeply falling \piz and \et spectrum, the bin center of each bin does not represent the \pT value of the measured y-value in the case of an unbinned spectrum~\cite{Lafferty:1994cj}. Hence, the bin centers are shifted in \pT by assuming the spectral shape obtained by a combined, modified \ac{TCM} parametrization~\cite{Bylinkin:2015xya}
	\begin{eqnarray}
		\label{eq:TCM}
		E \frac{\mbox{d}^3 \sigma}{\mbox{d}p^3} = 
		A_e \exp\left(-\frac{\sqrt{p_{\text{\tiny T}}^2 + M^2} - M}{T_{e}}\right)
		+ A\left( 1+ \frac{p_{\text{\tiny T}}^2}{T^2 n}\right)^{-n - m\cdot p_{\text{\tiny T}}}.
		\label{eq:TwoComponentModel}
	\end{eqnarray}
	Here, $M$ is the meson mass in \GeVmass,  $A_e$ and $A$ (in pb \GeV$^{-2}c^3$) are the normalization factors, $T_e$ and $T$ are the temperature parameters in \GeV, and $n$ is the power law order with $m\cdot$\pT being a \pT dependent term to account for deviations of the \piz spectrum from a pure power-law form at high transverse momenta. 
	The relative correction due to the finite bin width is below 1\% for the \piz and below 2\% for the \et above \pT~=~0.8~\GeVc while the first point has a correction of $\approx$~3.6\%. For the \etopi~ratio, a shift in \pT cannot be performed as the \piz and \et spectra may have different spectral shapes and therefore need slightly different corrections. Hence, for the \etopi~ratio, the shift is performed along the $y$-coordinate for both the \piz and \et spectra. The resulting correction is below 1\% for all \pT bins as the spectral shapes of \piz and \et are similar.
	\begin{figure}[t!]
		\centering
		\includegraphics[width=0.495\textwidth]{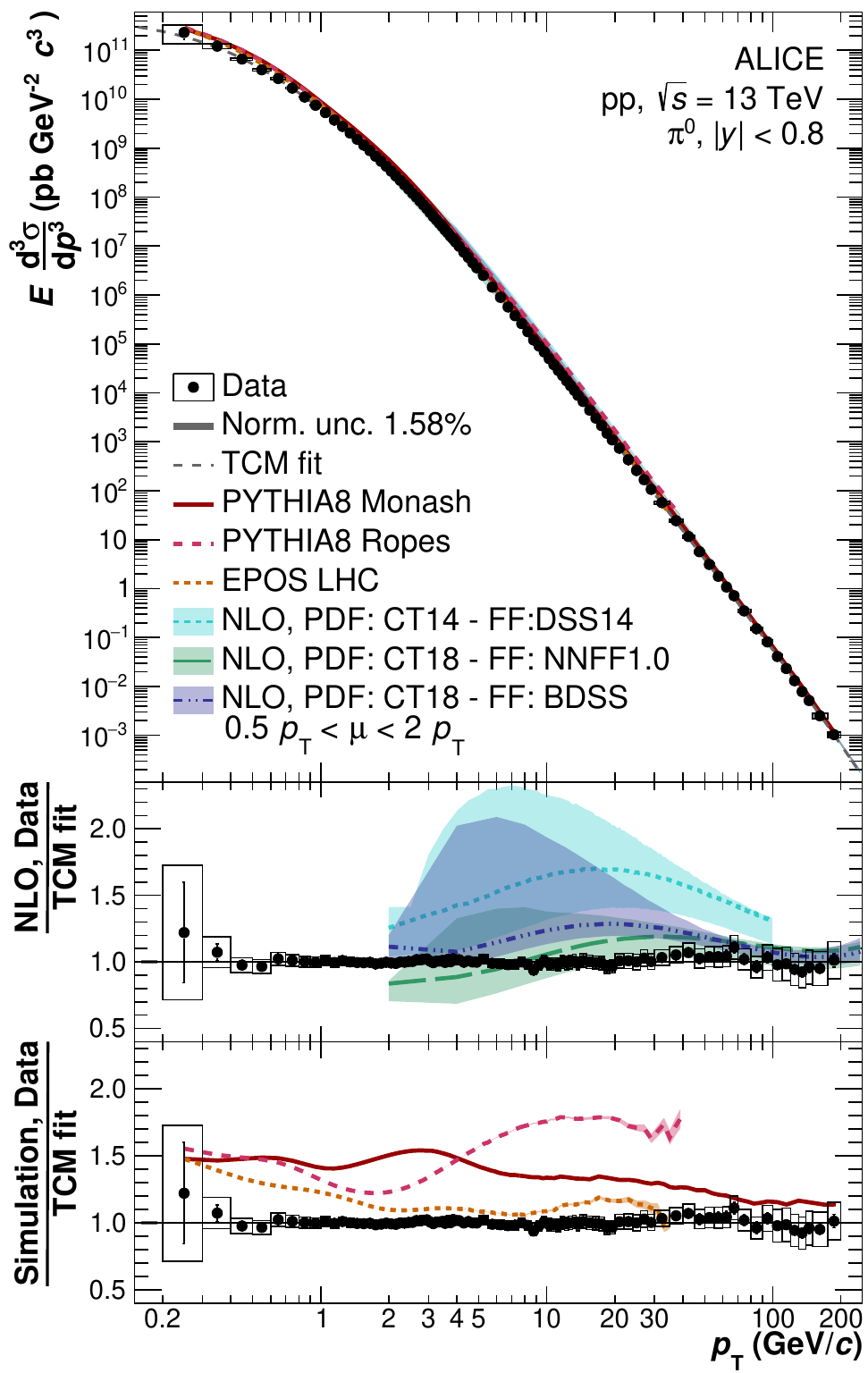}
		\includegraphics[width=0.495\textwidth]{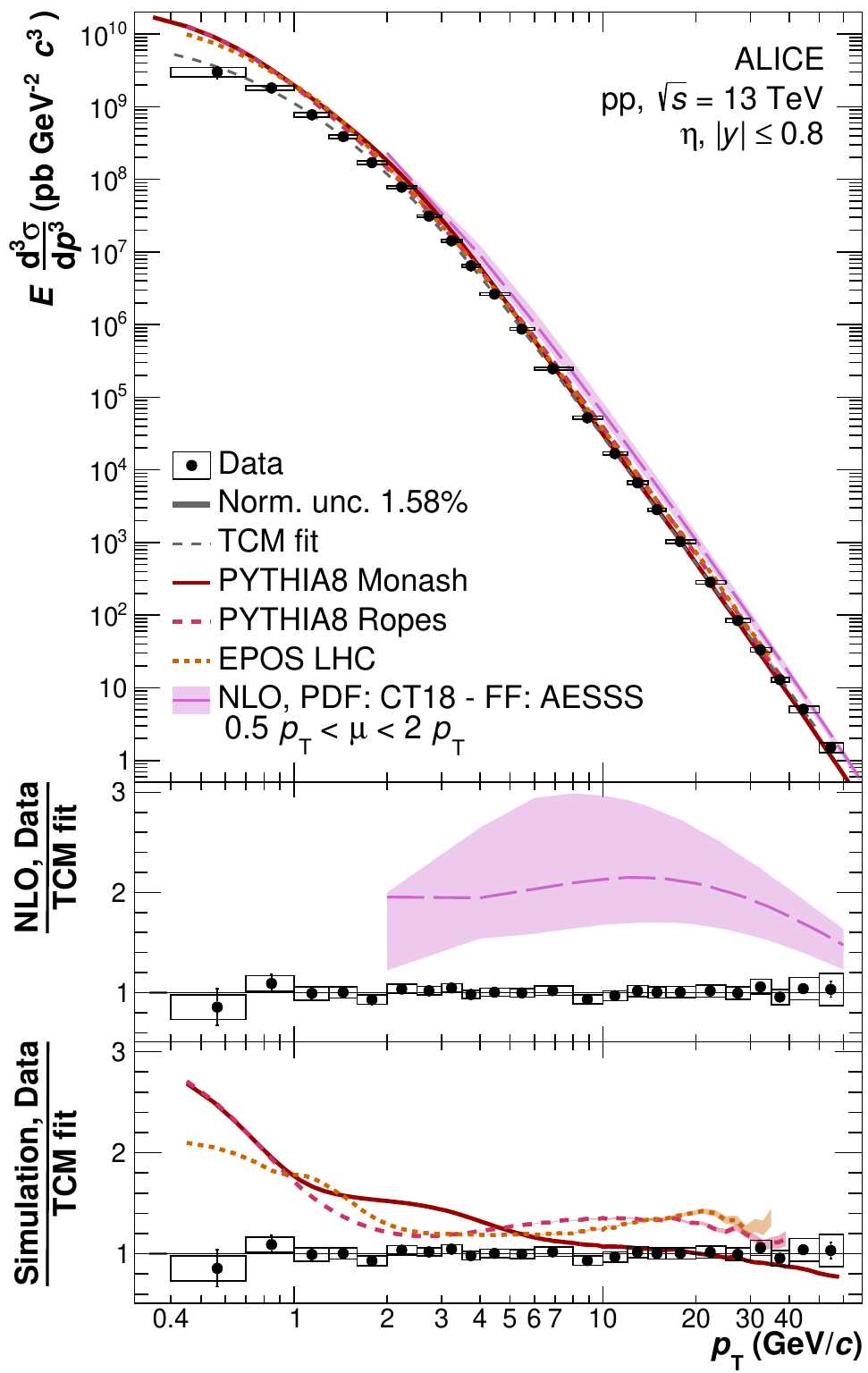}
		\caption{Invariant differential cross section of \piz (left) and \et (right)  versus transverse momentum for \pp collisions at \sT. The data are parametrized with a modified \ac{TCM} model (see \Eq{eq:TwoComponentModel})
			and compared to predictions from PYTHIA8 Monash, PYTHIA8 Ropes, EPOS LHC and predictions from NLO pQCD calculations using recent PDFs and FFs. Ratio plots of the data and model calculations to the modified \ac{TCM} fit of the data are shown in the lower panels. Statistical error bars are represented by vertical bars, and systematic uncertainties are shown as boxes. }
		\label{fig:Pi0EtaXSec}
	\end{figure}
	\begin{figure}[t!]
		\centering
		\includegraphics[width=0.495\textwidth]{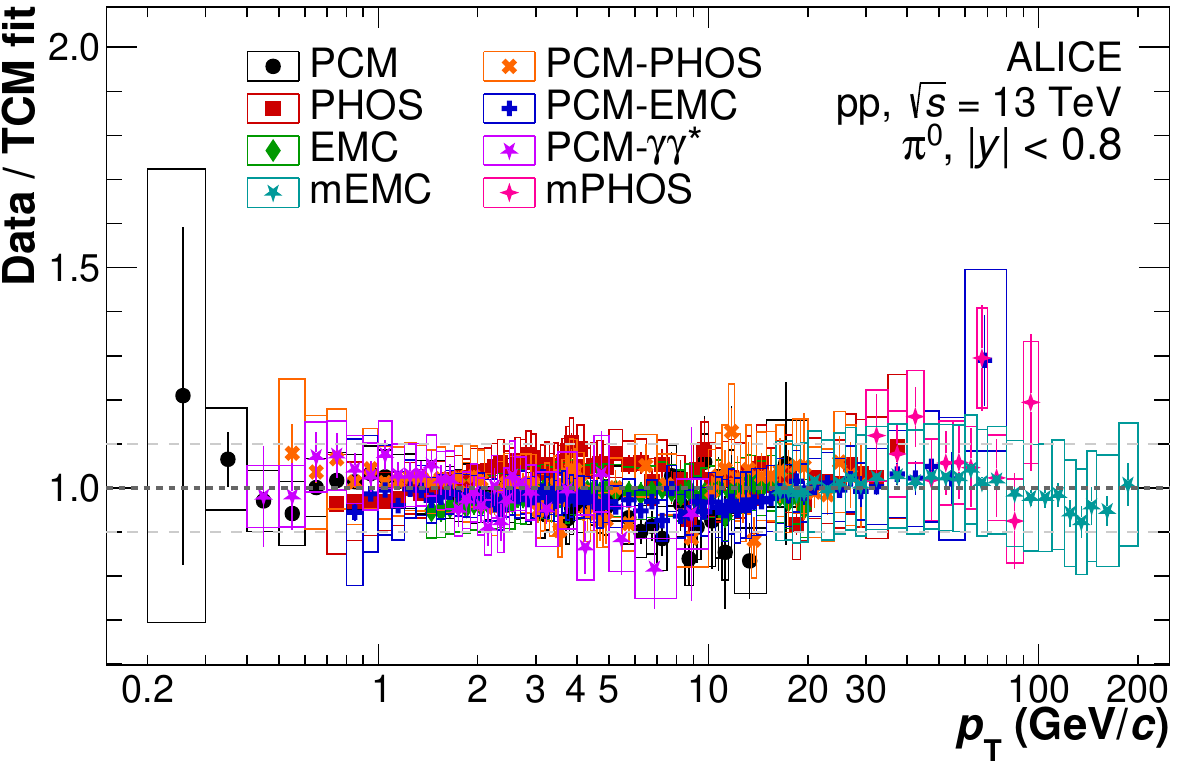}
		\includegraphics[width=0.495\textwidth]{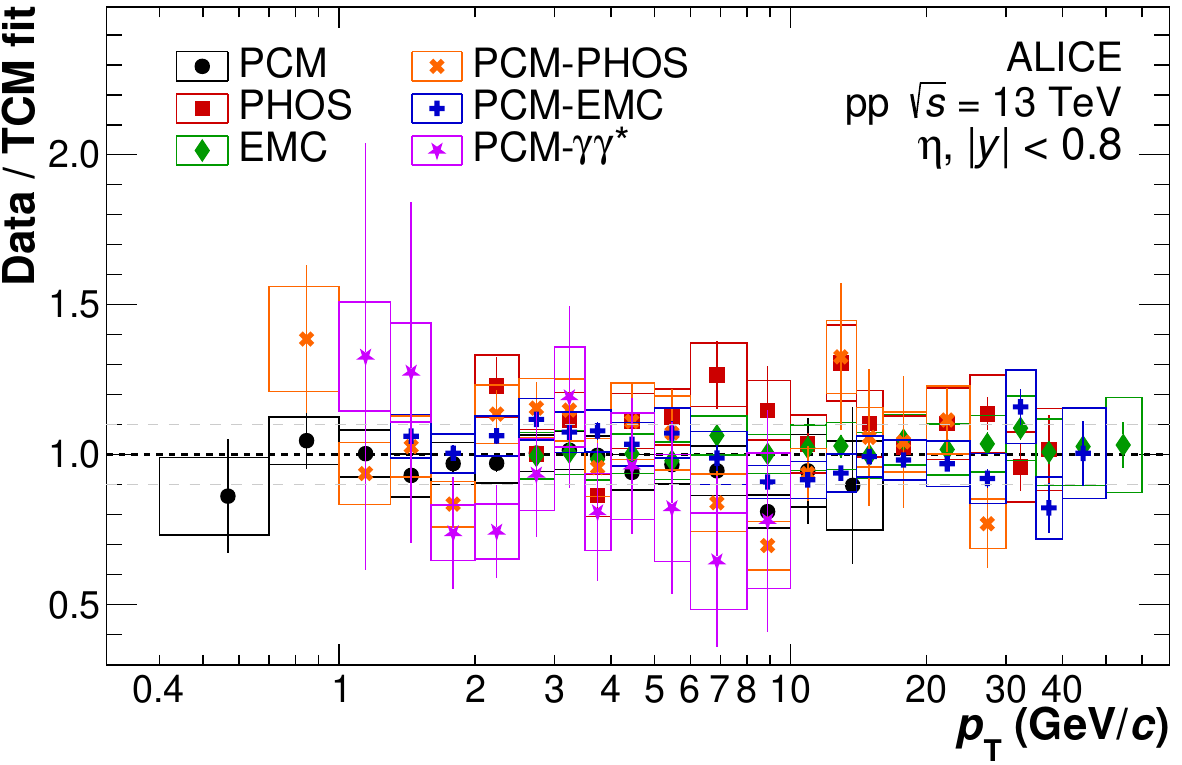}
		\caption{Ratio between each individual \piz (left) and \et (right) invariant differential cross section measurement, and the \ac{TCM} fit to the combined spectrum. The statistical uncertainties are represented as vertical error bars whereas the systematic uncertainties are shown as boxes. }
		\label{fig:Ratio2Ind}
	\end{figure}
	\\
	\Figure{fig:Pi0EtaXSec} shows the differential invariant cross section for the neutral pion (left) and the \et meson (right) for a transverse momentum range of $0.2 <$~\pT~$< 200$~\GeVc and $0.4 <$~\pT~$< 60$~\GeVc, respectively.
	The spectra are parameterized using a modified \ac{TCM} parametrization (see \Eq{eq:TwoComponentModel}).
	As~\cite{Sassot:2010bh} suggests, the deviation from the power-law shape at high \pT arises from the running coupling constant $\alpha_{s}$ and the scale evolution of the PDF and FF. 
	All free parameters of the fit are given in 
	\Tab{tab:TCMParams}.
	To compare the different measurements, the ratio of each one to the fit of the combined measurement is shown in 
	\Fig{fig:Ratio2Ind}. A very good agreement among the different measurement methods is obtained within the uncertainties over the full \pT range.

	\begin{table}[t!]
		\centering 
		\caption{Parameters of the modified \ac{TCM} parametrization given in \Eq{eq:TCM} for the neutral pion and the \et meson as shown in \Fig{fig:Pi0EtaXSec}.}
		\small 
		\begin{tabular}{c | c c c c c c}
			\toprule
			& $A_{e}$ $\times 10^{-9}$ $\left(\frac{\text{pb}\times c^{3}}{\text{\GeV}^{2}}\right)$ & $T_{e}$ (\GeV) & $A$ $\times 10^{-9}$ $\left(\frac{\text{pb}\times c^{3}}{\text{\GeV}^{2}}\right)$ & $T$ (\GeV) & $n$ & $m$ $\times 10^{3}$ $\left(\frac{c}{\text{\GeV}}\right)$ \\ \hline
			\piz &  $(427\pm 49) $ & $0.157 \pm 0.007$ & $(26\pm 2) $ & $0.65\pm0.01$ & $2.96\pm0.01$ & $(0.30\pm0.05)$ \\
			$\eta$ & $(5.95 \pm 3.61)$ & $0.173 \pm 0.046$ & $(3.15 \pm 0.56)$ & $0.81 \pm 0.03$ & $2.93\pm0.01$ & - \\
			\bottomrule
		\end{tabular}
		\label{tab:TCMParams}
	\end{table}
	The measured invariant differential cross sections are compared to predictions from the event generators PYTHIA8 with the Monash tune and the Ropes variant as well as EPOS LHC. NLO pQCD calculations for the meson production cross sections are also shown in \Fig{fig:Pi0EtaXSec}. The same factorization scale value $\mu$ ($0.5 p_{\mbox{\tiny T}} < \mu < 2p_{\mbox{\tiny T}}$) is used for the factorization, renormalization, and fragmentation scales in these calculations. 
	The spectra obtained with the event generators, as well as the predictions using NLO pQCD calculations, use the same definition of primary particles as is used for the presented data. Hence, the event generators as well as the fragmentation functions in the NLO pQCD calculations contain \piz and \et from strong and electromagnetic decays, but exclude those from weak decays.
	For the \piz, the calculation using CT18 PDF~\cite{Hou:2019jgw} and NNFF1.0 FF~\cite{Bertone:2017tyb} describes the measurement within the uncertainties. In contrast, the calculation using CT14 PDF~\cite{Dulat:2015mca} together with the DSS14 FF~\cite{deFlorian:2014xna} overestimates the production rate, as also seen in previous results of neutral pions at LHC energies~\cite{ALICE:2021est}. The calculation using the BDSS FF~\cite{FF_BDSS} describes the data well at low and large \pT and slightly overestimates the data for $5 \lessapprox$~\pT~$\lessapprox 50$~\GeVc. 
	It is noteworthy that the NNFF1.0 FF is tuned exclusively to data from electron-positron annihilation, showcasing the universality of the FF. Furthermore, it uses a neural network approach to describe the data, giving it more free parameters and fewer constraints on the shape of the FF than traditional approaches as used in the DSS and BDSS FF. The DSS14 FF incorporates data from RHIC and early LHC measurements, including the neutral pion measurement in pp collisions at $\sqrt{s}$~=~7~\TeV~\cite{ALICE:2012wos}. The BDSS FF additionally incorporates a large fraction of recent neutral and charged pion results from RHIC and LHC. In contrast to the DSS14 FF, the BDSS FF achieves a good description of the data with a consistent set of FF by taking the theoretical scale dependence in the global QCD analysis into account~\cite{FF_BDSS}. The NLO pQCD prediction of the \et meson production using CT18 PDF and the AESSS FF~\cite{Aidala:2010bn} overestimates the \et as also seen in~\cite{ALICE:2021est}.\\
	The prediction from PYTHIA8 overestimates the production of neutral pions over nearly the full \pT range for both variants. A similar result can be seen for charged pions~\cite{ALICE:2020jsh}. The prediction from EPOS LHC is in better agreement with the \piz data compared to PYTHIA8, especially for \pT~$>$~1~\GeVc.
	Predictions for the \et meson by both PYTHIA8 and EPOS LHC do not describe the spectrum and have a different \pT dependence than the measurement.

	\subsection{\et/\piz~ratio and \textit{m}$_{\text{\tiny{T}}}$ scaling}
	\label{sec:EtaToPi0}
	
\begin{figure}[!t]
		\centering
		\includegraphics[width=0.49\textwidth]{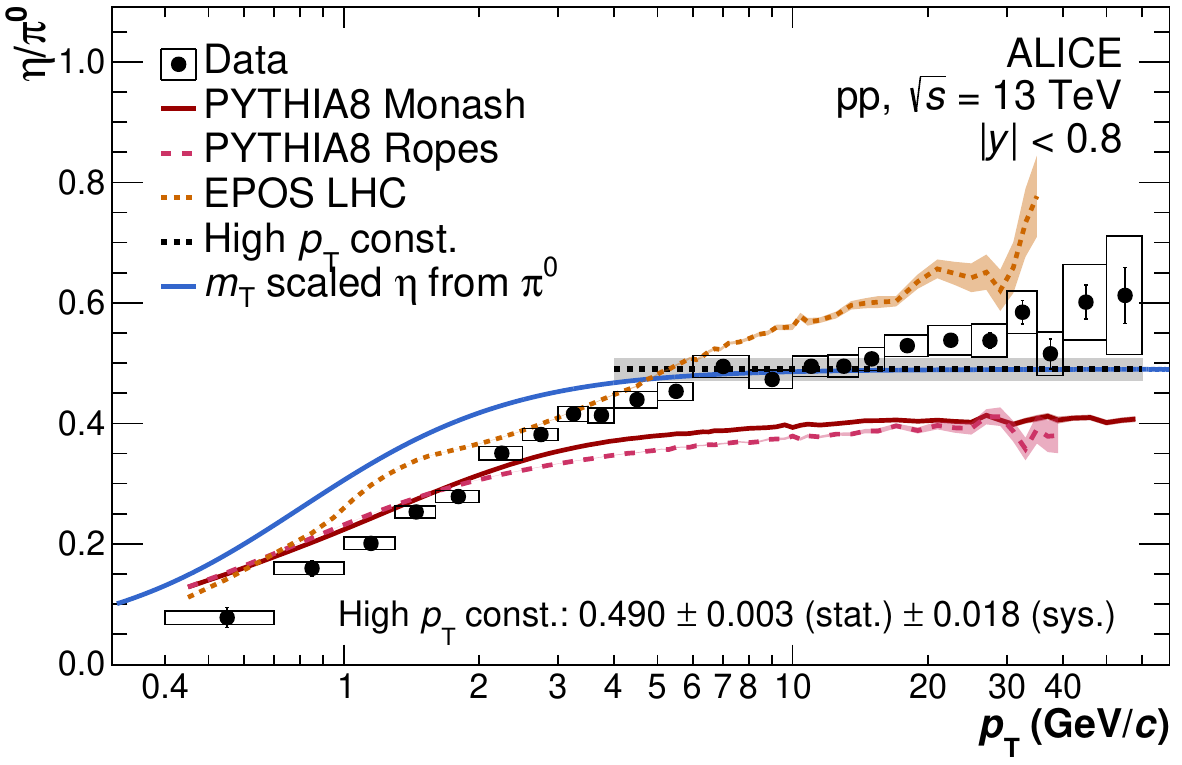}
		\includegraphics[width=0.49\textwidth]{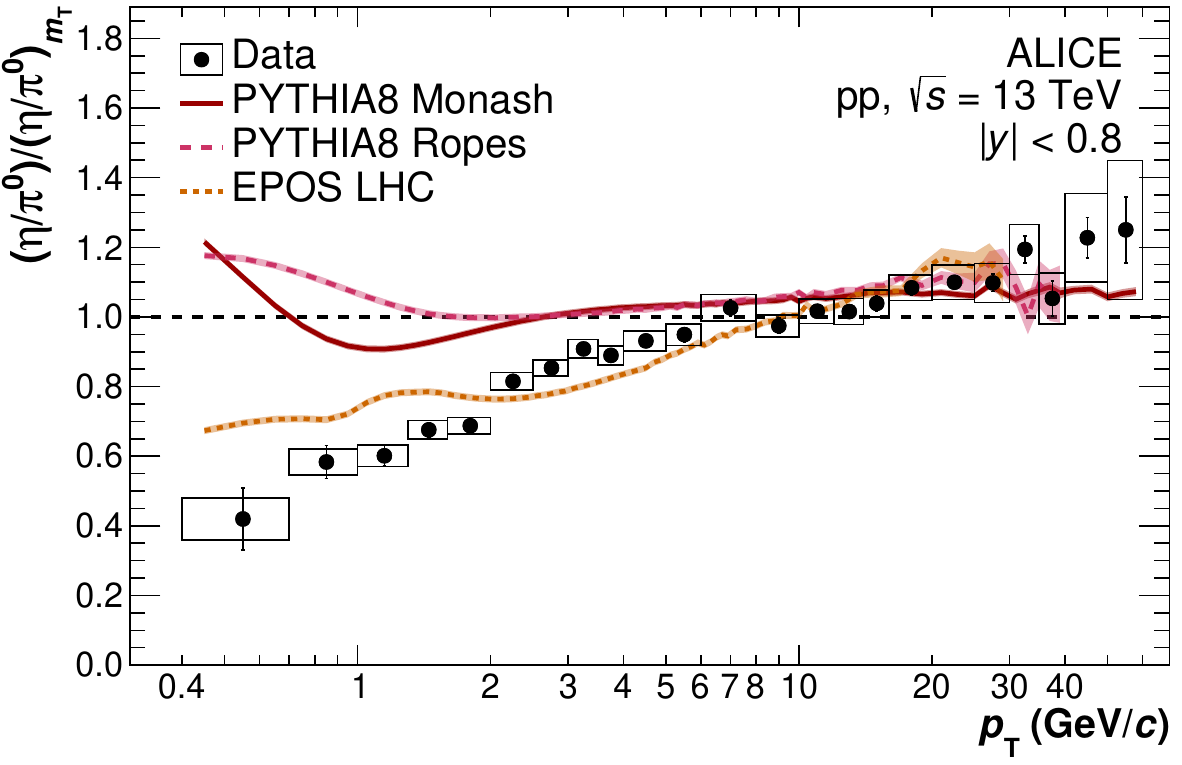}
		\caption{ Left: the \etopi~ratio as a function of \pT compared to expectations from PYTHIA8, EPOS LHC, and \mT scaling. Right: Ratio of data and model predictions to the respective \mT scaling prediction.}
		\label{fig:EtaToPi}
	\end{figure}

	\Figure{fig:EtaToPi} (left) shows the \etopi~ratio as a function of \pT from 0.4~\GeVc up to~60~\GeVc. The result is obtained by studying the \piz production cross section in the same \pT intervals as used for the \et meson for each reconstruction method. Systematic uncertainties are evaluated on the \etopi~ratio directly to cancel systematic uncertainties that are common for both the \et and the \piz. The combination of the \etopi~ratio follows the same principles as used in the combination of the production cross sections discussed before (see \Sec{sec:cross-section}).
	The \etopi~ratio significantly rises with \pT for \pT~$<$~4~\GeVc as expected from the mass difference between the mesons. According to \mT scaling, the ratio should saturate at high \pT. Hence, a constant fit above \pT~=~4~\GeVc is performed, giving a value of 0.490 $\pm$ 0.003 (stat) $\pm$ 0.018 (sys), compatible with a universal asymptotic value of 0.487 $\pm$ 0.024 obtained in pp collisions at LHC and RHIC energies~\cite{Ren:2021pzi}. To verify the validity of \mT scaling, the expected ratio given by \mT scaling is also shown in \Fig{fig:EtaToPi} (left) for comparison.
	It is calculated using the parametrization ($P_{\pi^{0}}$) of the \piz cross section (\Tab{tab:TCMParams}) evaluated at the transverse mass of the \et meson:
	$E\,\mbox{d}^3 N^{\eta}/\mbox{d}p^{3} = C_{m} P_{\pi^{0}}
	\left(\sqrt{p^2_{\rm{T}}+m^2_{\eta}-m^2_{\pi^0}}\right)$, 
	with $C_{m}$ being the aforementioned high-\pT constant. 
	For \pT~$> 4~\GeVc$, the measurement is roughly compatible with \mT scaling. However, the data suggest a rising \etopi ratio. This is quantified by a linear parametrization of the data for \pT~$> 4~\GeVc$, yielding a slope parameter of $0.0038 \pm 0.0016$ and hence a significance of about $2.4\sigma$, including both statistical and systematic uncertainties.
	For \pT~$< 4$~\GeVc the data is increasingly deviating from the \mT-scaling expectations as \pT decreases, reaching a deviation of 60\% at \pT~=~0.550 \GeVc. 
	This was observed before~\cite{Agakishiev:1998mw,ALICE:2017ryd,ALICE:2018vhm,ALICE:2021est,ALICE:2020jsh} and can be explained by a large contribution from feed-down to the \piz spectrum that is largest at low \pT~\cite{Agakishiev:1998mw,Altenkamper:2017qot,Ren:2021pzi}. 
	The predictions from PYTHIA8 underestimate the measured \etopi~ratio above \pT $\approx$ 3 \GeVc which was also observed in previous comparisons of the \etopi~ratio at lower center-of-mass energies~\cite{ALICE:2017ryd,ALICE:2012wos}.
	The prediction from the EPOS LHC event generator overestimates the measured \etopi~ratio over the whole \pT range and does not describe the shape of the data.
	
	To further investigate the validity of \mT~scaling, the ratio of the measured \etopi~ratio as well as the \etopi~ratio obtained from the event generators is compared to a \mT-scaling prediction, which are presented in \Fig{fig:EtaToPi} (right) as a function of \pT. The \mT-scaling prediction was determined for the data and the predictions from the \ac{MC} simulation event generators individually. While both PYTHIA8 tunes follow the \mT-scaling prediction within approximately 20\%, the deviation of both EPOS LHC and the measurement to the \mT-scaling prediction increases up to 40\% and 60\% respectively, at \pT~$\approx$~0.5~\GeVc. 
	
	\subsection{ \textit{x}$_{\text{\tiny{T}}}$ scaling}
	\label{subsec:xT}
	
	According to pQCD calculations, the production cross section, as presented in \Sec{sec:cross-section}, can be expressed as a function of \xT (\xT~=~2\pT/$\sqrt{s}$ at $y \approx 0$) and $\sqrt{s}$~\cite{Arleo:2010kw, Arleo:2009ch}:
	\begin{equation}
		\label{eq:xT}
		E \frac{\mbox{d}^3 \sigma}{\mbox{d}p^3} = 
		F(x_{\mbox{\tiny T}})/\sqrt{s}^{n(x_{\mbox{\tiny T}}, \sqrt{s})}
	\end{equation}
	The power-law exponent $n(x_{\text{T}}, \sqrt{s})$ in \Eq{eq:xT} is approximately constant in the perturbative region above \pT~$\approx$~3~\GeVc. A slight dependence is introduced by the running coupling of $\alpha_{s}$ and the scale evolution of the PDF and FF~\cite{Sassot:2010bh} as discussed previously. If \xT scaling holds, the production cross section of hadrons as a function of \xT is universal for different center-of-mass energies if scaled with $\sqrt{s}^{n(x_{\text{T}},\sqrt{s})}$, as \Eq{eq:xT} suggests. The exponent $n(x_{\text{T}},\sqrt{s})$ can be experimentally estimated as $n(x_{\text{T}},\sqrt{s_\text{1}},\sqrt{s_\text{2}})$ for all possible combinations of two hadron production cross sections ($\sigma_{\mathrm{inv}}(\xT)$) of the same species at two different center-of-mass energies ($\sqrt{s_{\text{1}}}$ and $\sqrt{s_{\text{2}}}$) using \Eq{eq:nxT}. \\
	\begin{equation}
		n(x_\text{\tiny T}, \sqrt{s_{1}}, \sqrt{s_{2}}) = -\frac{\ln(\sigma_{\mathrm{inv}}(\sqrt{s_{1}},x_\text{\tiny T})/\sigma_{\mathrm{inv}}(\sqrt{s_{2}},x_\text{\tiny T}) )}{\ln(\sqrt{s_{1}}/\sqrt{s_{2}})}.
		\label{eq:nxT}
	\end{equation}
	
	\begin{figure}[t!]
		\begin{minipage}[t]{.461\textwidth}
			\vspace{-\topskip}
			\includegraphics[width=0.99\textwidth]{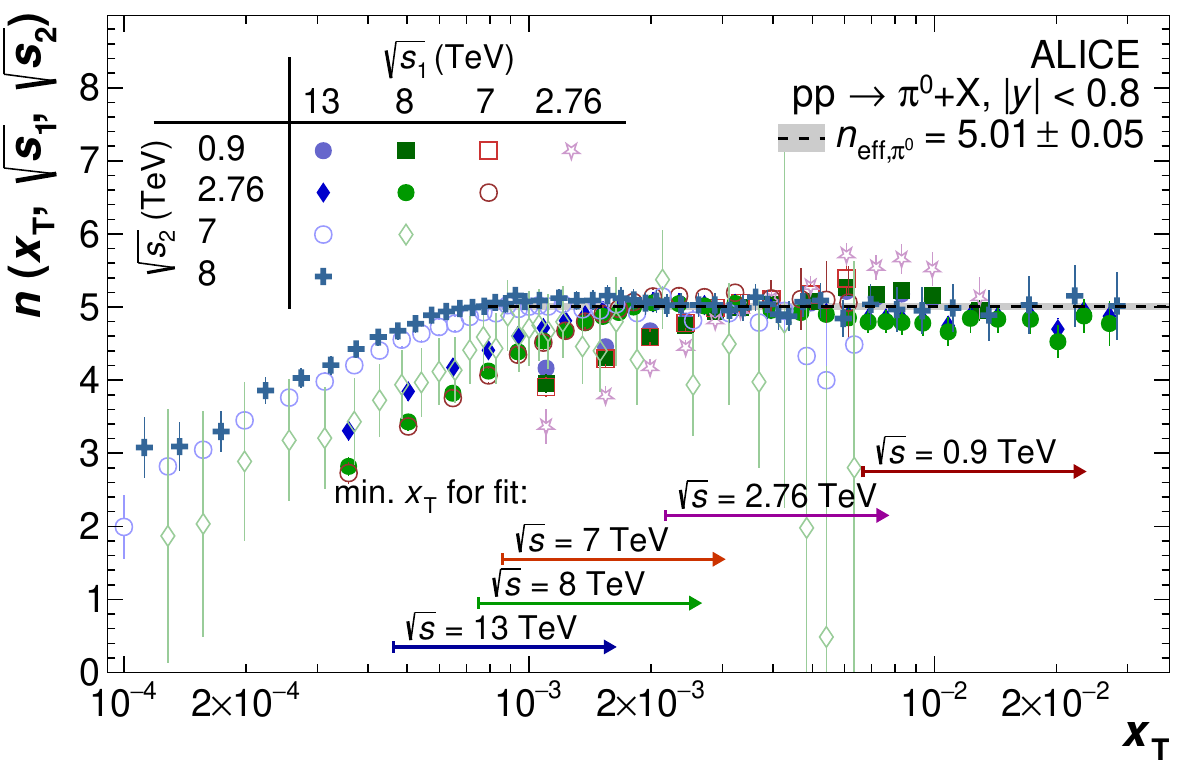}
			\includegraphics[width=0.99\textwidth]{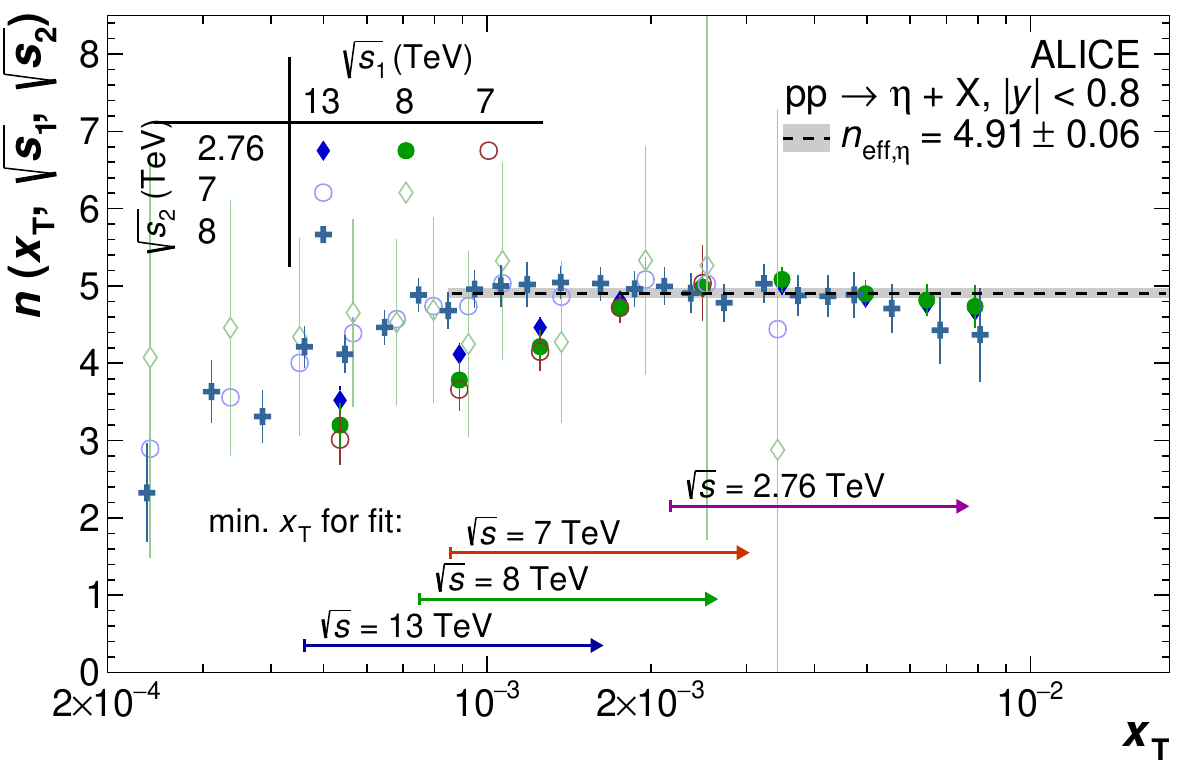}
		\end{minipage}%
		\hspace{0.1cm}
		\begin{minipage}[t]{.534\textwidth}
			\vspace{-\topskip}
			\includegraphics[width=0.99\textwidth]{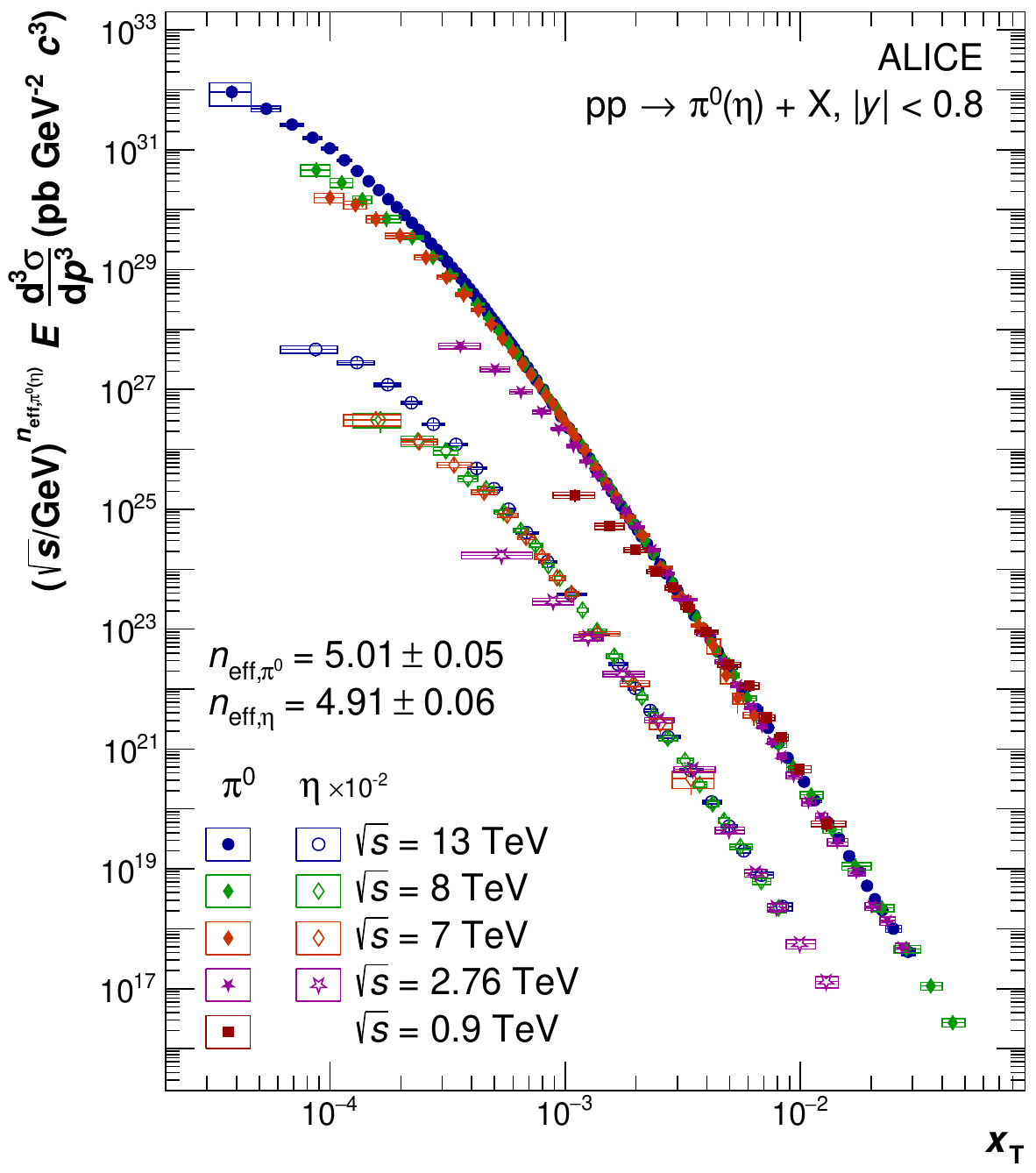}
			
		\end{minipage}%
		\caption{ (left) Parameter $n$ as a function of \xT for several \piz (top) and \et (bottom) spectra ratios at different collision energies. (right) Scaled differential invariant cross section of \piz (full markers) and \et (open markers) as a function of \xT at different collision energies from \s= 0.9 \TeV to \s = 13 \TeV~\cite{ALICE:2012wos,Abelev:2014ypa,Acharya:2017hyu,ALICE:2017ryd}.} 
		\label{fig:nPar_xT}
	\end{figure}
	
	The validity of \xT scaling was tested for recently published charged hadron measurements at LHC energies~\cite{ALICE:2020jsh,CMS:2011mry} and previously also at lower collision energies~\cite{CDF:2009cxa,CDF:1988evs,STAR:2006xud,PHENIX:2007kqm,PHENIX:2008sgl,PHENIX:2011rvu,UA1:1989bou}. Tests of \xT scaling, using the measured \piz and \et meson production cross sections at \sT, together with previous measurements of these particles in \pp collisions at $\sqrt{s}$ = 0.9, 2.76, 7 and 8 \TeV~\cite{ALICE:2012wos,ALICE:2017ryd,Acharya:2017hyu}, are presented in the \xT range from 6$\times10^{-4}$ to 3$\times10^{-2}$ and 6$\times10^{-4}$ to 9$\times10^{-3}$ for the \piz and the \et mesons, respectively. 
	\Figure{fig:nPar_xT}{ (left)} shows the exponent $n(x_{\text{T}}, \sqrt{s_{1}}, \sqrt{s_{2}})$ for all available combinations of measurements for the \piz (top) and the \et (bottom) as a function of \xT. Arrows indicate the perturbative regime for each center of mass energy which is expected to start at \pT = 3 \GeVc. Below that point, a decrease of $n$ can be perceived as soft physics, not calculable in pQCD, dominating the particle production. Above, a plateau region is observed that coincides for all pairs in the perturbative regime. For data points including the data from \eight and \sT, where the \pT~spectra reach up to \pT~=~200~\GeVc, a slight decrease of $n$ with rising \xT can be seen. This trend is expected, as a pure power-law term is not able to describe the spectrum for \pT~$\gtrapprox$~3~\GeVc as shown in \Sec{sec:cross-section}. However, as the dependence on \xT is only mild, an approximate scaling still holds true.
	To estimate the effective power-law scale ($n_{\text{eff}}$), a combined constant fit is performed using all data points in their respective perturbative regime. For the neutral pion, a value of $n_{\text{eff}, \pi^{0}} = 5.01 \pm 0.05$, while for the \et meson $n_{\text{eff}, \eta} = 4.91 \pm 0.06$ is observed. The two values are in agreement within their respective uncertainties, even though their \xT range as well as the available number of input spectra are different. The fit uncertainty is estimated from the uncertainties of the fit parameters themselves, as well as by varying the assumption on the start of the perturbative region between \pT~=~2~\GeVc and \pT~=~4~\GeVc.
	\Figure{fig:nPar_xT}{ (right)} shows the scaled \xT spectra for the \piz and the \et. As expected, all spectra follow the same trend in their respective perturbative region, demonstrating the validity of \xT scaling over the \xT range 6$\times10^{-4}$ to 3$\times10^{-2}$ covered by the ALICE neutral meson measurements. The agreement of the scaled \xT spectra is shown in~\cite{ALICE:LNM}, where the ratio of these spectra to the parametrization of the spectrum at \sT is presented. The majority of the data agree within about 10\% in the perturbative region.
	The value of $n_{\text{eff}, \pi^{0}} = 5.01 \pm 0.05$ is found to be in agreement with the value of $n_{\text{eff}, \pi^{\pm}} = 5.04 \pm 0.02$ for charged pions reported in~\cite{ALICE:2020jsh} where similar center-of-mass energies are used. Additionally, it is in agreement with the expectations from pQCD calculations at LHC energies~\cite{Sassot:2010bh}. On the other hand, the exponent $n$ is about 20\% smaller than the one at lower collision energies~\cite{CDF:2009cxa,CDF:1988evs,STAR:2006xud,PHENIX:2007kqm,PHENIX:2008sgl,PHENIX:2011rvu,UA1:1989bou}, in agreement with an increase of hard scattering processes at LHC energies.

	\subsection{Cross section dependence on the charged-particle multiplicity}
	\label{sec:MultDep}

	\begin{figure}[t!]
		\centering
		\includegraphics[width=0.49\textwidth]{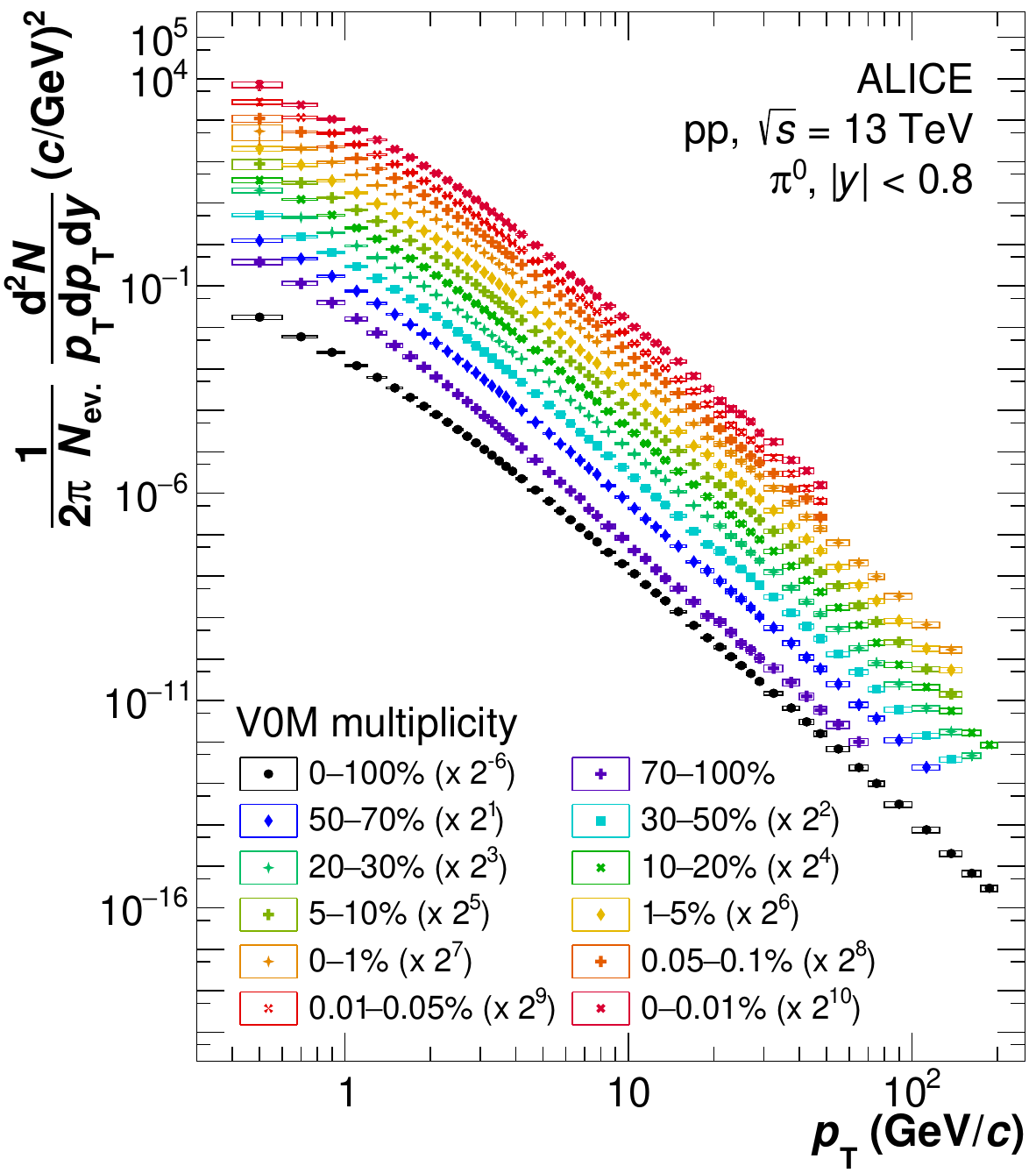}
		\includegraphics[width=0.49\textwidth]{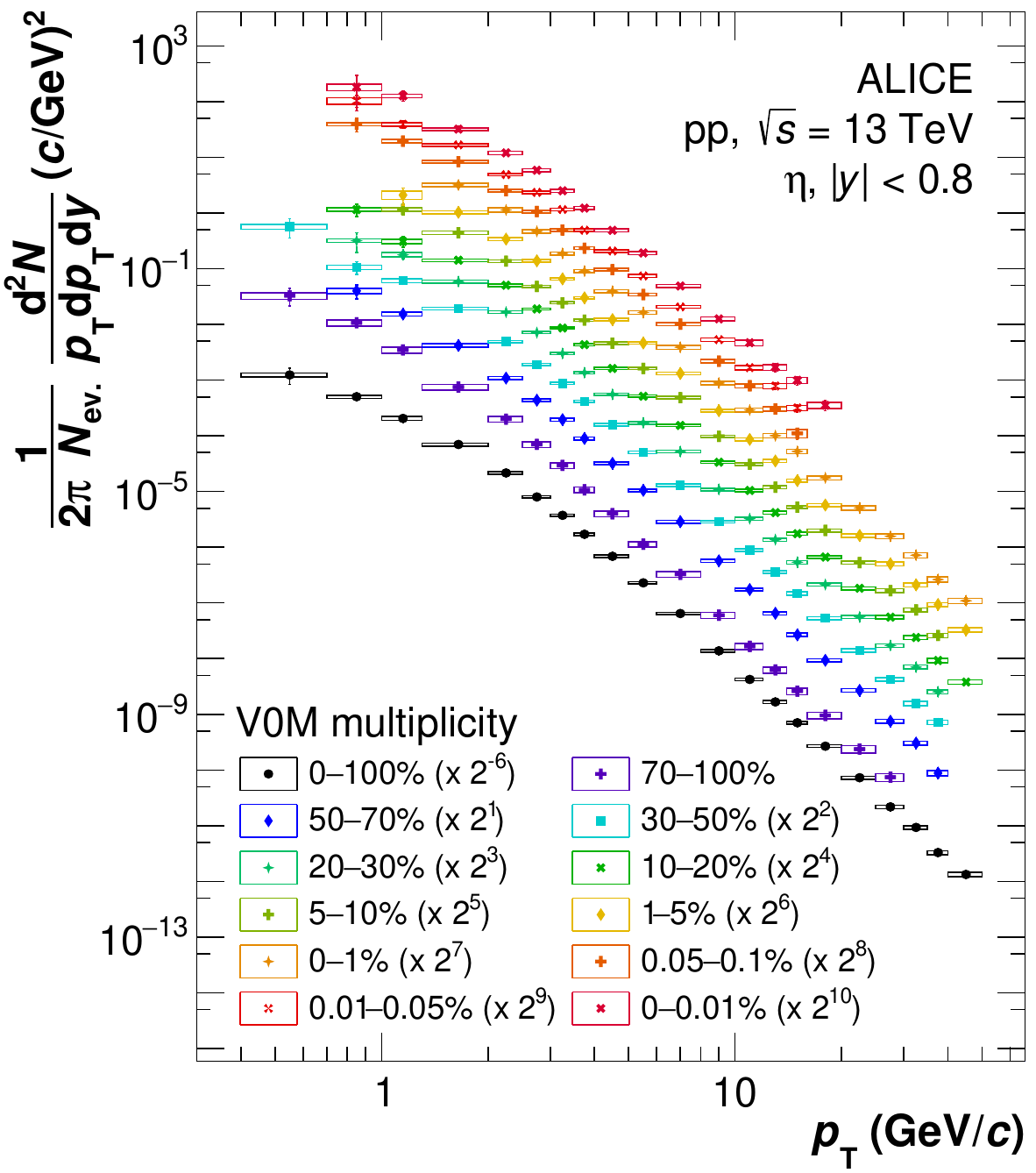}
		\caption{Invariant differential yields of \piz (left) and \et mesons (right) in each selected multiplicity class as defined 
			in \Tab{tab:multClass} and also for the inclusive measurement (0-100\%). The spectra are scaled for better visibility.} 
		\label{fig:Yields_Mult}
	\end{figure}

	\begin{figure}[t!]
		\centering
		\begin{tabular}{lr}
			\includegraphics[width=0.47\textwidth]{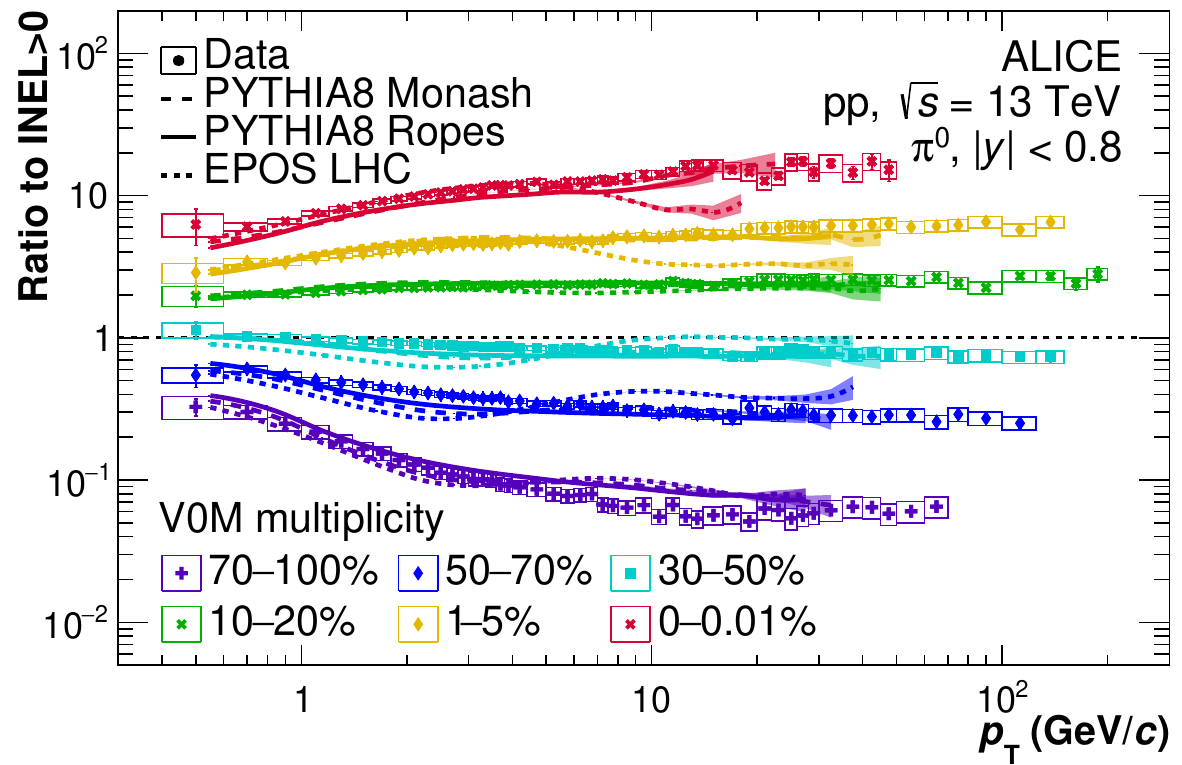}&
			\includegraphics[width=0.47\textwidth]{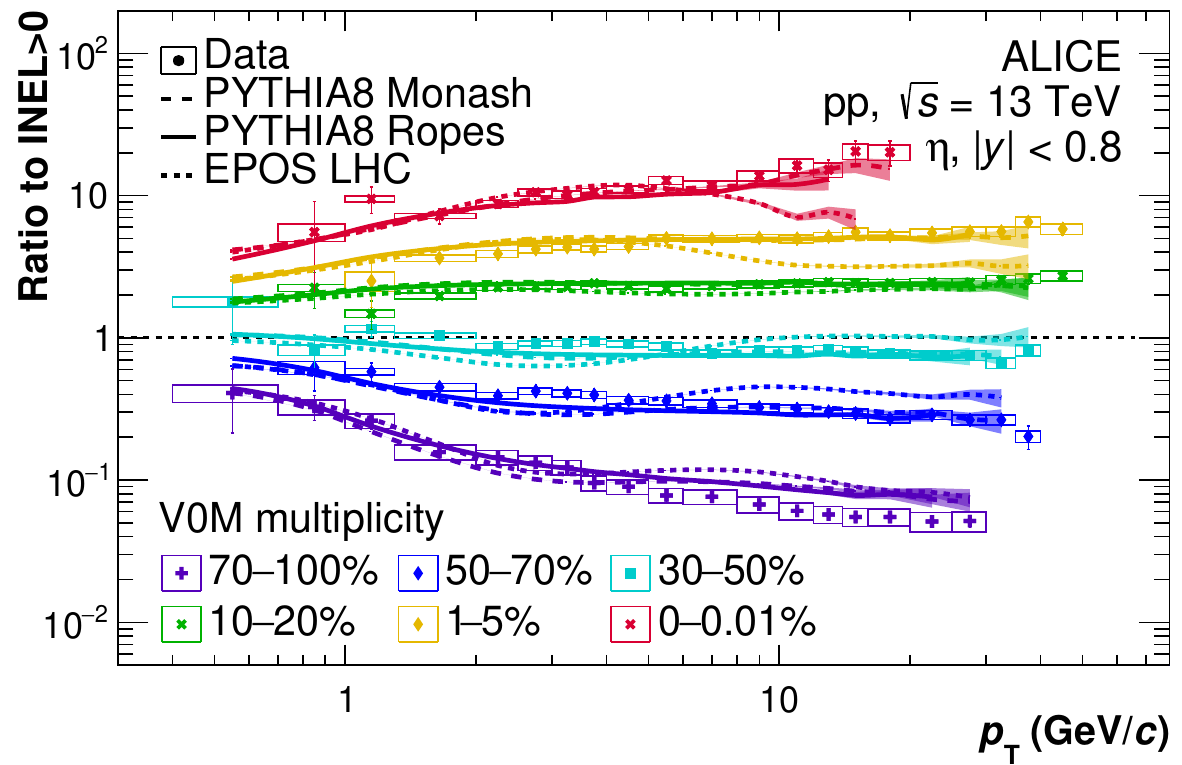}\\
		\end{tabular}
		\caption{Ratios of the invariant differential yields of \piz (left) and \et mesons (right) to the spectra obtained in the INEL$>$0 event class together with predictions from PYTHIA8 with the Monash tune and Ropes variant as well as EPOS LHC.  }
		\label{fig:Mult_Ratios}
	\end{figure}
	
	Neutral-meson production is measured in eleven charged-particle multiplicity intervals (see \Tab{tab:multClass}). The multiplicity is estimated using the V0M detector system at forward and backward rapidity to avoid autocorrelations with the presented neutral meson measurement at midrapidity. The highest multiplicity interval covers the 0.01\% collisions with the highest forward charged-particle multiplicities. The extraction of the neutral meson corrected yield follows the same procedure as described for the inclusive measurement, with slight adjustments as mentioned in \Sec{sec:Corrections}.
	\Figure{fig:Yields_Mult} shows the \piz and \et invariant yield as a function of \pT for the eleven V0M charged-particle multiplicity classes as well as for the inclusive case. The \pT coverage for the low and high charged-particle multiplicity intervals is limited compared to the inclusive spectrum as the statistics in each multiplicity interval are lower compared to the inclusive data. The individual spectra are scaled with a constant to allow a visible separation. 
	
	To further investigate differences between the spectra obtained in different charged-particle multiplicity classes, \Fig{fig:Mult_Ratios}
	shows the ratios of the \piz (left) and \et (right) spectra to the \pT~spectra obtained in the full INEL$>$0 event class. The ratios have a strong dependence on \pT and the charged-particle multiplicity, showing a clear rise and hardening of the meson production with multiplicity.
	The hardening of the spectra was previously reported for charged hadrons~\cite{ALICE:2019dfi} as well as for charged pions, kaons, and protons~\cite{Acharya:2020zji}. The presented results for the \piz exhibit the same behavior and extend these findings up to \pt = 50 \GeVc and 200 \GeVc depending on the multiplicity class.
	Comparisons to PYTHIA8 with the Monash tune and the Ropes variant show a reasonable description by the models of both the hardening and the ordering of the \piz and \et spectra ratios. However, the Ropes variant performs slightly better than the Monash tune, as already observed for other multiplicity-dependent measurements~\cite{Acharya:2020zji,ALICERT:2023}. In contrast to PYTHIA8, EPOS LHC fails to describe the spectral shape at high \pT, where it deviates from the hardening with rising charged-particle multiplicity.
	
\begin{figure}[t]
		\centering
		\includegraphics[width=0.48\textwidth]{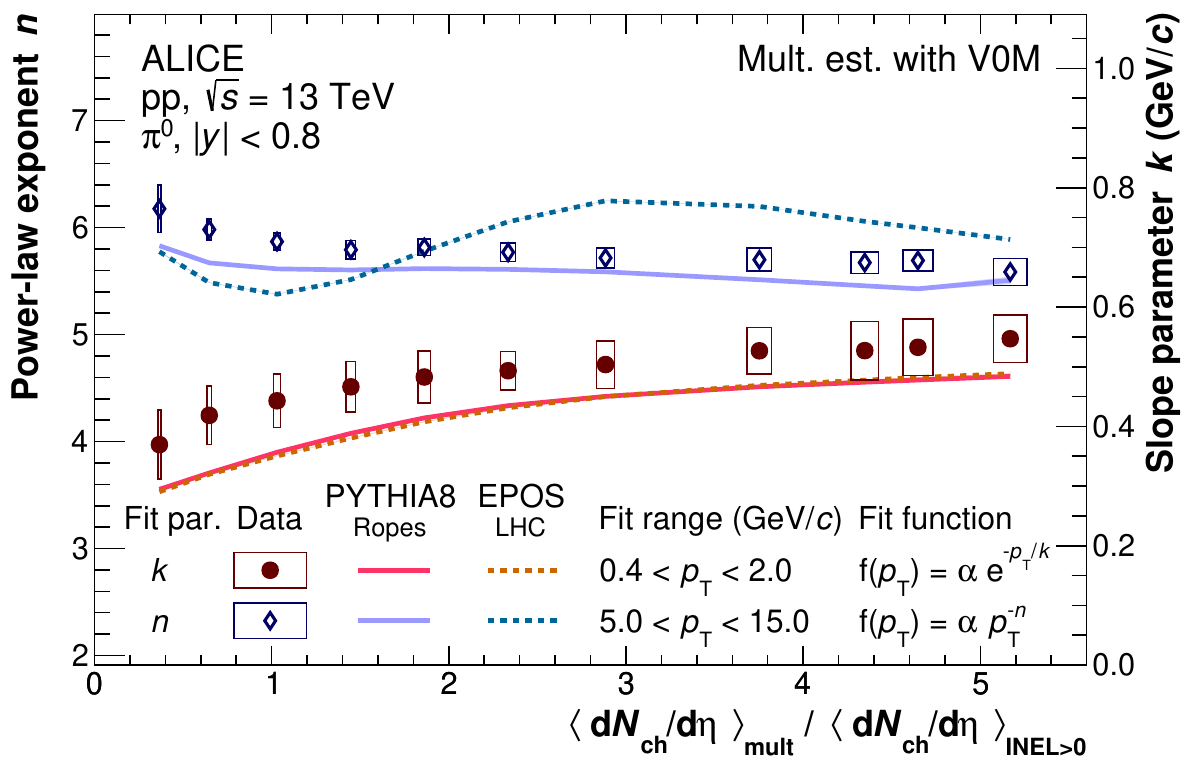}
		\includegraphics[width=0.48\textwidth]{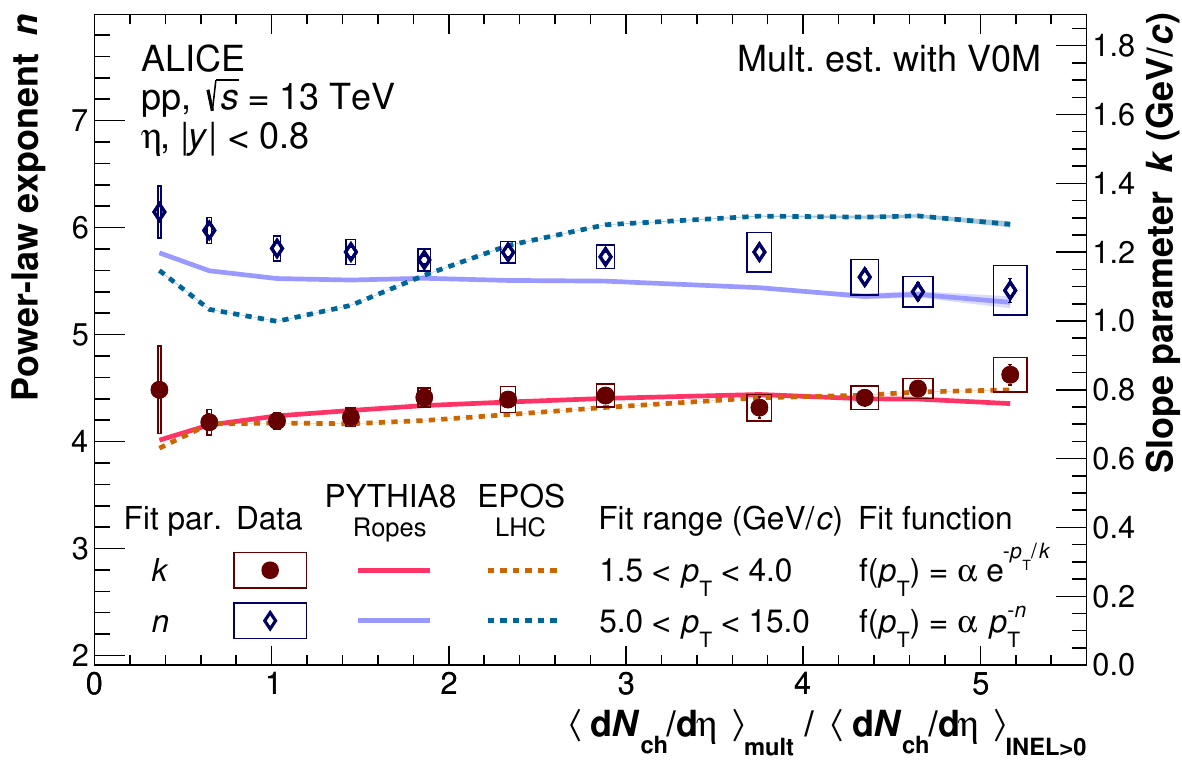}
		\caption{Parameters of a power-law fit and an exponential fit as a function the charged-particle multiplicity density in units of the average multiplicity density for the INEL$>$0 event class for the neutral pion (left) and the \et meson (right). The data are compared to predictions from PYTHIA8 Ropes and EPOS LHC.}
		\label{fig:MultPowerLawNvsMult}
	\end{figure}

	To further quantify the changing shape of the \piz and \et spectra with the charged-particle multiplicity, an exponential fit is performed at low \pT (0.4–2 \GeVc for \piz and 1.5–4 \GeVc for \et), while at high \pT the spectrum is parametrized using a power-law function (5–15 \GeVc for both \piz and \et). The characteristic parameter for each of these two functions is shown in \Fig{fig:MultPowerLawNvsMult} as a function of the normalized mean charged-particle  multiplicity $\langle \text{d}N_{\text{ch}}/\text{d}\eta \rangle_{\text{mult}}$/$\langle \text{d}N_{\text{ch}}/\text{d}\eta \rangle_{\text{INEL}>\text{0}}$ measured at mid rapidity ($|\eta|$~$<$~0.5) for data together with the prediction from PYTHIA8 Ropes and EPOS. The exponential term is underestimated by both models for the \piz, whereas the general trend of the power-law term is described by PYTHIA while EPOS shows a different trend with multiplicity. Comparisons including the PYTHIA8 Monash tune can be found in~\cite{ALICE:LNM}. Discrepancies in the exponential term between the \piz and \et mainly arise from the different \pT intervals as the \piz can be measured to much lower \pT compared to the \et.
	The power-law exponents are in agreement with the measured ones for charged hadrons~\cite{ALICE:2019dfi}.

	\subsection{ Multiplicity dependence of the \texorpdfstring{\etopi}~ratio}
\begin{figure}[t!]
		\begin{minipage}[t]{.48\textwidth}
			\vspace{-\topskip}
			\vspace{0.1cm}
			\includegraphics[width=0.99\textwidth]{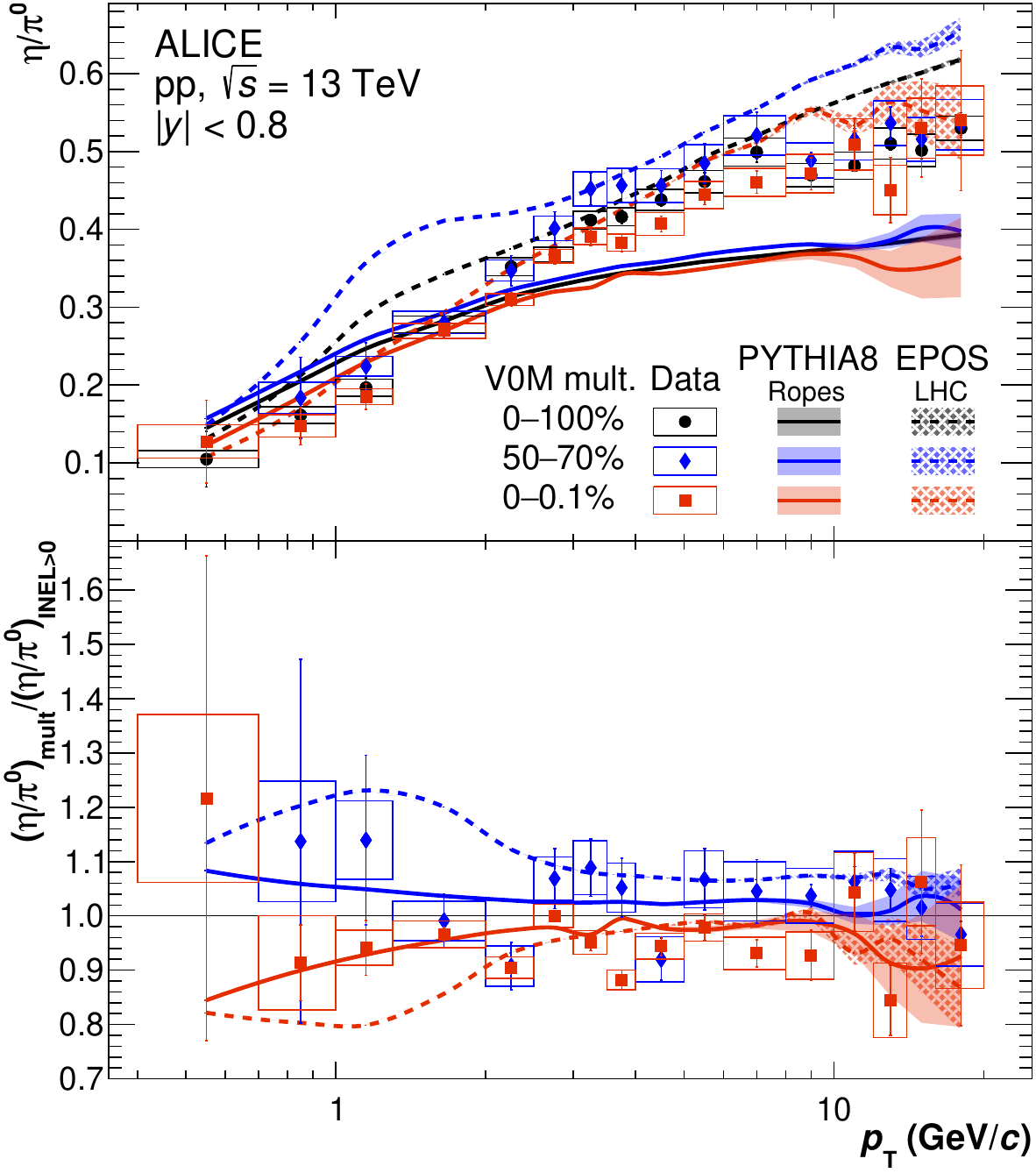}
		\end{minipage}%
		\hspace{0.3cm}
		\begin{minipage}[t]{.48\textwidth}
			\vspace{-\topskip}
			\includegraphics[width=0.99\textwidth]{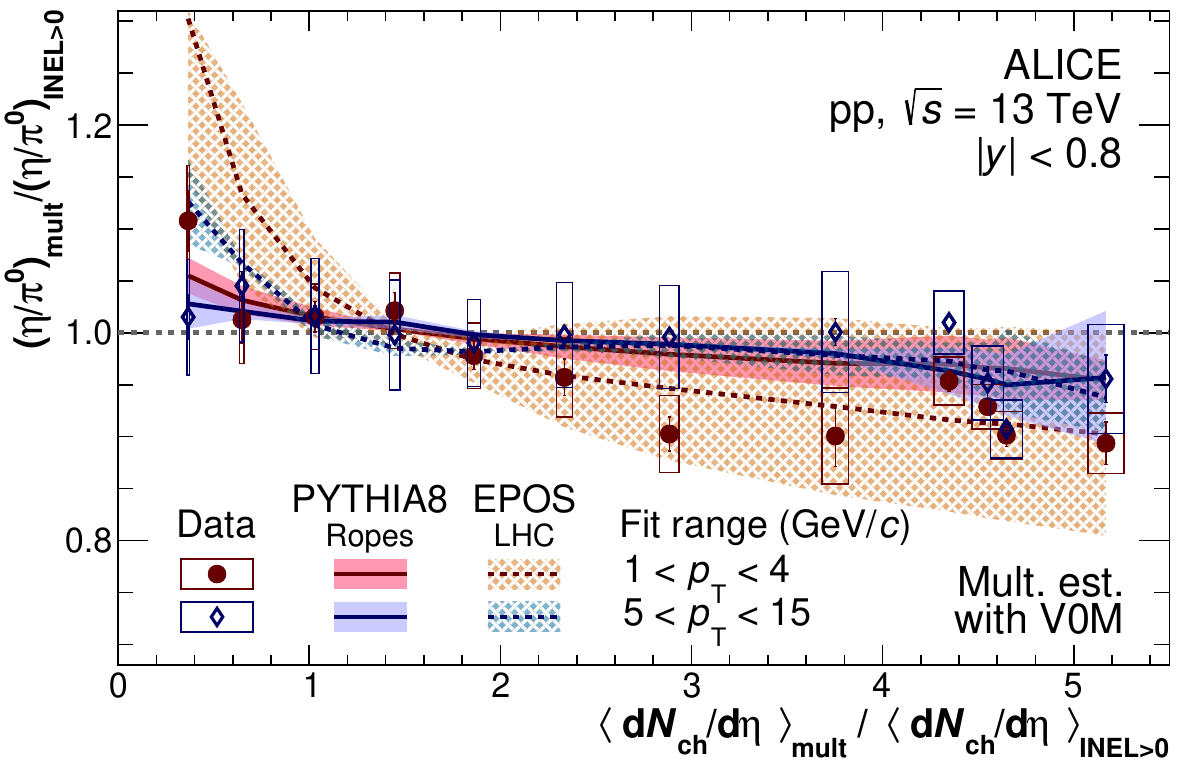}
			\caption{ (left) \etopi~ratio for a low and high multiplicity interval, together with the inclusive \etopi~ratio. Predictions from EPOS LHC and PYTHIA8 Ropes are also shown. The lower panel displays the ratio of the \etopi~ratio in each multiplicity interval to the inclusive \etopi~ratio. (right) Mean values of the \etopi~double ratios as a function of the normalized charged-particle multiplicity for two \pT intervals. The depicted model predictions are shown as lines with bands representing the minimum and maximum values of the double ratio from the left panel in the given \pT interval.}
			\label{fig:Eta2Pi0RatioMult_and_R2individual}
		\end{minipage}%
	\end{figure}

	The \etopi~ratio is calculated in the same eleven charged-particle multiplicity intervals as the individual \piz and \et spectra, following the same procedure as described in \Sec{sec:EtaToPi0} for \ac{MB} collisions. \Figure{fig:Eta2Pi0RatioMult_and_R2individual}{ (left)} shows the \etopi-ratio for a high (0-0.1\%) and low (50-70\%) multiplicity interval, as well as for the inclusive data. To investigate a possible modification of the \etopi~ratio with the charged-particle multiplicity, the ratio to the measurement in the integrated INEL$>$0 event class is performed for each multiplicity interval, obtaining the \etopi double ratios shown in \Fig{fig:Eta2Pi0RatioMult_and_R2individual} for a high and low multiplicity interval. 
	The sources of systematic uncertainties of these ratios include the signal extraction as well as the uncertainty of the efficiency correction. These ratios are calculated for each reconstruction method and then combined via a weighted average using the BLUE algorithm~\cite{Lyons:1988rp,Valassi:2003mu}. 
	
	A quantitative study of the different \etopi double ratios is done by calculating the average value of the ratio in two \pT intervals representing the low (1 $<$ \pT $<$ 4 \GeVc) and high \pT (5 $<$ \pT $<$ 15 \GeVc) regions. The results are shown in  
	\Fig{fig:Eta2Pi0RatioMult_and_R2individual}{ (right)}
	as a function of the normalized mean charged-particle multiplicity measured at midrapidity ($|\eta|$~$<$~0.5). The systematic uncertainties for the data are calculated assuming fully correlated systematic uncertainties of the data points shown in the lower panel of \Fig{fig:Eta2Pi0RatioMult_and_R2individual}. The uncertainty band for the generator curves is evaluated by taking the difference between the mean value and the largest and smallest value in the inspected \pT interval.
	A dependence of the \etopi~ratio as a function of the normalized charged-particle multiplicity is observed for the low \pT interval 1~$<$~\pT~$<$~4~\GeVc: an enhancement of the double ratio can be seen at low multiplicity while a suppression is visible for high multiplicities. The observed low \pT suppression at high multiplicity (0-0.1\%) is about 8\% with a significance of about 3.1$\sigma$. 
	The \etopi~ratio in the high-\pT interval is in agreement with unity over the full multiplicity range, taking into account statistical and systematic uncertainties.
	
	PYTHIA8 Ropes and EPOS LHC describe the behavior qualitatively, however with different magnitudes. The PYTHIA8 Monash tune shows the same trend as the Ropes variant as shown in~\cite{ALICE:LNM}. In PYTHIA8, the dependence of the \etopi~ratio is driven by an enhancement of \piz mesons from feed-down of heavier particles, primarily $\rho^{\pm}$, $\omega$ and \et mesons, with rising multiplicity. The fraction of these non-prompt (originating from hadronic decays with decay vertices indistinguishable from the primary vertex) \piz is largest at low \pT, resulting in the multiplicity dependence seen in \Fig{fig:Eta2Pi0RatioMult_and_R2individual}{ (right)} for the low-\pT interval, while the high-\pT interval is nearly unaffected. A similar effect is also seen in the EPOS model. However, in contrast to PYTHIA8, where the ratio of prompt \et to \piz is approximately constant with multiplicity, EPOS shows a more pronounced dependence of the prompt and non-prompt \etopi~ratio with multiplicity.\\
	The results of the multiplicity dependence of the \etopi~ratio can be compared to the $K^{\pm}/\pi^{\pm}$~ratio due to the similar masses of the \et and the \piz to the charged kaons and charged pions, respectively. The $K^{\pm}/\pi^{\pm}$~ratio as a function of the charged-particle multiplicity was reported in~\cite{Acharya:2020zji} where a small enhancement of the \pT-integrated $K^{\pm}/\pi^{\pm}$~ratio with rising multiplicity was found. However, for comparable \pT intervals (0.5~$<$~\pT~$<$~0.55~\GeVc, 2.4~$<$~\pT~$<$~2.6~\GeVc) as presented in this paper for the \etopi~ratio, no significant modification of the $K^{\pm}/\pi^{\pm}$~ratio is observed. These results suggest that the enhancement for the kaon, containing a strange quark, is stronger than for the \et, which only contains hidden strangeness. 
	

	\section{Conclusion}
	\label{sec:conclusions}
	
	We have presented the measurement of \piz and \et meson production at midrapidity in inelastic \pp collisions at \sT as well as the production of these mesons as a function of the charged-particle multiplicity. The \pT invariant differential cross sections are extracted in the kinematic range of $|y|<$ 0.8 and 0.2~$<$~\pT~$<$~200~\GeVc for the \piz and
	0.4~$<$~\pT~$<$~50~\GeVc for the \et. These results provide constraints to PDF and FF over an unprecedented kinematic range. A violation of \mT scaling is confirmed at \sT for \pT~$<$~4~\GeVc as observed at lower energies, as well as a hint of \mT scaling violation at higher transverse momenta. The data are in agreement with \xT scaling predictions for LHC energies over the \xT range from 6$\times$10$^{-4}$ to 3$\times$10$^{-2}$. The exponent $n$ for \piz \xT spectra agrees with the values obtained at LHC energies for charged pions and is about 20\% smaller than the one at RHIC energies.
	
	Next-to-leading order pQCD calculations using the NNFF1.0 or the BDSS FF describe the \piz transverse momentum spectrum better than calculations using the older DSS14 FF. Calculations for the \et meson based on pre-LHC FF (AESSS) overestimate the data over the complete \pT range. Predictions from PYTHIA8 overestimate the \piz spectrum and miss the slope and the yield of the \et spectrum. The \piz and \et spectrum from EPOS LHC overestimate the data by about 10--20\% at \pT = 2 \GeVc, while at lower \pT the deviation becomes larger.
	
	The transverse momentum spectra show a hardening with increasing charged-particle multiplicity, as previously observed for other particle species. PYTHIA8 qualitatively describes the dependence of the spectra on the charged-particle multiplicity, while EPOS fails to describe the spectra above \pT $\approx$ 3 \GeVc.
	The \etopi~ratio shows a slight dependence on the charged-particle multiplicity. In high multiplicity events, the \etopi~ratio shows a depletion of up to 10\% at low \pT with a significance of about 3.1$\sigma$. For \pT $>$ 4 \GeVc, no such dependence was found. This dependence is attributed to contributions from feed-down from heavier particles into the \piz spectrum, which is most relevant at low transverse momenta. Hence, both PYTHIA8 and EPOS LHC are able to describe this dependence qualitatively.

	
	
	\newenvironment{acknowledgement}{\relax}{\relax}
	\begin{acknowledgement}
		\section*{Acknowledgements}
		We would like to thank W. Vogelsang for providing the NLO pQCD calculations used in this article.

The ALICE Collaboration would like to thank all its engineers and technicians for their invaluable contributions to the construction of the experiment and the CERN accelerator teams for the outstanding performance of the LHC complex.
The ALICE Collaboration gratefully acknowledges the resources and support provided by all Grid centres and the Worldwide LHC Computing Grid (WLCG) collaboration.
The ALICE Collaboration acknowledges the following funding agencies for their support in building and running the ALICE detector:
A. I. Alikhanyan National Science Laboratory (Yerevan Physics Institute) Foundation (ANSL), State Committee of Science and World Federation of Scientists (WFS), Armenia;
Austrian Academy of Sciences, Austrian Science Fund (FWF): [M 2467-N36] and Nationalstiftung f\"{u}r Forschung, Technologie und Entwicklung, Austria;
Ministry of Communications and High Technologies, National Nuclear Research Center, Azerbaijan;
Conselho Nacional de Desenvolvimento Cient\'{\i}fico e Tecnol\'{o}gico (CNPq), Financiadora de Estudos e Projetos (Finep), Funda\c{c}\~{a}o de Amparo \`{a} Pesquisa do Estado de S\~{a}o Paulo (FAPESP) and Universidade Federal do Rio Grande do Sul (UFRGS), Brazil;
Bulgarian Ministry of Education and Science, within the National Roadmap for Research Infrastructures 2020-2027 (object CERN), Bulgaria;
Ministry of Education of China (MOEC) , Ministry of Science \& Technology of China (MSTC) and National Natural Science Foundation of China (NSFC), China;
Ministry of Science and Education and Croatian Science Foundation, Croatia;
Centro de Aplicaciones Tecnol\'{o}gicas y Desarrollo Nuclear (CEADEN), Cubaenerg\'{\i}a, Cuba;
Ministry of Education, Youth and Sports of the Czech Republic, Czech Republic;
The Danish Council for Independent Research | Natural Sciences, the VILLUM FONDEN and Danish National Research Foundation (DNRF), Denmark;
Helsinki Institute of Physics (HIP), Finland;
Commissariat \`{a} l'Energie Atomique (CEA) and Institut National de Physique Nucl\'{e}aire et de Physique des Particules (IN2P3) and Centre National de la Recherche Scientifique (CNRS), France;
Bundesministerium f\"{u}r Bildung und Forschung (BMBF) and GSI Helmholtzzentrum f\"{u}r Schwerionenforschung GmbH, Germany;
General Secretariat for Research and Technology, Ministry of Education, Research and Religions, Greece;
National Research, Development and Innovation Office, Hungary;
Department of Atomic Energy Government of India (DAE), Department of Science and Technology, Government of India (DST), University Grants Commission, Government of India (UGC) and Council of Scientific and Industrial Research (CSIR), India;
National Research and Innovation Agency - BRIN, Indonesia;
Istituto Nazionale di Fisica Nucleare (INFN), Italy;
Japanese Ministry of Education, Culture, Sports, Science and Technology (MEXT) and Japan Society for the Promotion of Science (JSPS) KAKENHI, Japan;
Consejo Nacional de Ciencia (CONACYT) y Tecnolog\'{i}a, through Fondo de Cooperaci\'{o}n Internacional en Ciencia y Tecnolog\'{i}a (FONCICYT) and Direcci\'{o}n General de Asuntos del Personal Academico (DGAPA), Mexico;
Nederlandse Organisatie voor Wetenschappelijk Onderzoek (NWO), Netherlands;
The Research Council of Norway, Norway;
Pontificia Universidad Cat\'{o}lica del Per\'{u}, Peru;
Ministry of Science and Higher Education, National Science Centre and WUT ID-UB, Poland;
Korea Institute of Science and Technology Information and National Research Foundation of Korea (NRF), Republic of Korea;
Ministry of Education and Scientific Research, Institute of Atomic Physics, Ministry of Research and Innovation and Institute of Atomic Physics and Universitatea Nationala de Stiinta si Tehnologie Politehnica Bucuresti, Romania;
Ministry of Education, Science, Research and Sport of the Slovak Republic, Slovakia;
National Research Foundation of South Africa, South Africa;
Swedish Research Council (VR) and Knut \& Alice Wallenberg Foundation (KAW), Sweden;
European Organization for Nuclear Research, Switzerland;
Suranaree University of Technology (SUT), National Science and Technology Development Agency (NSTDA) and National Science, Research and Innovation Fund (NSRF via PMU-B B05F650021), Thailand;
Turkish Energy, Nuclear and Mineral Research Agency (TENMAK), Turkey;
National Academy of  Sciences of Ukraine, Ukraine;
Science and Technology Facilities Council (STFC), United Kingdom;
National Science Foundation of the United States of America (NSF) and United States Department of Energy, Office of Nuclear Physics (DOE NP), United States of America.
In addition, individual groups or members have received support from:
Czech Science Foundation (grant no. 23-07499S), Czech Republic;
FORTE project, reg.\ no.\ CZ.02.01.01/00/22\_008/0004632, Czech Republic, co-funded by the European Union, Czech Republic;
European Research Council (grant no. 950692), European Union;
ICSC - Centro Nazionale di Ricerca in High Performance Computing, Big Data and Quantum Computing, European Union - NextGenerationEU;
Academy of Finland (Center of Excellence in Quark Matter) (grant nos. 346327, 346328), Finland;
Deutsche Forschungs Gemeinschaft (DFG, German Research Foundation) ``Neutrinos and Dark Matter in Astro- and Particle Physics'' (grant no. SFB 1258), Germany.

	\end{acknowledgement}
	
	\bibliographystyle{utphys}   
	\bibliography{bibliography}
	
	\newpage
	\appendix
	
	%
	%
	
	\section{The ALICE Collaboration}
	\label{app:collab}
\begin{flushleft} 
\small

S.~Acharya\,\orcidlink{0000-0002-9213-5329}\,$^{\rm 126}$, 
A.~Agarwal$^{\rm 134}$, 
G.~Aglieri Rinella\,\orcidlink{0000-0002-9611-3696}\,$^{\rm 32}$, 
L.~Aglietta\,\orcidlink{0009-0003-0763-6802}\,$^{\rm 24}$, 
M.~Agnello\,\orcidlink{0000-0002-0760-5075}\,$^{\rm 29}$, 
N.~Agrawal\,\orcidlink{0000-0003-0348-9836}\,$^{\rm 25}$, 
Z.~Ahammed\,\orcidlink{0000-0001-5241-7412}\,$^{\rm 134}$, 
S.~Ahmad\,\orcidlink{0000-0003-0497-5705}\,$^{\rm 15}$, 
S.U.~Ahn\,\orcidlink{0000-0001-8847-489X}\,$^{\rm 71}$, 
I.~Ahuja\,\orcidlink{0000-0002-4417-1392}\,$^{\rm 36}$
% ...

\section*{Affiliation Notes}

$^{\rm I}$ Deceased\\
$^{\rm II}$ Also at: Max-Planck-Institut fur Physik, Munich, Germany\\
$^{\rm III}$ Also at: Italian National Agency for New Technologies, Energy and Sustainable Economic Development (ENEA), Bologna, Italy\\
$^{\rm IV}$ Also at: Dipartimento DET del Politecnico di Torino, Turin, Italy\\
$^{\rm V}$ Also at: Department of Applied Physics, Aligarh Muslim University, Aligarh, India\\
$^{\rm VI}$ Also at: Institute of Theoretical Physics, University of Wroclaw, Poland\\
$^{\rm VII}$ Also at: Facultad de Ciencias, Universidad Nacional Autónoma de México, Mexico City, Mexico\\

\section*{Collaboration Institutes}

$^{1}$ A.I. Alikhanyan National Science Laboratory (Yerevan Physics Institute) Foundation, Yerevan, Armenia\\
$^{2}$ AGH University of Krakow, Cracow, Poland\\
$^{3}$ Bogolyubov Institute for Theoretical Physics, National Academy of Sciences of Ukraine, Kiev, Ukraine\\
$^{4}$ Bose Institute, Department of Physics  and Centre for Astroparticle Physics and Space Science (CAPSS), Kolkata, India\\
$^{5}$ California Polytechnic State University, San Luis Obispo, California, United States\\
$^{6}$ Central China Normal University, Wuhan, China\\
$^{7}$ Centro de Aplicaciones Tecnol\'{o}gicas y Desarrollo Nuclear (CEADEN), Havana, Cuba\\
$^{8}$ Centro de Investigaci\'{o}n y de Estudios Avanzados (CINVESTAV), Mexico City and M\'{e}rida, Mexico\\
$^{9}$ Chicago State University, Chicago, Illinois, United States\\
$^{10}$ China Institute of Atomic Energy, Beijing, China\\
$^{11}$ China University of Geosciences, Wuhan, China\\
$^{12}$ Chungbuk National University, Cheongju, Republic of Korea\\
$^{13}$ Comenius University Bratislava, Faculty of Mathematics, Physics and Informatics, Bratislava, Slovak Republic\\
$^{14}$ Creighton University, Omaha, Nebraska, United States\\
$^{15}$ Department of Physics, Aligarh Muslim University, Aligarh, India\\
$^{16}$ Department of Physics, Pusan National University, Pusan, Republic of Korea\\
$^{17}$ Department of Physics, Sejong University, Seoul, Republic of Korea\\
$^{18}$ Department of Physics, University of California, Berkeley, California, United States\\
$^{19}$ Department of Physics, University of Oslo, Oslo, Norway\\
$^{20}$ Department of Physics and Technology, University of Bergen, Bergen, Norway\\
$^{21}$ Dipartimento di Fisica, Universit\`{a} di Pavia, Pavia, Italy\\
$^{22}$ Dipartimento di Fisica dell'Universit\`{a} and Sezione INFN, Cagliari, Italy\\
$^{23}$ Dipartimento di Fisica dell'Universit\`{a} and Sezione INFN, Trieste, Italy\\
$^{24}$ Dipartimento di Fisica dell'Universit\`{a} and Sezione INFN, Turin, Italy\\
$^{25}$ Dipartimento di Fisica e Astronomia dell'Universit\`{a} and Sezione INFN, Bologna, Italy\\
$^{26}$ Dipartimento di Fisica e Astronomia dell'Universit\`{a} and Sezione INFN, Catania, Italy\\
$^{27}$ Dipartimento di Fisica e Astronomia dell'Universit\`{a} and Sezione INFN, Padova, Italy\\
$^{28}$ Dipartimento di Fisica `E.R.~Caianiello' dell'Universit\`{a} and Gruppo Collegato INFN, Salerno, Italy\\
$^{29}$ Dipartimento DISAT del Politecnico and Sezione INFN, Turin, Italy\\
$^{30}$ Dipartimento di Scienze MIFT, Universit\`{a} di Messina, Messina, Italy\\
$^{31}$ Dipartimento Interateneo di Fisica `M.~Merlin' and Sezione INFN, Bari, Italy\\
$^{32}$ European Organization for Nuclear Research (CERN), Geneva, Switzerland\\
$^{33}$ Faculty of Electrical Engineering, Mechanical Engineering and Naval Architecture, University of Split, Split, Croatia\\
$^{34}$ Faculty of Nuclear Sciences and Physical Engineering, Czech Technical University in Prague, Prague, Czech Republic\\
$^{35}$ Faculty of Physics, Sofia University, Sofia, Bulgaria\\
$^{36}$ Faculty of Science, P.J.~\v{S}af\'{a}rik University, Ko\v{s}ice, Slovak Republic\\
$^{37}$ Faculty of Technology, Environmental and Social Sciences, Bergen, Norway\\
$^{38}$ Frankfurt Institute for Advanced Studies, Johann Wolfgang Goethe-Universit\"{a}t Frankfurt, Frankfurt, Germany\\
$^{39}$ Fudan University, Shanghai, China\\
$^{40}$ Gangneung-Wonju National University, Gangneung, Republic of Korea\\
$^{41}$ Gauhati University, Department of Physics, Guwahati, India\\
$^{42}$ Helmholtz-Institut f\"{u}r Strahlen- und Kernphysik, Rheinische Friedrich-Wilhelms-Universit\"{a}t Bonn, Bonn, Germany\\
$^{43}$ Helsinki Institute of Physics (HIP), Helsinki, Finland\\
$^{44}$ High Energy Physics Group,  Universidad Aut\'{o}noma de Puebla, Puebla, Mexico\\
$^{45}$ Horia Hulubei National Institute of Physics and Nuclear Engineering, Bucharest, Romania\\
$^{46}$ HUN-REN Wigner Research Centre for Physics, Budapest, Hungary\\
$^{47}$ Indian Institute of Technology Bombay (IIT), Mumbai, India\\
$^{48}$ Indian Institute of Technology Indore, Indore, India\\
$^{49}$ INFN, Laboratori Nazionali di Frascati, Frascati, Italy\\
$^{50}$ INFN, Sezione di Bari, Bari, Italy\\
$^{51}$ INFN, Sezione di Bologna, Bologna, Italy\\
$^{52}$ INFN, Sezione di Cagliari, Cagliari, Italy\\
$^{53}$ INFN, Sezione di Catania, Catania, Italy\\
$^{54}$ INFN, Sezione di Padova, Padova, Italy\\
$^{55}$ INFN, Sezione di Pavia, Pavia, Italy\\
$^{56}$ INFN, Sezione di Torino, Turin, Italy\\
$^{57}$ INFN, Sezione di Trieste, Trieste, Italy\\
$^{58}$ Inha University, Incheon, Republic of Korea\\
$^{59}$ Institute for Gravitational and Subatomic Physics (GRASP), Utrecht University/Nikhef, Utrecht, Netherlands\\
$^{60}$ Institute of Experimental Physics, Slovak Academy of Sciences, Ko\v{s}ice, Slovak Republic\\
$^{61}$ Institute of Physics, Homi Bhabha National Institute, Bhubaneswar, India\\
$^{62}$ Institute of Physics of the Czech Academy of Sciences, Prague, Czech Republic\\
$^{63}$ Institute of Space Science (ISS), Bucharest, Romania\\
$^{64}$ Institut f\"{u}r Kernphysik, Johann Wolfgang Goethe-Universit\"{a}t Frankfurt, Frankfurt, Germany\\
$^{65}$ Instituto de Ciencias Nucleares, Universidad Nacional Aut\'{o}noma de M\'{e}xico, Mexico City, Mexico\\
$^{66}$ Instituto de F\'{i}sica, Universidade Federal do Rio Grande do Sul (UFRGS), Porto Alegre, Brazil\\
$^{67}$ Instituto de F\'{\i}sica, Universidad Nacional Aut\'{o}noma de M\'{e}xico, Mexico City, Mexico\\
$^{68}$ iThemba LABS, National Research Foundation, Somerset West, South Africa\\
$^{69}$ Jeonbuk National University, Jeonju, Republic of Korea\\
$^{70}$ Johann-Wolfgang-Goethe Universit\"{a}t Frankfurt Institut f\"{u}r Informatik, Fachbereich Informatik und Mathematik, Frankfurt, Germany\\
$^{71}$ Korea Institute of Science and Technology Information, Daejeon, Republic of Korea\\
$^{72}$ Laboratoire de Physique Subatomique et de Cosmologie, Universit\'{e} Grenoble-Alpes, CNRS-IN2P3, Grenoble, France\\
$^{73}$ Lawrence Berkeley National Laboratory, Berkeley, California, United States\\
$^{74}$ Lund University Department of Physics, Division of Particle Physics, Lund, Sweden\\
$^{75}$ Nagasaki Institute of Applied Science, Nagasaki, Japan\\
$^{76}$ Nara Women{'}s University (NWU), Nara, Japan\\
$^{77}$ National and Kapodistrian University of Athens, School of Science, Department of Physics , Athens, Greece\\
$^{78}$ National Centre for Nuclear Research, Warsaw, Poland\\
$^{79}$ National Institute of Science Education and Research, Homi Bhabha National Institute, Jatni, India\\
$^{80}$ National Nuclear Research Center, Baku, Azerbaijan\\
$^{81}$ National Research and Innovation Agency - BRIN, Jakarta, Indonesia\\
$^{82}$ Niels Bohr Institute, University of Copenhagen, Copenhagen, Denmark\\
$^{83}$ Nikhef, National institute for subatomic physics, Amsterdam, Netherlands\\
$^{84}$ Nuclear Physics Group, STFC Daresbury Laboratory, Daresbury, United Kingdom\\
$^{85}$ Nuclear Physics Institute of the Czech Academy of Sciences, Husinec-\v{R}e\v{z}, Czech Republic\\
$^{86}$ Oak Ridge National Laboratory, Oak Ridge, Tennessee, United States\\
$^{87}$ Ohio State University, Columbus, Ohio, United States\\
$^{88}$ Physics department, Faculty of science, University of Zagreb, Zagreb, Croatia\\
$^{89}$ Physics Department, Panjab University, Chandigarh, India\\
$^{90}$ Physics Department, University of Jammu, Jammu, India\\
$^{91}$ Physics Program and International Institute for Sustainability with Knotted Chiral Meta Matter (WPI-SKCM$^{2}$), Hiroshima University, Hiroshima, Japan\\
$^{92}$ Physikalisches Institut, Eberhard-Karls-Universit\"{a}t T\"{u}bingen, T\"{u}bingen, Germany\\
$^{93}$ Physikalisches Institut, Ruprecht-Karls-Universit\"{a}t Heidelberg, Heidelberg, Germany\\
$^{94}$ Physik Department, Technische Universit\"{a}t M\"{u}nchen, Munich, Germany\\
$^{95}$ Politecnico di Bari and Sezione INFN, Bari, Italy\\
$^{96}$ Research Division and ExtreMe Matter Institute EMMI, GSI Helmholtzzentrum f\"ur Schwerionenforschung GmbH, Darmstadt, Germany\\
$^{97}$ Saga University, Saga, Japan\\
$^{98}$ Saha Institute of Nuclear Physics, Homi Bhabha National Institute, Kolkata, India\\
$^{99}$ School of Physics and Astronomy, University of Birmingham, Birmingham, United Kingdom\\
$^{100}$ Secci\'{o}n F\'{\i}sica, Departamento de Ciencias, Pontificia Universidad Cat\'{o}lica del Per\'{u}, Lima, Peru\\
$^{101}$ Stefan Meyer Institut f\"{u}r Subatomare Physik (SMI), Vienna, Austria\\
$^{102}$ SUBATECH, IMT Atlantique, Nantes Universit\'{e}, CNRS-IN2P3, Nantes, France\\
$^{103}$ Sungkyunkwan University, Suwon City, Republic of Korea\\
$^{104}$ Suranaree University of Technology, Nakhon Ratchasima, Thailand\\
$^{105}$ Technical University of Ko\v{s}ice, Ko\v{s}ice, Slovak Republic\\
$^{106}$ The Henryk Niewodniczanski Institute of Nuclear Physics, Polish Academy of Sciences, Cracow, Poland\\
$^{107}$ The University of Texas at Austin, Austin, Texas, United States\\
$^{108}$ Universidad Aut\'{o}noma de Sinaloa, Culiac\'{a}n, Mexico\\
$^{109}$ Universidade de S\~{a}o Paulo (USP), S\~{a}o Paulo, Brazil\\
$^{110}$ Universidade Estadual de Campinas (UNICAMP), Campinas, Brazil\\
$^{111}$ Universidade Federal do ABC, Santo Andre, Brazil\\
$^{112}$ Universitatea Nationala de Stiinta si Tehnologie Politehnica Bucuresti, Bucharest, Romania\\
$^{113}$ University of Cape Town, Cape Town, South Africa\\
$^{114}$ University of Derby, Derby, United Kingdom\\
$^{115}$ University of Houston, Houston, Texas, United States\\
$^{116}$ University of Jyv\"{a}skyl\"{a}, Jyv\"{a}skyl\"{a}, Finland\\
$^{117}$ University of Kansas, Lawrence, Kansas, United States\\
$^{118}$ University of Liverpool, Liverpool, United Kingdom\\
$^{119}$ University of Science and Technology of China, Hefei, China\\
$^{120}$ University of South-Eastern Norway, Kongsberg, Norway\\
$^{121}$ University of Tennessee, Knoxville, Tennessee, United States\\
$^{122}$ University of the Witwatersrand, Johannesburg, South Africa\\
$^{123}$ University of Tokyo, Tokyo, Japan\\
$^{124}$ University of Tsukuba, Tsukuba, Japan\\
$^{125}$ Universit\"{a}t M\"{u}nster, Institut f\"{u}r Kernphysik, M\"{u}nster, Germany\\
$^{126}$ Universit\'{e} Clermont Auvergne, CNRS/IN2P3, LPC, Clermont-Ferrand, France\\
$^{127}$ Universit\'{e} de Lyon, CNRS/IN2P3, Institut de Physique des 2 Infinis de Lyon, Lyon, France\\
$^{128}$ Universit\'{e} de Strasbourg, CNRS, IPHC UMR 7178, F-67000 Strasbourg, France, Strasbourg, France\\
$^{129}$ Universit\'{e} Paris-Saclay, Centre d'Etudes de Saclay (CEA), IRFU, D\'{e}partment de Physique Nucl\'{e}aire (DPhN), Saclay, France\\
$^{130}$ Universit\'{e}  Paris-Saclay, CNRS/IN2P3, IJCLab, Orsay, France\\
$^{131}$ Universit\`{a} degli Studi di Foggia, Foggia, Italy\\
$^{132}$ Universit\`{a} del Piemonte Orientale, Vercelli, Italy\\
$^{133}$ Universit\`{a} di Brescia, Brescia, Italy\\
$^{134}$ Variable Energy Cyclotron Centre, Homi Bhabha National Institute, Kolkata, India\\
$^{135}$ Warsaw University of Technology, Warsaw, Poland\\
$^{136}$ Wayne State University, Detroit, Michigan, United States\\
$^{137}$ Yale University, New Haven, Connecticut, United States\\
$^{138}$ Yildiz Technical University, Istanbul, Turkey\\
$^{139}$ Yonsei University, Seoul, Republic of Korea\\
$^{140}$ Affiliated with an institute covered by a cooperation agreement with CERN\\
$^{141}$ Affiliated with an international laboratory covered by a cooperation agreement with CERN.\\

\end{flushleft}

\end{document}